\documentclass[twocolumn]{aastex63}
\accepted{for Publication in ApJ}

\usepackage{natbib,aas_macros,amsmath}
\usepackage{multirow,color}
\usepackage{natbib}

\newcommand{\m}[1]{\mathrm{#1}}

\newcommand{\redc}[1]{\textcolor{black}{#1}}
\newcommand{\redcc}[1]{\textcolor{black}{#1}}

\usepackage{booktabs}

\begin{document}
\shortauthors{Harikane et al.}

\shorttitle{
JWST/NIRSpec Faint AGNs at $z=4-7$
}

\title{
A JWST/NIRSpec First Census of Broad-Line AGNs at $\mathbf{z=4-7}$:\\ Detection of 10 Faint AGNs with $\mathbf{M_\mathrm{BH}\sim10^6-10^8\ M_\odot}$ and Their Host Galaxy Properties
}

\email{hari@icrr.u-tokyo.ac.jp}
\author[0000-0002-6047-430X]{Yuichi Harikane}
\affiliation{Institute for Cosmic Ray Research, The University of Tokyo, 5-1-5 Kashiwanoha, Kashiwa, Chiba 277-8582, Japan}

\author[0000-0003-3817-8739]{Yechi Zhang}
\affiliation{Institute for Cosmic Ray Research, The University of Tokyo, 5-1-5 Kashiwanoha, Kashiwa, Chiba 277-8582, Japan}
\affiliation{Department of Astronomy, Graduate School of Science, The University of Tokyo, 7-3-1 Hongo, Bunkyo, Tokyo 113-0033, Japan}

\author[0000-0003-2965-5070]{Kimihiko Nakajima}
\affiliation{National Astronomical Observatory of Japan, 2-21-1 Osawa, Mitaka, Tokyo 181-8588, Japan}

\author[0000-0002-1049-6658]{Masami Ouchi}
\affiliation{National Astronomical Observatory of Japan, 2-21-1 Osawa, Mitaka, Tokyo 181-8588, Japan}
\affiliation{Institute for Cosmic Ray Research, The University of Tokyo, 5-1-5 Kashiwanoha, Kashiwa, Chiba 277-8582, Japan}
\affiliation{Kavli Institute for the Physics and Mathematics of the Universe (WPI), University of Tokyo, Kashiwa, Chiba 277-8583, Japan}

\author[0000-0001-7730-8634]{Yuki Isobe}
\affiliation{Institute for Cosmic Ray Research, The University of Tokyo, 5-1-5 Kashiwanoha, Kashiwa, Chiba 277-8582, Japan}
\affiliation{Department of Physics, Graduate School of Science, The University of Tokyo, 7-3-1 Hongo, Bunkyo, Tokyo 113-0033, Japan}

\author[0000-0001-9011-7605]{Yoshiaki Ono}
\affiliation{Institute for Cosmic Ray Research, The University of Tokyo, 5-1-5 Kashiwanoha, Kashiwa, Chiba 277-8582, Japan}

\author[0000-0002-5816-4660]{Shun Hatano}
\affiliation{National Astronomical Observatory of Japan, 2-21-1 Osawa, Mitaka, Tokyo 181-8588, Japan}
\affiliation{Department of Astronomical Science, The Graduate University for Advanced Studies, SOKENDAI, 2-21-1 Osawa, Mitaka, Tokyo, 181-8588, Japan}

\author[0000-0002-5768-8235]{Yi Xu}
\affiliation{Institute for Cosmic Ray Research, The University of Tokyo, 5-1-5 Kashiwanoha, Kashiwa, Chiba 277-8582, Japan}
\affiliation{Department of Astronomy, Graduate School of Science, The University of Tokyo, 7-3-1 Hongo, Bunkyo, Tokyo 113-0033, Japan}

\author[0009-0008-0167-5129]{Hiroya Umeda}
\affiliation{Institute for Cosmic Ray Research, The University of Tokyo, 5-1-5 Kashiwanoha, Kashiwa, Chiba 277-8582, Japan}
\affiliation{Department of Physics, Graduate School of Science, The University of Tokyo, 7-3-1 Hongo, Bunkyo, Tokyo 113-0033, Japan}

\begin{abstract}
We present a first statistical sample of faint type-1 AGNs at $z>4$ identified by JWST/NIRSpec deep spectroscopy.
Among the 185 galaxies at $z_\mathrm{spec}=3.8-8.9$ confirmed with NIRSpec, our systematic search for broad-line emission reveals 10 type-1 AGNs at $z=4.015-6.936$ whose broad component is only seen in the permitted H$\alpha$ line and not in the forbidden {\sc[Oiii]}$\lambda$5007 line that is detected with greater significance than H$\alpha$.
The broad H$\alpha$ line widths of $\mathrm{FWHM}\simeq1000-6000\ \mathrm{km\ s^{-1}}$ suggest that the AGNs have low-mass black holes with $M_\mathrm{BH}\sim10^6-10^8\ M_\odot$, remarkably lower than those of low-luminosity quasars previously identified at $z>4$ with ground-based telescopes.
JWST and HST high-resolution images reveal that the majority of them show extended morphologies indicating significant contribution to the total lights from their host galaxies, except for three compact objects two of which show red SEDs, probably in a transition phase from faint AGNs to low luminosity quasars.
Careful AGN-host decomposition analyses show that their host's stellar masses are systematically lower than the local relation between the black hole mass and the stellar mass, implying a fast black hole growth consistent with predictions from theoretical simulations.
A high fraction of the broad-line AGNs ($\sim5\%$), higher than $z\sim0$, indicates that a number density of such faint AGNs is higher than an extrapolation of the quasar luminosity function, implying a large population of AGNs in the early universe.
Such faint AGNs contribute to cosmic reionization, while the total contribution is not large, up to $\sim50\%$ at $z\sim6$, because of their faint nature.
\end{abstract}


\section{Introduction}\label{ss_intro}

It has been known for over two decades that the mass of supermassive black holes (SMBHs) at $z\sim0$ are tightly correlated with the bulge properties of their host galaxies, such as velocity dispersion and bulge mass \citep[e.g.,][]{1998AJ....115.2285M,2000ApJ...539L..13G,2013ARA&A..51..511K,2015ApJ...813...82R}.
This tight correlation indicates the strong connection between the growth of central SMBHs and their host galaxies, known as the galaxy-SMBH coevolution.
Although the underlying physical mechanisms are still under debate, theoretical models suggest that feedback of active galactic nuclei (AGNs) connected to galaxy's merger histories play an important role \citep[e.g.,][]{2004ApJ...600..580G,2005Natur.433..604D,2006ApJS..163....1H,2007ApJ...665..187L}.
Since theoretical models usually make predictions for the time evolution of the systems \citep[e.g.,][]{2017MNRAS.472L.109A,2021ApJ...907...74T,2021MNRAS.507....1V,2022MNRAS.514.5583Z,2022MNRAS.511.3751H,2022MNRAS.511..616T,2022ApJ...927..237I,2022ApJ...935..140H,2023MNRAS.518.2123Z,2023arXiv230308150Z}, observations of both SMBHs and their host galaxies over cosmic time are essential to test and/or refine our current understandings of their build-up \citep[e.g.,][]{2017PASA...34...22G,2017PASA...34...31V}.

Observations of AGNs at the high-redshift universe are thus crucial to understanding the evolution of the SMBH growth in cosmic history, but previous observations have been limited to bright quasars identified in surveys with ground-based telescopes (e.g., \citealt{2016ApJS..227...11B,2019ApJ...873...35S}, see \citealt{2020ARA&A..58...27I} for a review) including the most distant quasars at $z>7.5$ \citep{2018Natur.553..473B,2020ApJ...897L..14Y,2021ApJ...907L...1W}.
Recent surveys using 4-8 m-class telescopes including the Subaru/Hyper Suprime-Cam survey \citep{2018PASJ...70S...4A} are finding low-luminosity quasars at $z\sim4-7$ \citep[e.g.,][]{2010AJ....139..906W,2015ApJ...798...28K,2016ApJ...828...26M,2017ApJ...847L..15O,2018PASJ...70S..34A,2020ApJ...904...89N}, but these low-luminosity quasars have moderately massive black hole masses with $M_\m{BH}\sim10^8-10^9\ M_\odot$ compared to AGNs found in the local universe \cite[e.g.,][]{2019ApJS..243...21L}.
Although intensive spectroscopic observations targeting high redshift galaxies at $z\gtrsim7$ have identified high ionization emission lines indicative of AGN activity \citep[e.g.,][]{2016ApJ...827L..14T,2017ApJ...851...40L,2018MNRAS.479.1180M}, the physical properties of their central SMBHs are unclear.

The James Webb Space Telescope (JWST) was launched at the end of 2021 and started its operation in early 2022.
JWST observations are now beginning to improve our understanding of the connection between high redshift AGNs and their host galaxies, especially in the low-luminosity and low-mass regimes.
JWST/NIRCam deep and high-resolution images allow us to detect stellar lights from host galaxies of low-luminosity quasars at $z\sim6$ \citep{2022arXiv221114329D}, and to identify the least-massive black hole candidate at $z\sim5$ \citep{2023ApJ...942L..17O} and a triply-lensed red-quasar candidate at $z\sim8$ \citep{2022arXiv221210531F}.
JWST/NIRSpec deep-spectroscopic observations have confirmed the candidate in \citet{2023ApJ...942L..17O} at $z_\m{spec}=5.2$ by detecting a broad H$\alpha$ emission line that indicates a black hole mass of $M_\m{BH}\sim10^7\ M_\odot$, with an additional finding of a red AGN at $z_\m{spec}=5.6$ \citep{2023ApJ...954L...4K}.
NIRSpec spectroscopy has also identified a supermassive black hole in a quiescent galaxy at $z=4.658$ \citep{2023Natur.619..716C}. 
NIRSpec IFU observations have revealed broad H$\alpha$ and H$\beta$ emission lines in a galaxy at $z=5.55$ \citep{2023arXiv230206647U}, which is interpreted as a type 1.8 AGN with a black hole mass of $M_\m{BH}\sim10^8\ M_\odot$.
Very recently, \citet{2023arXiv230308918L} have reported a broad H$\beta$ emission line in a galaxy at $z_\m{spec}=8.7$, suggesting a SMBH whose mass is $M_\m{BH}\sim10^7\ M_\odot$ 570 Myrs after the Big Bang \redcc{(see also \citealt{2023arXiv230515458B,2023arXiv230802750G,2023arXiv230905714G,2023arXiv230805735F,2023arXiv230811609F,2023arXiv230811610K,2023arXiv230607320L,2023arXiv230801230M,2023arXiv230605448M} for studies that appeared after our initial submission of this paper)}.

Motivated by these recent JWST findings of high redshift AGNs, we systematically search for broad-line AGNs at $z>4$ using the JWST/NIRSpec spectroscopic data presented in \citet{2023arXiv230112825N}.
Our sample of low-luminosity AGNs with low-mass SMBHs allows us to obtain the first statistical view on the physical properties of faint AGNs and their host galaxies at $z>4$ such as black hole masses, Eddington ratios, number densities, host's stellar mass, and morphologies, crucial to understand the galaxy-SMBH co-evolution in the early universe and implication for cosmic reionization.
Moreover, the black hole masses of our low-luminosity AGNs can be directly compared with AGNs at $z\sim0$, allowing us to understand the redshift evolution over cosmic time.

This paper is organized as follows.
Section \ref{ss_data} presents the JWST and HST observational data sets used in this study.
In Section \ref{ss_selection} we explain our systematic sample selection and the final AGN sample.
We show our main results in Section \ref{ss_result} and discuss the contribution to cosmic reionization and the nature of our AGNs in Section \ref{ss_dis}.
Section \ref{ss_summary} summarizes our findings.
Throughout this paper, we use the Planck cosmological parameter sets of the TT, TE, EE+lowP+lensing+BAO result \citep{2020A&A...641A...6P}: $\Omega_\m{m}=0.3111$, $\Omega_\Lambda=0.6899$, $\Omega_\m{b}=0.0489$, $h=0.6766$, and $\sigma_8=0.8102$.
All magnitudes are in the AB system \citep{1983ApJ...266..713O}.

\section{Observational Dataset}\label{ss_data}

\subsection{JWST/NIRSpec Spectra}
We use JWST/NIRSpec datasets reduced in \citet{2023arXiv230112825N}.
Here we briefly describe the observations and the data reduction.
Please see \citet{2023arXiv230112825N} for details.

The data sets used in this study were obtained in the Early Release Observations (EROs; \citealt{2022ApJ...936L..14P}) targeting the SMACS 0723 lensing cluster field (ERO-2736) and the Early Release Science (ERS) observations of GLASS (ERS-1324, PI: T. Treu; \citealt{2022arXiv220607978T}) and the Cosmic Evolution Early Release Science (CEERS; ERS-1345, PI: S. Finkelstein; \citealt{2022arXiv221105792F}).
The ERO data were taken in the medium resolution ($R\sim1000$) filter-grating pairs F170LP-G235M and F290LP-G395M covering the wavelength ranges of $1.7-3.1$ and $2.9-5.1$ $\mu$m, respectively.
The total exposure time of the ERO data is 4.86 hours for each filter-grating pair.
The GLASS data were taken with high resolution ($R\sim2700$) filter-grating pairs of F100LP-G140H, F170LP-G235H, and F290LP-G395H covering the wavelength ranges of $1.0-1.6$, $1.7-3.1$ and $2.9-5.1$ $\mu$m, respectively.
The total exposure time of the GLASS data is 4.9 hours for each filter-grating pair.
CEERS data were taken with the Prism ($R\sim100$) that covers $0.6-5.3$ and medium-resolution filter-grating pairs of F100LP-G140M, F170LP-G235M, and F290LP-G395M covering the wavelength ranges of $1.0-1.6$, $1.7-3.1$ and $2.9-5.1$ $\mu$m, respectively.
The total exposure time of the CEERS data is 0.86 hours for each filter-grating pair.
These data were reduced in \citet{2023arXiv230112825N} with the JWST pipeline version 1.8.5 with the Calibration Reference Data System (CRDS) context file of {\tt jwst\_1028.pmap or jwst\_1027.pmap} with additional processes improving the flux calibration, noise estimate, and the composition.
Reduced spectra are corrected for slit loss (see \citealt{2023arXiv230112825N} for details).
Finally, we obtain NIRSpec spectra of a total of 185 galaxies at $z_\m{spec}=3.8-8.9$.

\subsection{JWST/NIRCam and HST/ACS\&WFC3 Images}

For the GLASS field, we use the JWST/NIRCam and HST/ACS\&WFC3 images produced by the Ultradeep NIRSpec and NIRCam ObserVations before the Epoch of Reionization (UNCOVER) team.\footnote{https://jwst-uncover.github.io}
In the GLASS NIRSpec field around the center of the Abell 2744 lensing cluster, the JWST/NIRCam images were taken in the F115W, F150W, F200W, F277W, F356W, and F444W bands in the UNCOVER program (GO-2561, PIs: I. Labbe and R. Bezanson; \citealt{2022arXiv221204026B}), and the HST/ACS and WCS3 images were taken in the F435W, F606W, F814W, F105W, F125W, F140W, and F160W-bands in the Hubble Frontier Field program (PI: J. Lotz; \redc{\citealt{2017ApJ...837...97L}}) and the Beyond Ultra-deep Frontier Fields And Legacy Observations (BUFFALO) program (PI: C. Steinhardt; \citealt{2020ApJS..247...64S}).

For the CEERS field, we use the JWST/NIRCam and HST/ACS\&WFC3 official images produced by the CEERS team (version 0.5; \citealt{2022arXiv221102495B}).\footnote{https://ceers.github.io}
In the CEERS field, the JWST/NIRCam images were taken in the F115W, F150W, F200W, F277W, F356W, F410M, and F444W bands in the CEERS program, and the HST/ACS and WCS3 images were taken in the F606W, F814W, F105W, F125W, F140W, and F160W-bands in the Cosmic Assembly Near-infrared Deep Extragalactic Legacy Survey (CANDELS; \citealt{2011ApJS..197...35G}; \citealt{2011ApJS..197...36K}, see also \citealt{2022ApJ...928...52F}).
For part of the CEERS field where the official images are not available (i.e., fields whose data were taken in December 2022), we reduced the NIRCam data in the same manner as \citet{2023ApJS..265....5H} using the JWST pipeline version 1.8.5 with the CRDS context file of {\tt jwst\_1027.pmap}.

\section{Broad-Line AGN Selection}\label{ss_selection}

\subsection{Emission Line Fitting}\label{ss_line_fitting}

From 185 galaxies at $z_\m{spec}=3.8-8.9$ in \citet{2023arXiv230112825N}, we search for broad-line AGNs that show broad permitted lines such as H$\alpha$ and H$\beta$.
We fit rest-frame optical emission lines such as H$\beta$, {\sc[Oiii]}$\lambda\lambda$4959,5007, H$\alpha$, and {\sc[Nii]}$\lambda\lambda$6548,6584.
We use the line spread functions made from spectra of a planetary nebula in \citet{2023arXiv230106811I} for spectra taken with medium and high-resolution gratings.
For spectra taken with the Prism, we use the FWHM measured in \citet{2023arXiv230106811I} from the planetary nebula's spectra assuming a Gaussian profile because Prism's line spread function made from the spectra is severely contaminated by nearby lines due to its low resolution.

First, we fit the H$\beta$ and [OIII]$\lambda\lambda$4959,5007 lines with an assumed line ratio of $f_\m{\sc[OIII]5007}/f_\m{[OIII]4959}=2.98$ that is taken from theoretical calculations \citep{2000MNRAS.312..813S}.
Initially, we fit the emission lines with a single component.
Most sources can be fitted with a single component whose line width is narrow ($\m{FWHM}<500\ \m{km\ s^{-1}}$), but some sources show a weak moderately broad component in {\sc[Oiii]}$\lambda$5007 that may be due to outflow \redc{based on visual inspection (see \citealt{2023arXiv230607940Z} for statistics)}.
For such sources, we simultaneously fit the [OIII]$\lambda\lambda$4959,5007 with the narrow ($\m{FWHM_{narrow}}<500\ \m{km\ s^{-1}}$) and outflow ($\m{FWHM_{outflow}}>\m{FWHM_{narrow}}$) components.
Then we fit the H$\alpha$ and {\sc[Nii]}$\lambda\lambda$6548,6584 emission line with an assume line ratio of $f_\m{\sc[NII]6584}/f_\m{[NII]6548}=2.94$.
We find that the {\sc[Nii]} and most of the H$\alpha$ lines can be fitted with a single narrow component ($\m{FWHM}<500\ \m{km\ s^{-1}}$), but some sources show a broad component in H$\alpha$.
For such sources, we simultaneously fit the narrow ($\m{FWHM_{narrow}}<500\ \m{km\ s^{-1}}$) and broad ($\m{FWHM_{broad}}>500\ \m{km\ s^{-1}}$) components for H$\alpha$, and the narrow components for {\sc[Nii]}$\lambda\lambda$6548,6584.
For sources showing the outflow components in the {\sc[Oiii]} line, we add an outflow component ($\m{FWHM_{narrow}}<\m{FWHM_{outflow}}<\m{FWHM_{broad}}$) in H$\alpha$.

\subsection{Selection Criteria}

Based on these fitting results, we select type 1 AGNs showing broad emission only in a permitted line (i.e., H$\alpha$ \redc{and/or H$\beta$}).
We select sources that show
\begin{enumerate}
\item broad ($\m{FWHM}>1000\ \m{km\ s^{-1}}$) and significant ($\m{SNR}>5$) permitted H$\alpha$ and/or H$\beta$ emission line, and
\item narrow ($\m{FWHM}<\redc{700}\ \m{km\ s^{-1}}$) forbidden {\sc[Oiii]} and {\sc[Nii]} emission lines, even if an outflow component is seen.
\end{enumerate}
\redc{Here we evaluate the signal-to-noise ratio of the broad line using the ratio of the total broad-line flux to the error from our fitting results in Section \ref{ss_line_fitting}.}
For galaxies at $z\lesssim7$ whose H$\alpha$ line is detected, we select galaxies with the following criteria:
\begin{align}
&(\m{FWHM}_\m{H\alpha,broad}>1000\ \m{km\ s^{-1}})\ \land\\
&(\m{SNR}_\m{H\alpha,broad}>5)\ \land\\
&(\m{FWHM}_\m{[OIII],narrow}<700\ \m{km\ s^{-1}})\ \land\\
&((\m{FWHM}_\m{[NII]}<700\ \m{km\ s^{-1}}) \lor (\m{SNR}_\m{[NII]6584}<2))\ \land\\
&[(\m{FWHM}_\m{[OIII],outflow}<700\ \m{km\ s^{-1}})\ \lor\\
&(\m{SNR}_\m{[OIII],outflow}<2)]\ \land\\
&[(\m{FWHM}_\m{H\alpha,outflow}<700\ \m{km\ s^{-1}})\ \lor\\
&(\m{SNR}_\m{H\alpha,outflow}<2)].
\end{align}
For galaxies at $z\gtrsim7$ whose H$\alpha$ line is not detected but H$\beta$ is detected, we select galaxies with the following criteria:
\begin{align}
&(\m{FWHM}_\m{H\beta,broad}>1000\ \m{km\ s^{-1}})\ \land\\
&(\m{SNR}_\m{H\beta,broad}>5)\ \land\\
&(\m{FWHM}_\m{[OIII],narrow}<700\ \m{km\ s^{-1}})\ \land\\
&[(\m{FWHM}_\m{[OIII],outflow}<700\ \m{km\ s^{-1}})\ \lor\\
&(\m{SNR}_\m{[OIII],outflow}<2)]\ \land\\
&[(\m{FWHM}_\m{H\beta,outflow}<700\ \m{km\ s^{-1}})\ \lor\\
&(\m{SNR}_\m{H\beta,outflow}<2)].
\end{align}
We choose the threshold value of $>1000\ \m{km\ s^{-1}}$ for the definition of the broad line made by an AGN, because such a high-velocity component is seen in AGNs \citep[e.g.,][]{2001AJ....122..549V,2015ApJ...813...82R} but not seen in star-forming galaxies (typically $\m{FWHM}<400\ \m{km\ s^{-1}}$, e.g., \citealt{2019ApJ...873..102F,2019MNRAS.487..381S,2022ApJ...929..134X}).
This threshold value is also used in \citet{2012MNRAS.423..600S}, and is more stringent than those used in other studies at $z\sim0$ \citep[e.g.,][]{2015ApJ...813...82R,2019ApJS..243...21L}.

\begin{figure*}
\centering
\begin{minipage}{0.47\hsize}
\begin{center}
\includegraphics[width=0.99\hsize, bb=2 9 707 430]{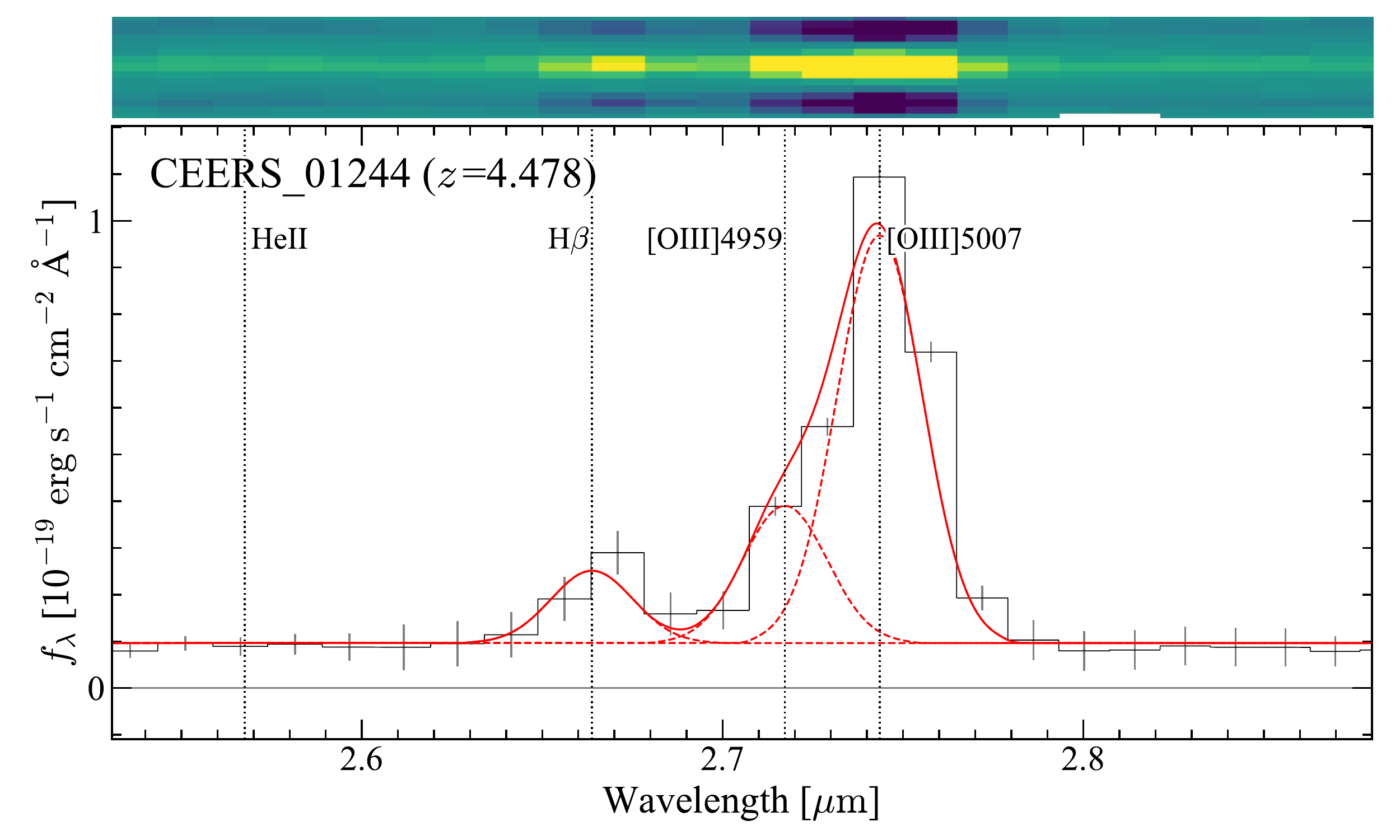}
\end{center}
\end{minipage}
\begin{minipage}{0.277\hsize}
\begin{center}
\includegraphics[width=0.99\hsize, bb=9 9 425 430]{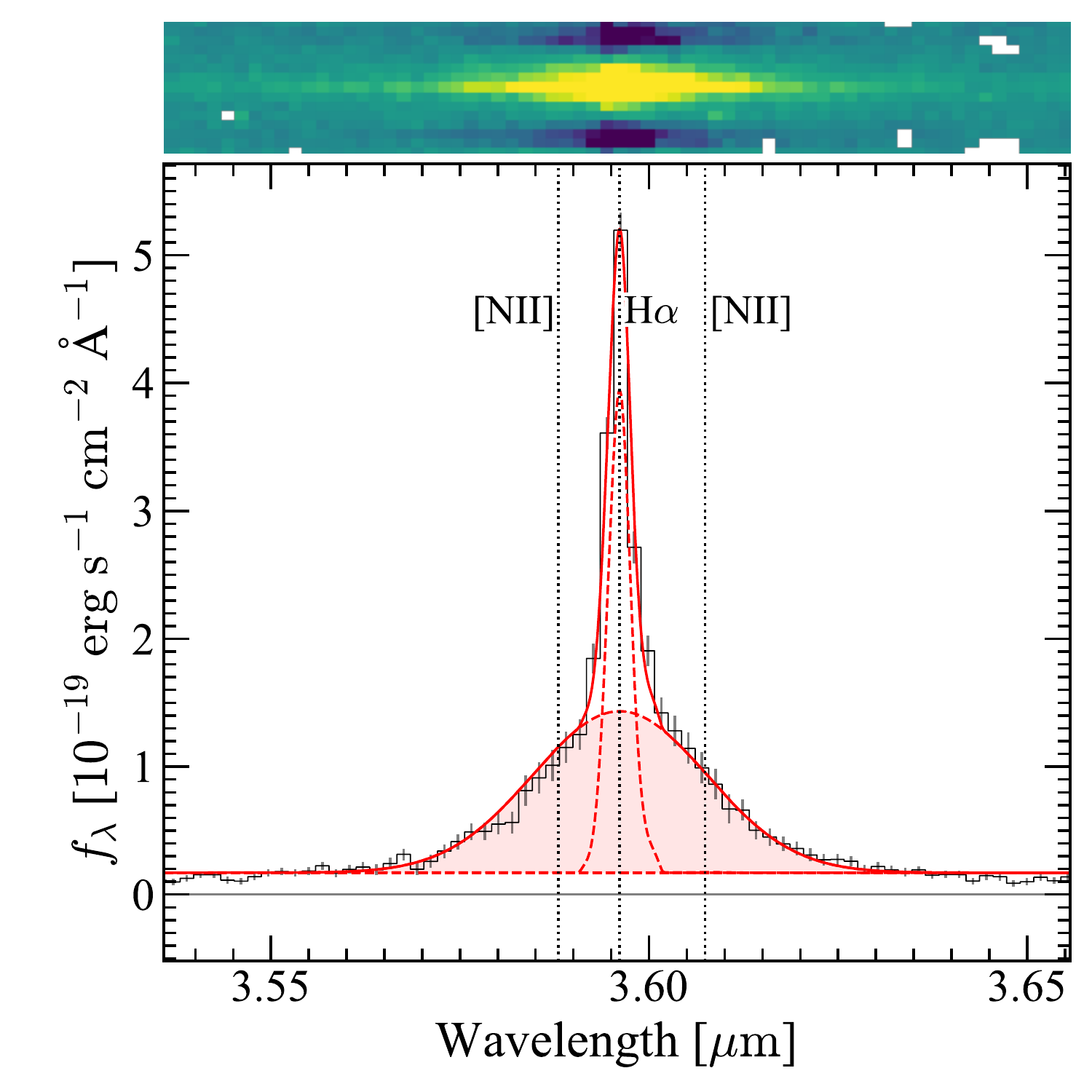}
\end{center}
\end{minipage}
\begin{minipage}{0.227\hsize}
\begin{center}
\includegraphics[width=0.99\hsize, bb=7 9 354 425]{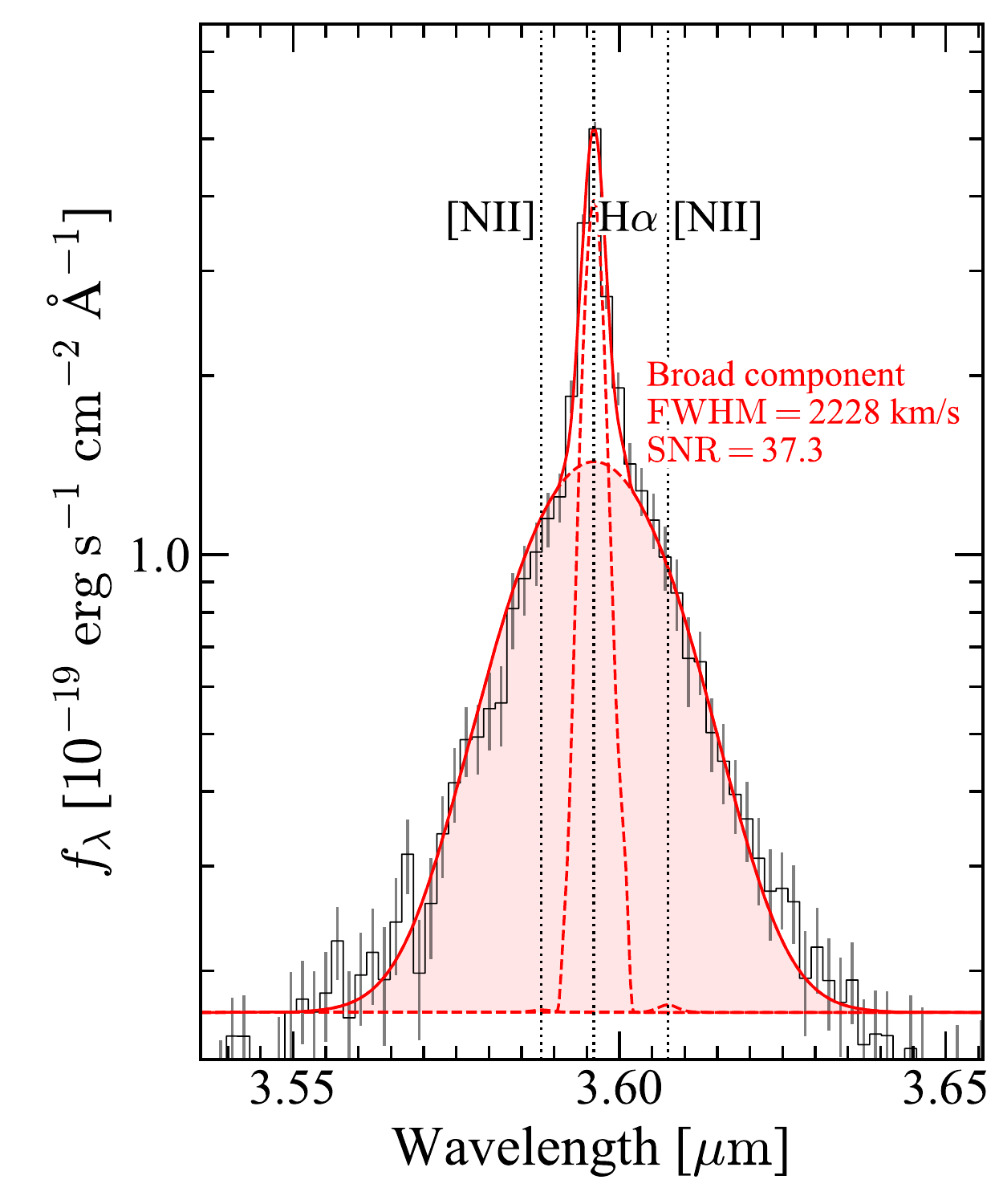}
\end{center}
\end{minipage}
\begin{minipage}{0.47\hsize}
\begin{center}
\includegraphics[width=0.99\hsize, bb=2 9 707 430]{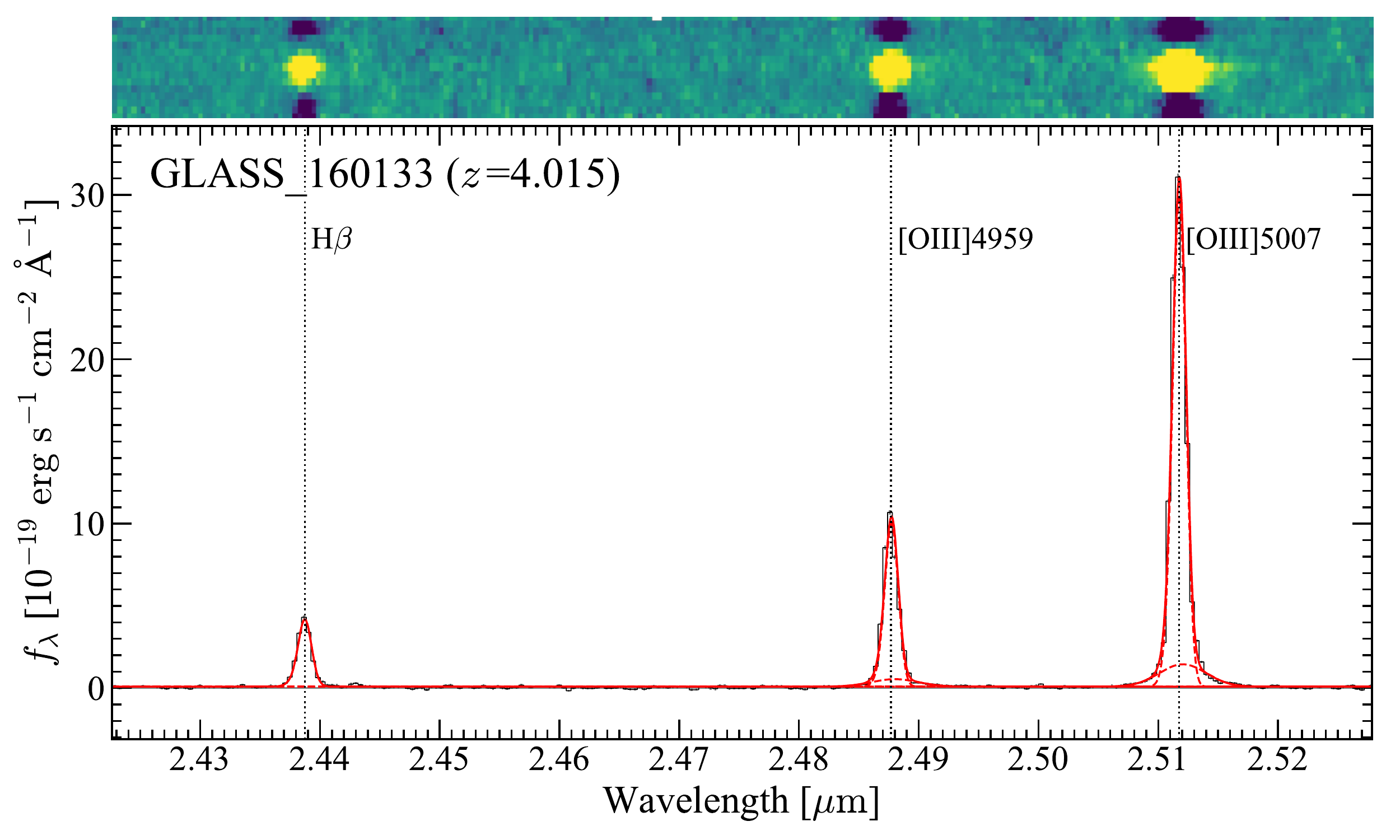}
\end{center}
\end{minipage}
\begin{minipage}{0.277\hsize}
\begin{center}
\includegraphics[width=0.99\hsize, bb=9 9 425 430]{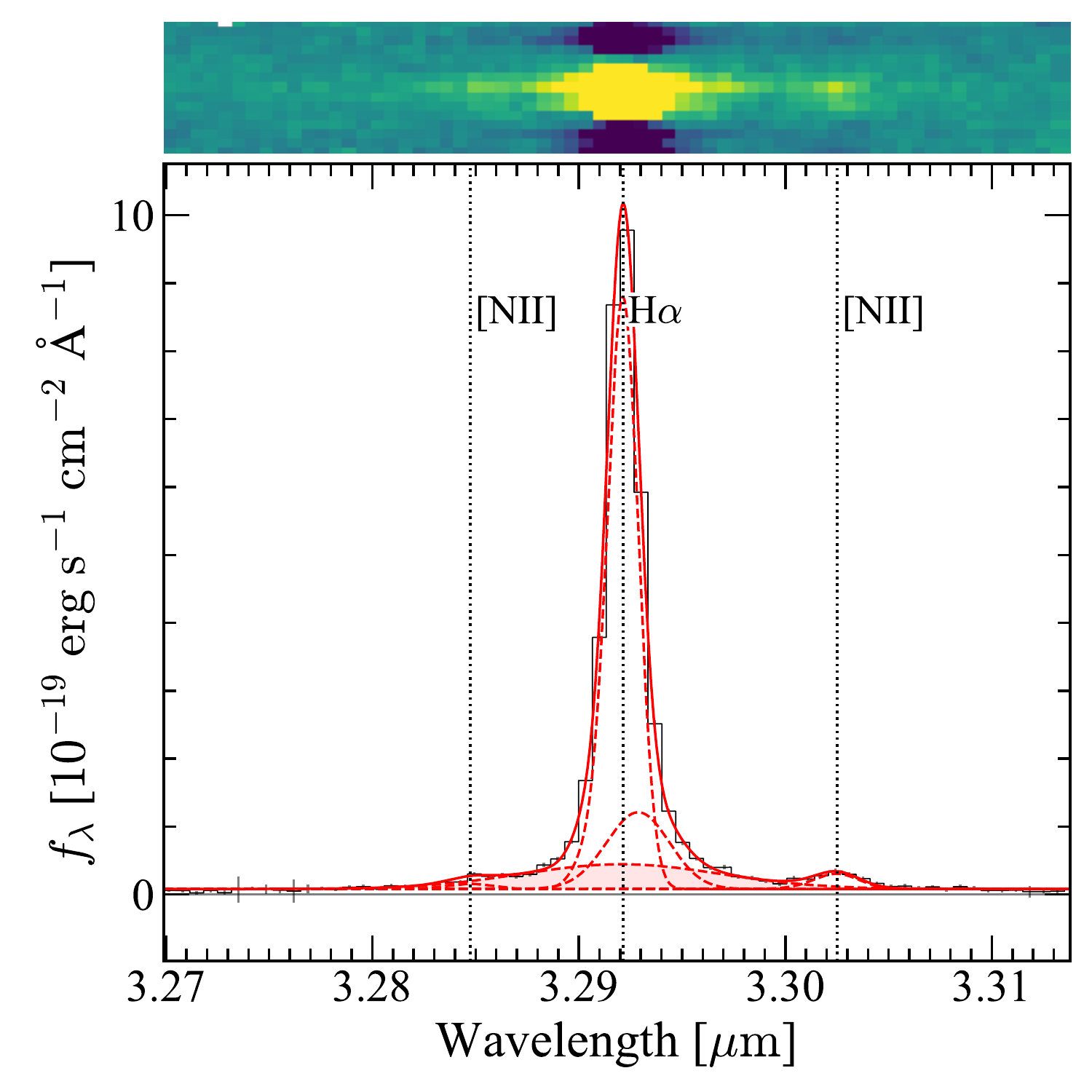}
\end{center}
\end{minipage}
\begin{minipage}{0.227\hsize}
\begin{center}
\includegraphics[width=0.99\hsize, bb=7 9 354 425]{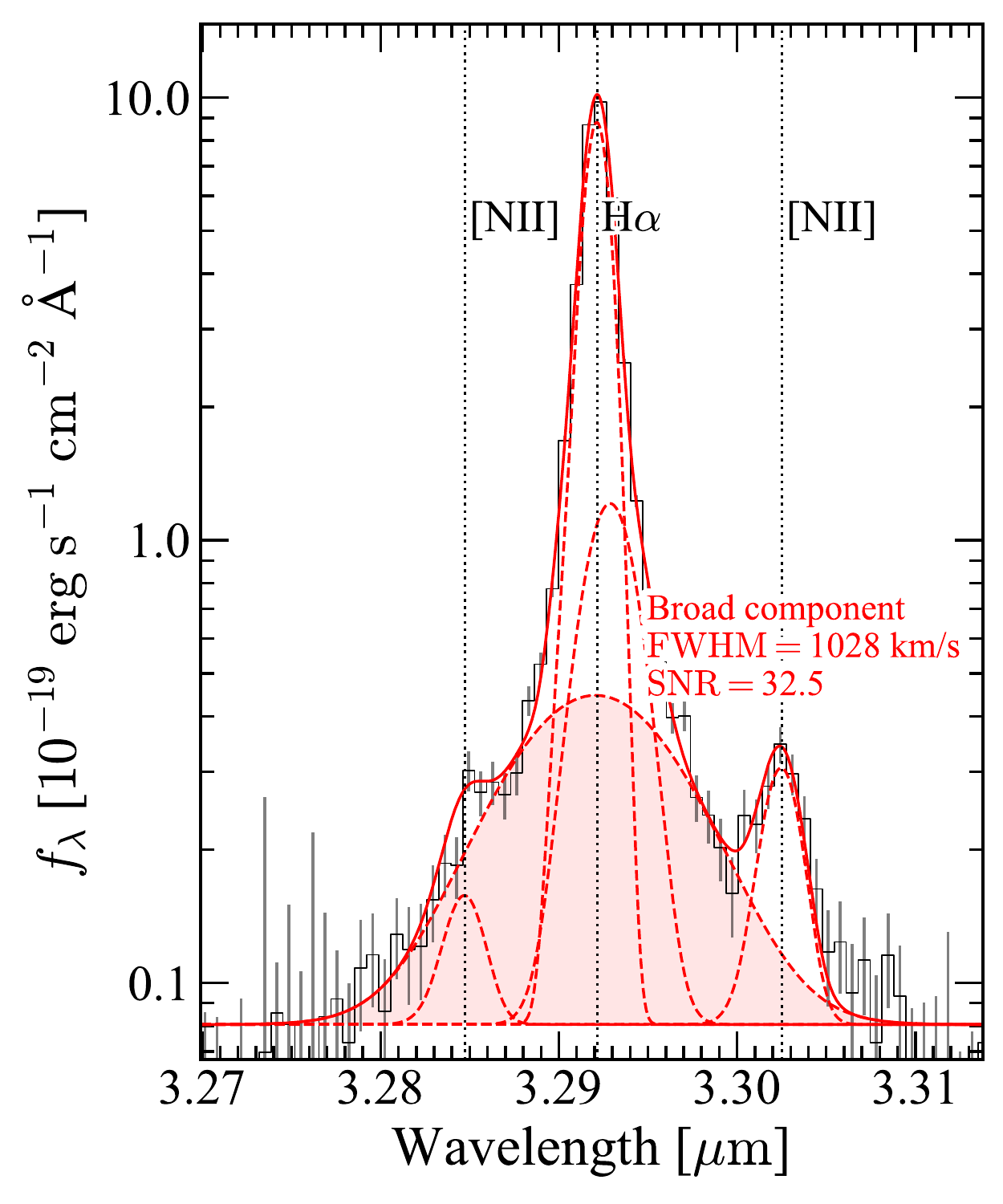}
\end{center}
\end{minipage}
\begin{minipage}{0.47\hsize}
\begin{center}
\includegraphics[width=0.99\hsize, bb=2 9 707 430]{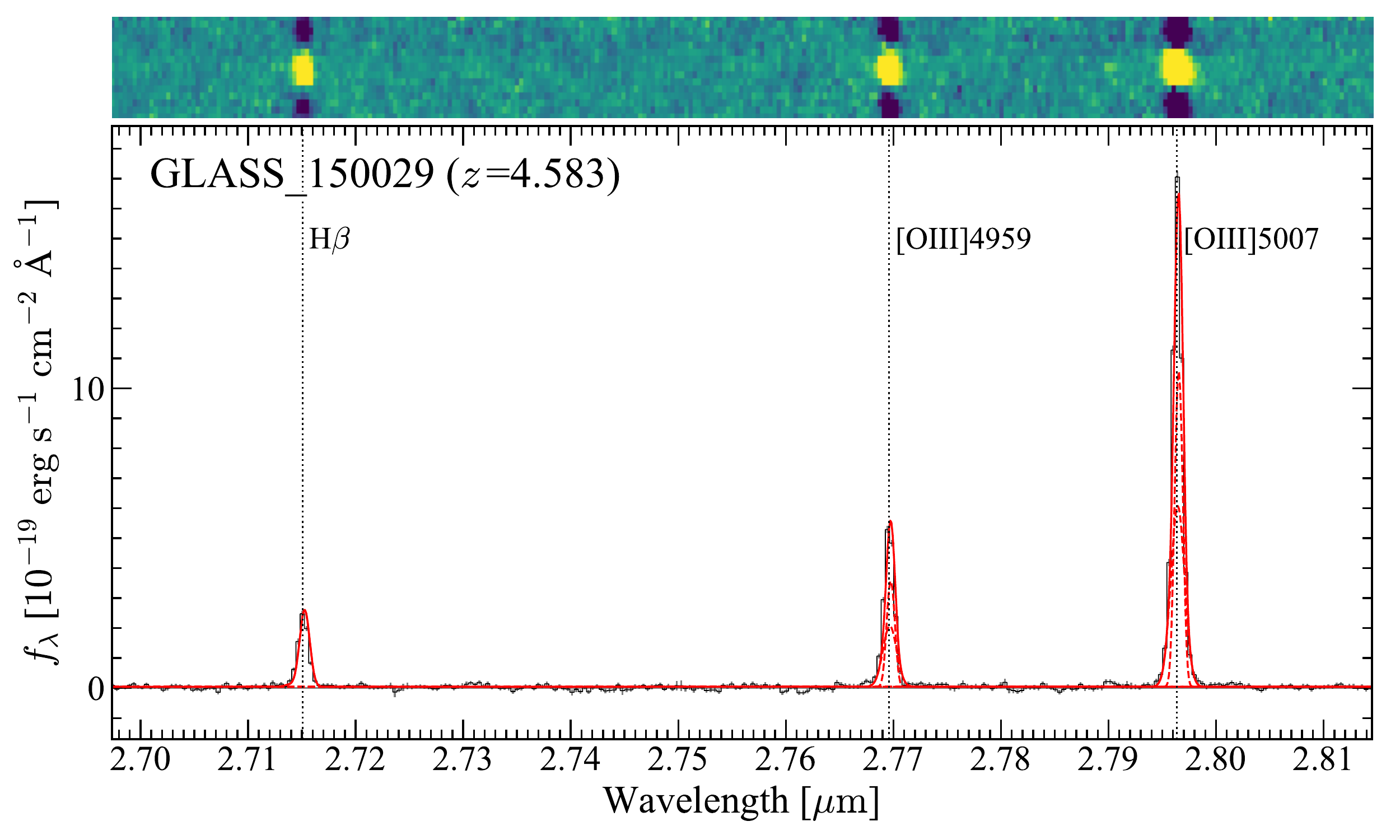}
\end{center}
\end{minipage}
\begin{minipage}{0.277\hsize}
\begin{center}
\includegraphics[width=0.99\hsize, bb=9 9 425 430]{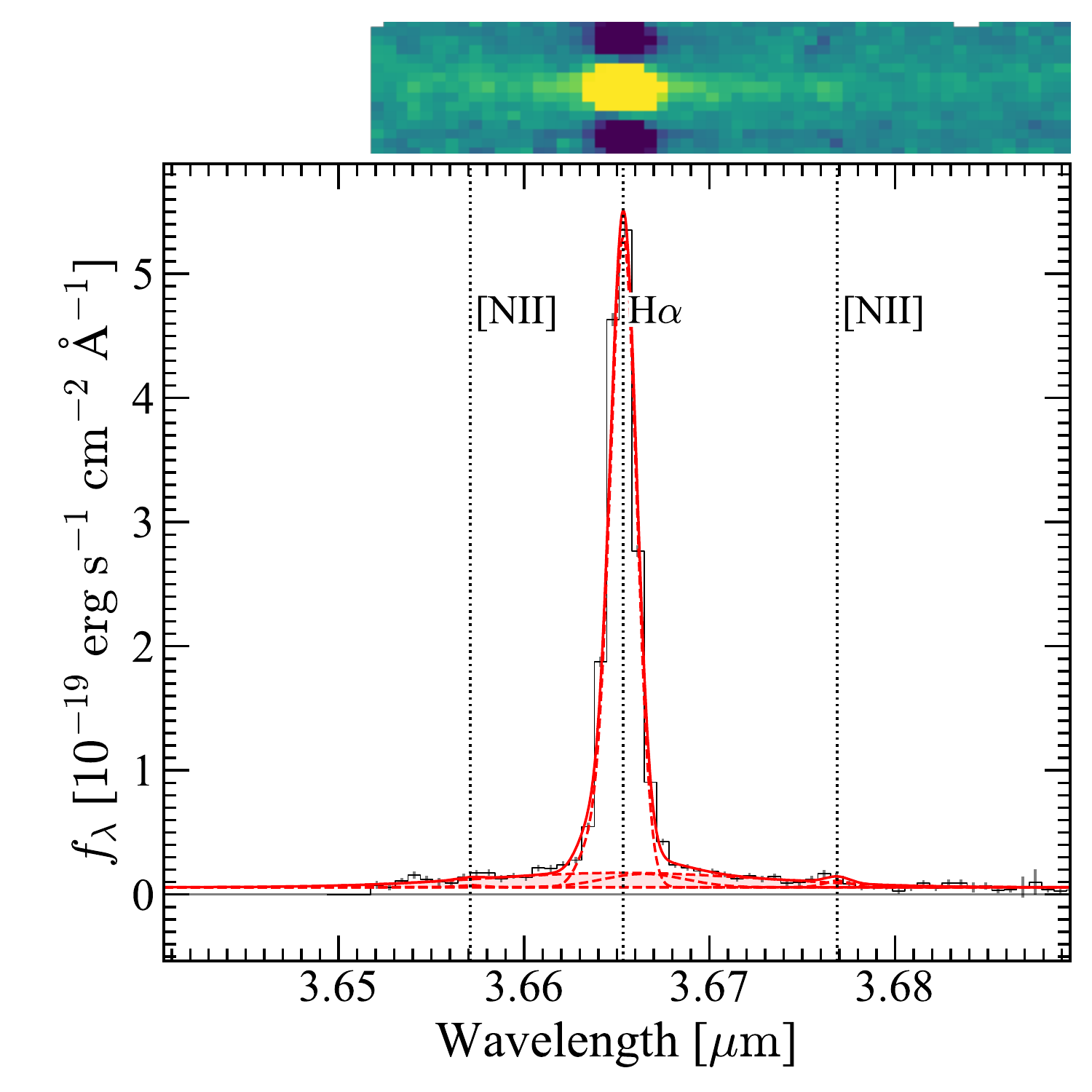}
\end{center}
\end{minipage}
\begin{minipage}{0.227\hsize}
\begin{center}
\includegraphics[width=0.99\hsize, bb=7 9 354 425]{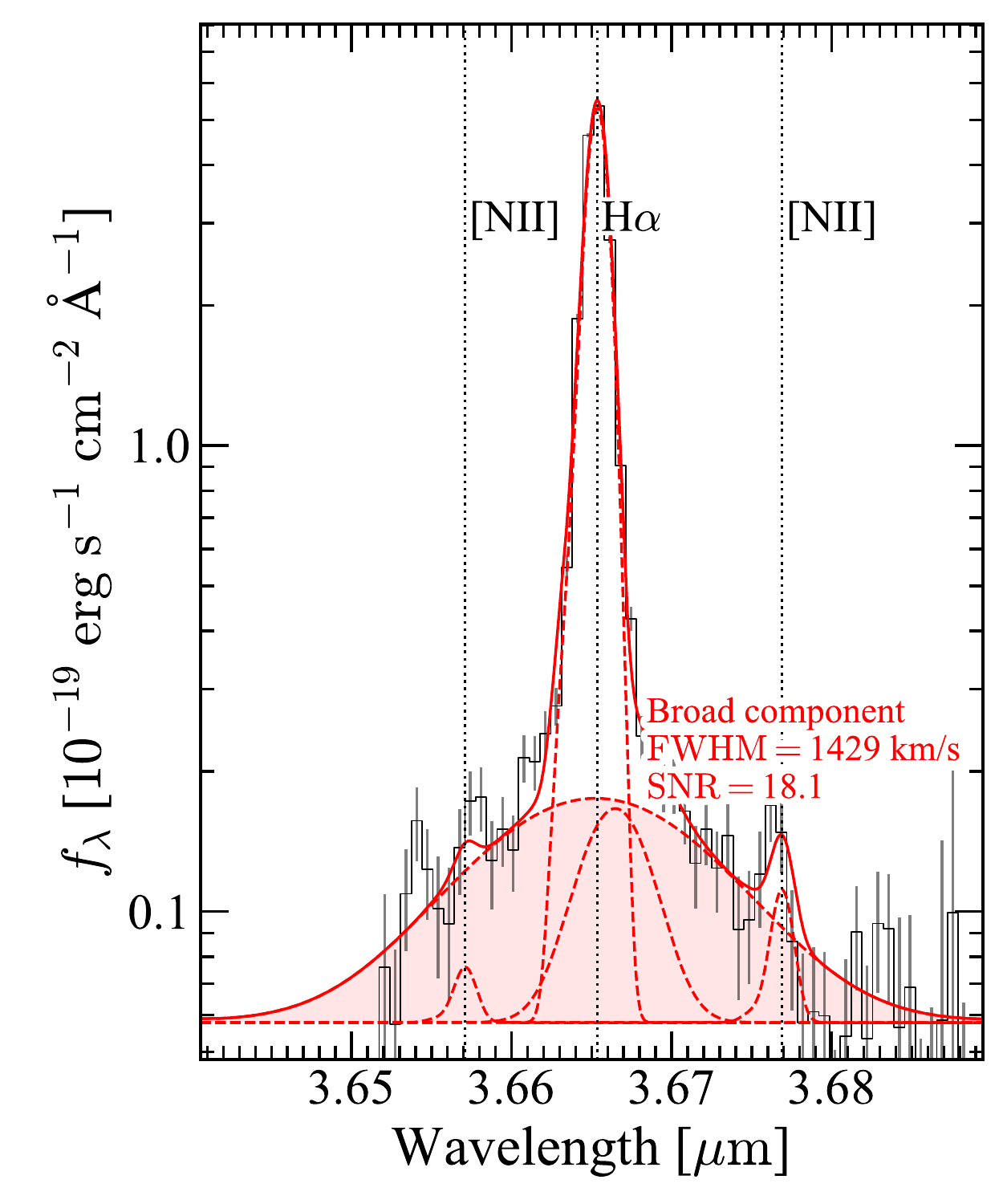}
\end{center}
\end{minipage}
\begin{minipage}{0.47\hsize}
\begin{center}
\includegraphics[width=0.99\hsize, bb=2 9 707 430]{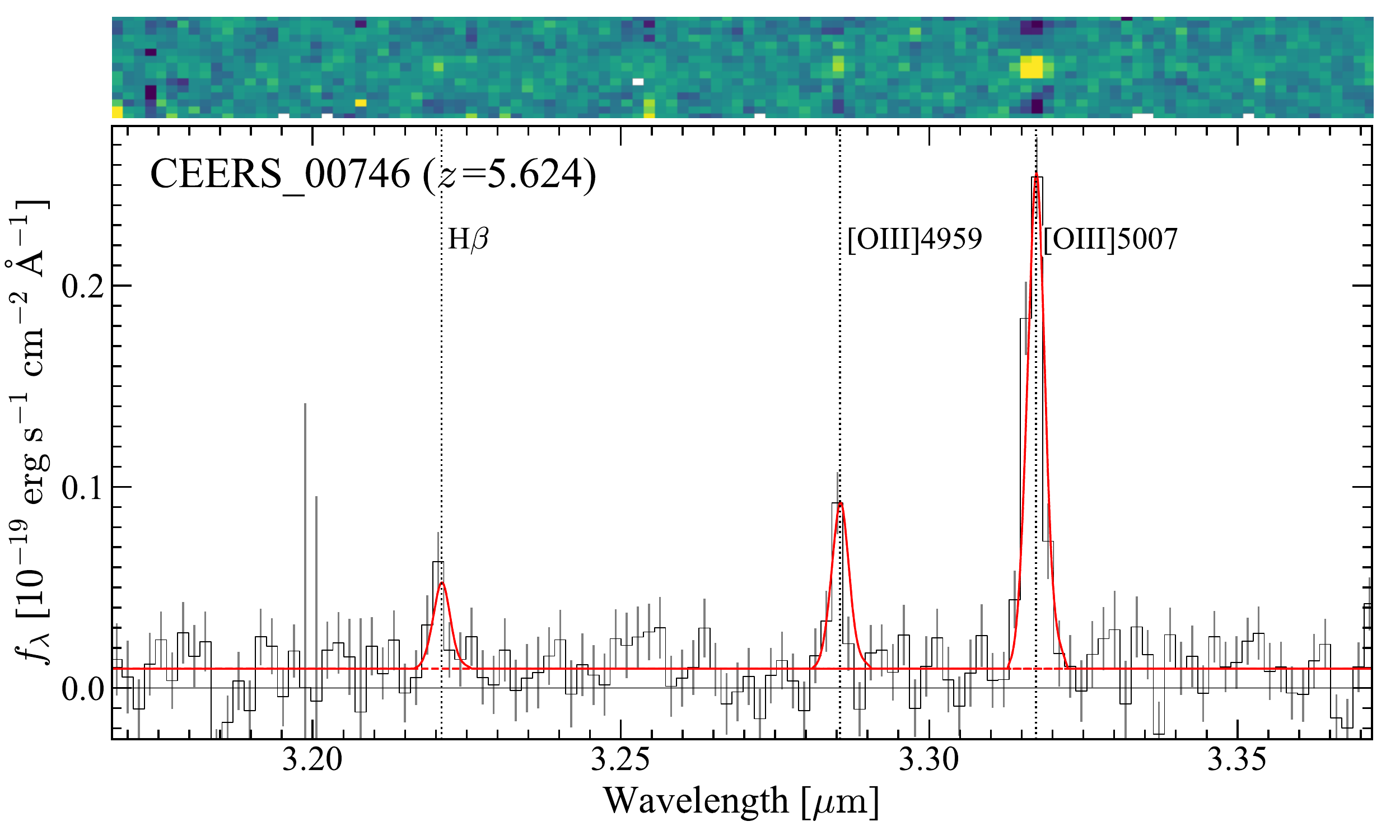}
\end{center}
\end{minipage}
\begin{minipage}{0.277\hsize}
\begin{center}
\includegraphics[width=0.99\hsize, bb=9 9 425 430]{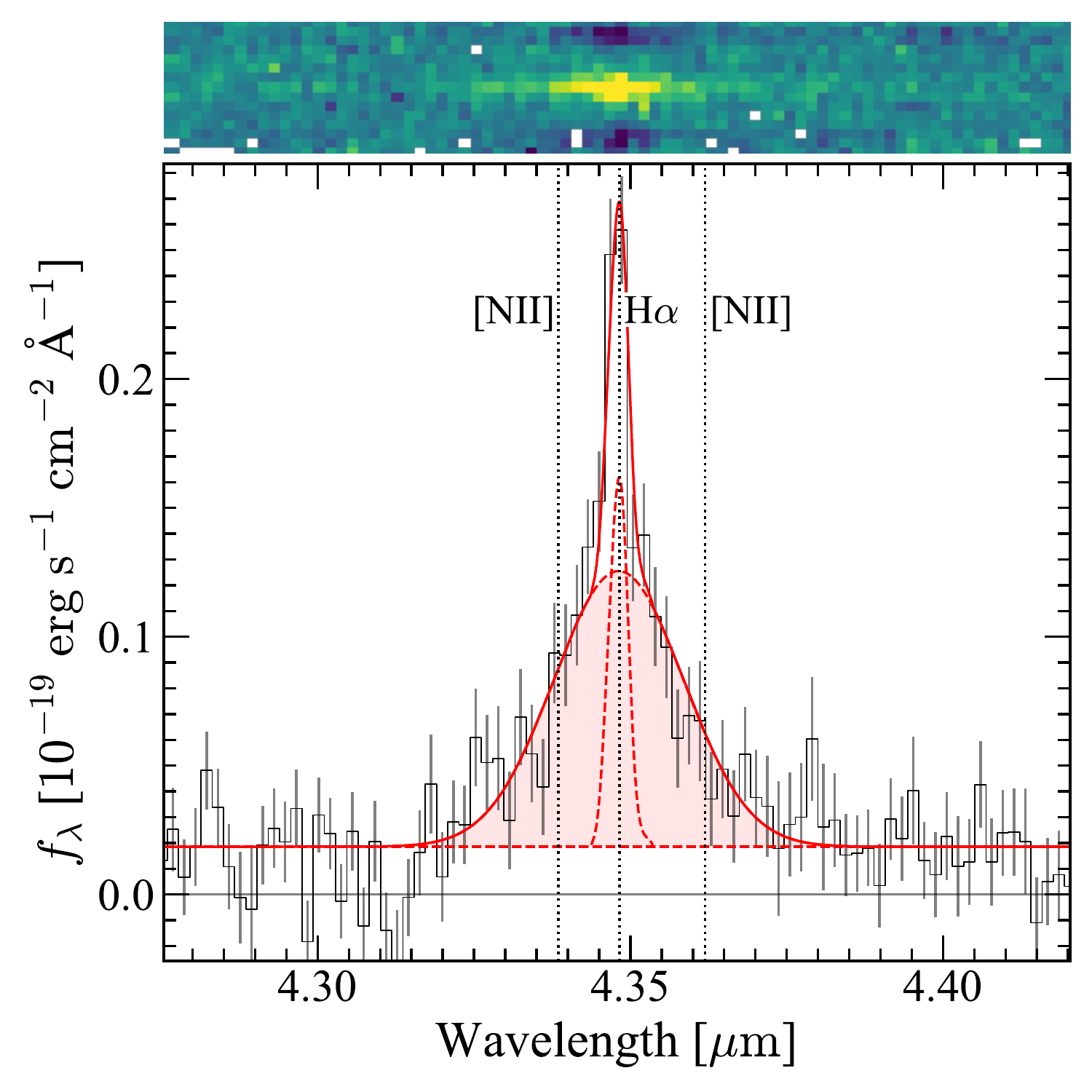}
\end{center}
\end{minipage}
\begin{minipage}{0.227\hsize}
\begin{center}
\includegraphics[width=0.99\hsize, bb=7 9 354 425]{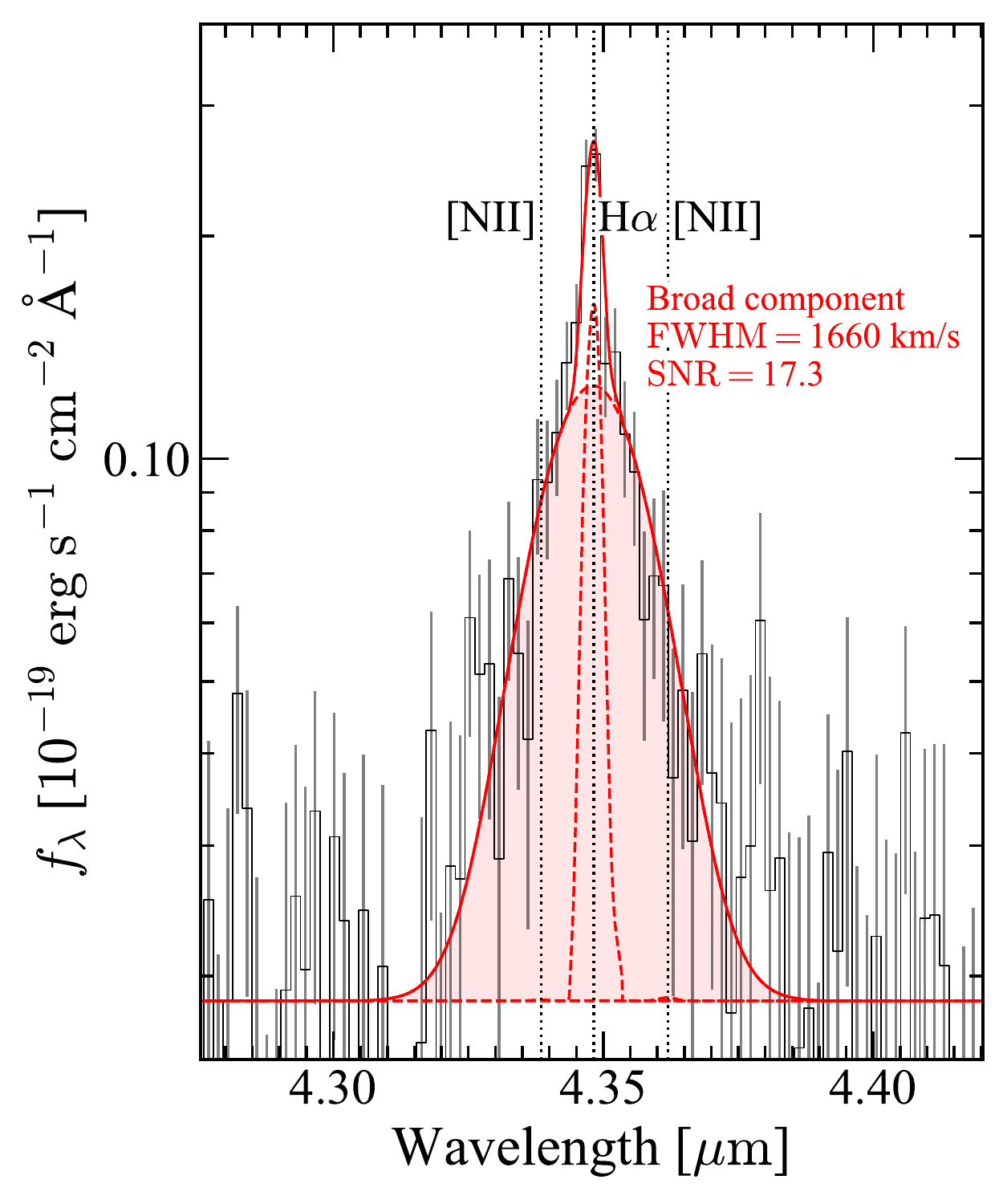}
\end{center}
\end{minipage}
\caption{
NIRSpec spectra of CEERS\_01244, GLASS\_160133, GLASS\_150029, and CEERS\_00746.
For each object, the left and middle panels show spectra around H$\beta$+[{\sc Oiii}]$\lambda\lambda$4959,5007 and H$\alpha$+[{\sc Nii}]$\lambda\lambda$6548,6584, respectively.
The 2D and 1D spectra are shown in the top and bottom panels, respectively.
The red dashed line with the shaded region shows the best-fit broad-line component ($\m{FWHM}>1000\ \m{km\ s^{-1}}$) and other red dashed lines show the best-fit narrow components ($\m{FWHM}<500\ \m{km\ s^{-1}}$).
For GLASS\_160133 and GLASS\_150029, we also show the outflow components with $\m{FWHM}\lesssim500\ \m{km\ s^{-1}}$.
The right panels show the spectra around H$\alpha$+[{\sc Nii}]$\lambda\lambda$6548,6584 with the logarithmic scale.
The broad-line components only seen in H$\alpha$, which are detected with a higher signal-to-noise ratio than {\sc[Oiii]}$\lambda$5007, indicates that these objects are type-1 AGNs.
}
\label{fig_spec_line_1}
\end{figure*}

\begin{figure*}
\centering
\begin{minipage}{0.47\hsize}
\begin{center}
\includegraphics[width=0.99\hsize, bb=2 9 707 430]{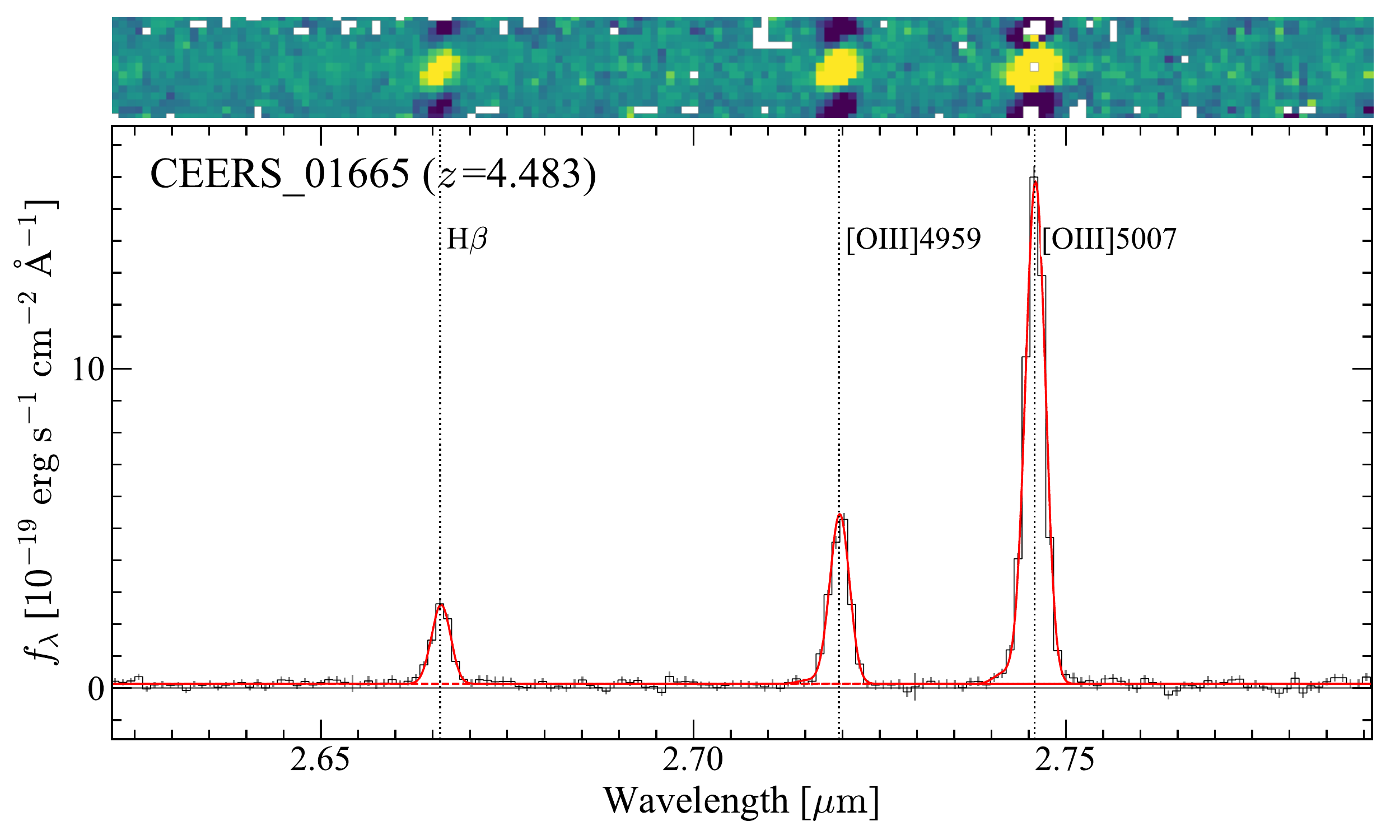}
\end{center}
\end{minipage}
\begin{minipage}{0.277\hsize}
\begin{center}
\includegraphics[width=0.99\hsize, bb=9 9 425 430]{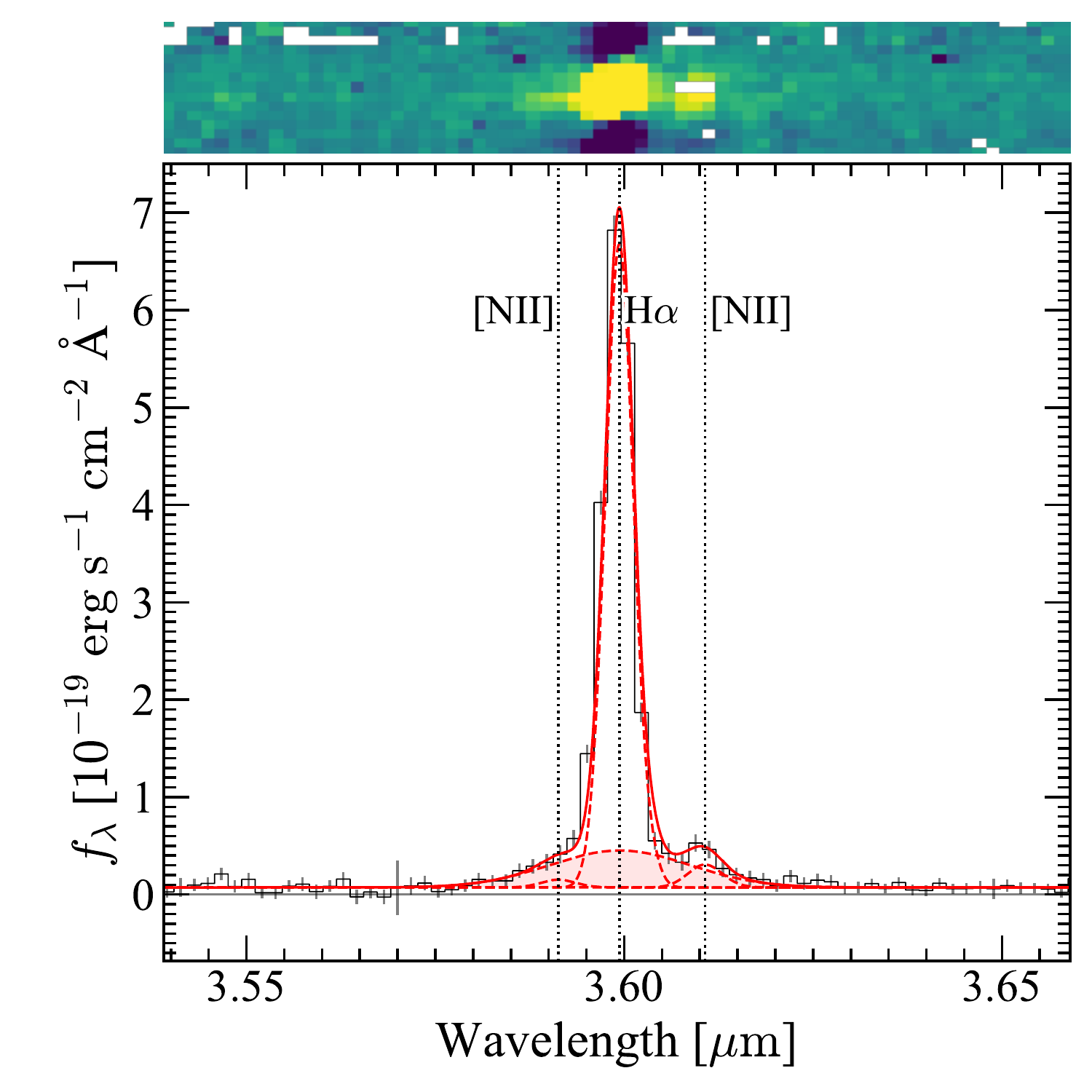}
\end{center}
\end{minipage}
\begin{minipage}{0.227\hsize}
\begin{center}
\includegraphics[width=0.99\hsize, bb=7 9 354 425]{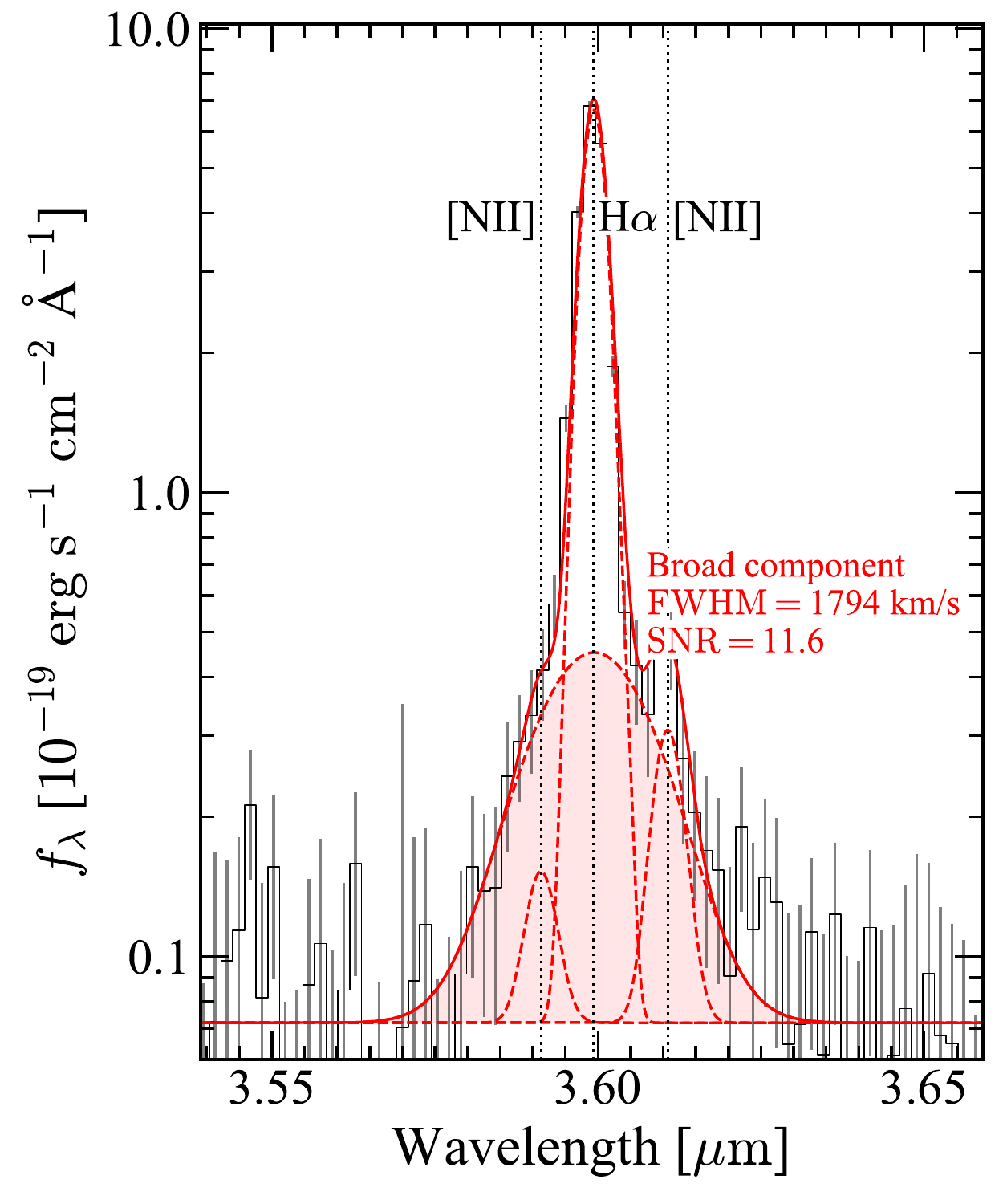}
\end{center}
\end{minipage}
\begin{minipage}{0.47\hsize}
\begin{center}
\includegraphics[width=0.99\hsize, bb=2 9 707 430]{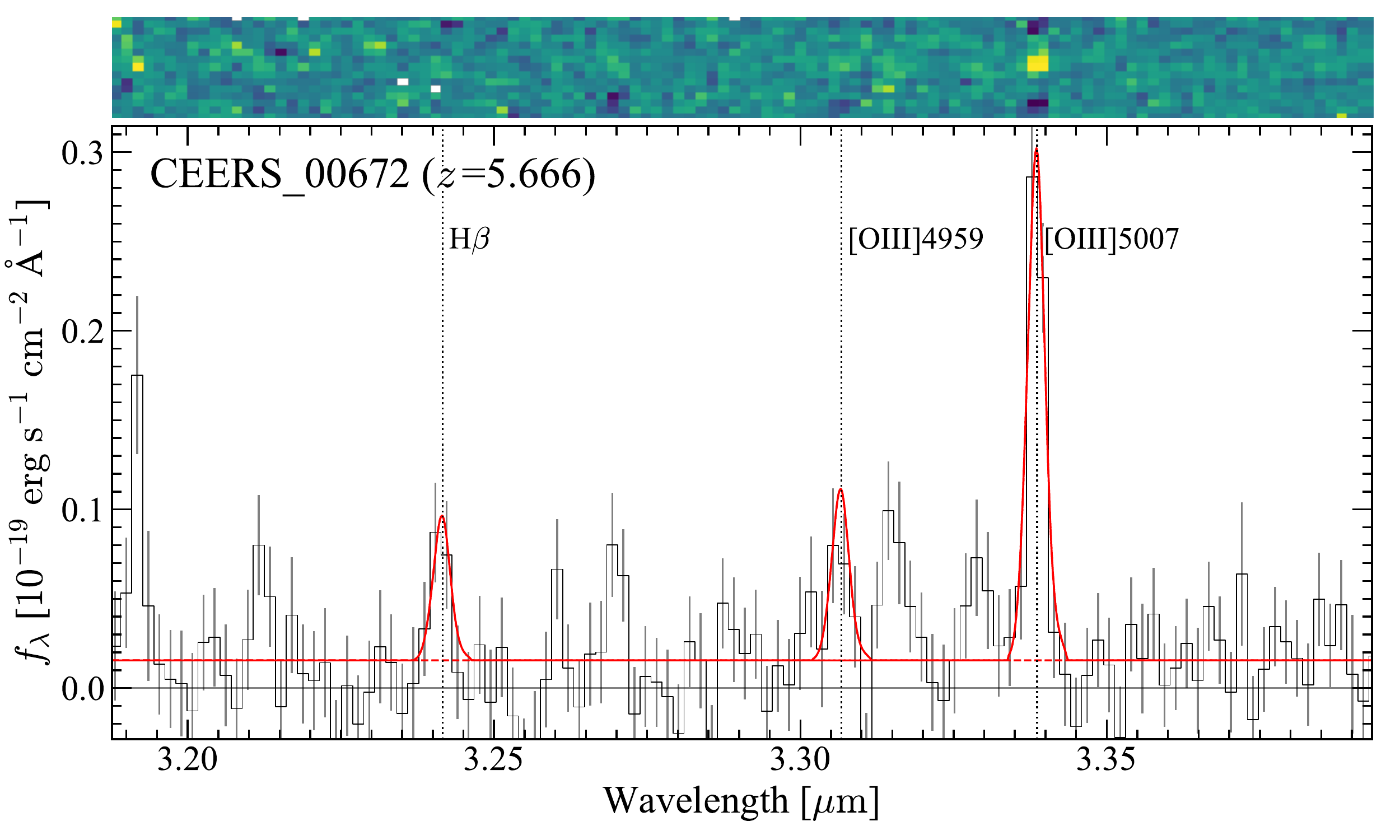}
\end{center}
\end{minipage}
\begin{minipage}{0.277\hsize}
\begin{center}
\includegraphics[width=0.99\hsize, bb=9 9 425 430]{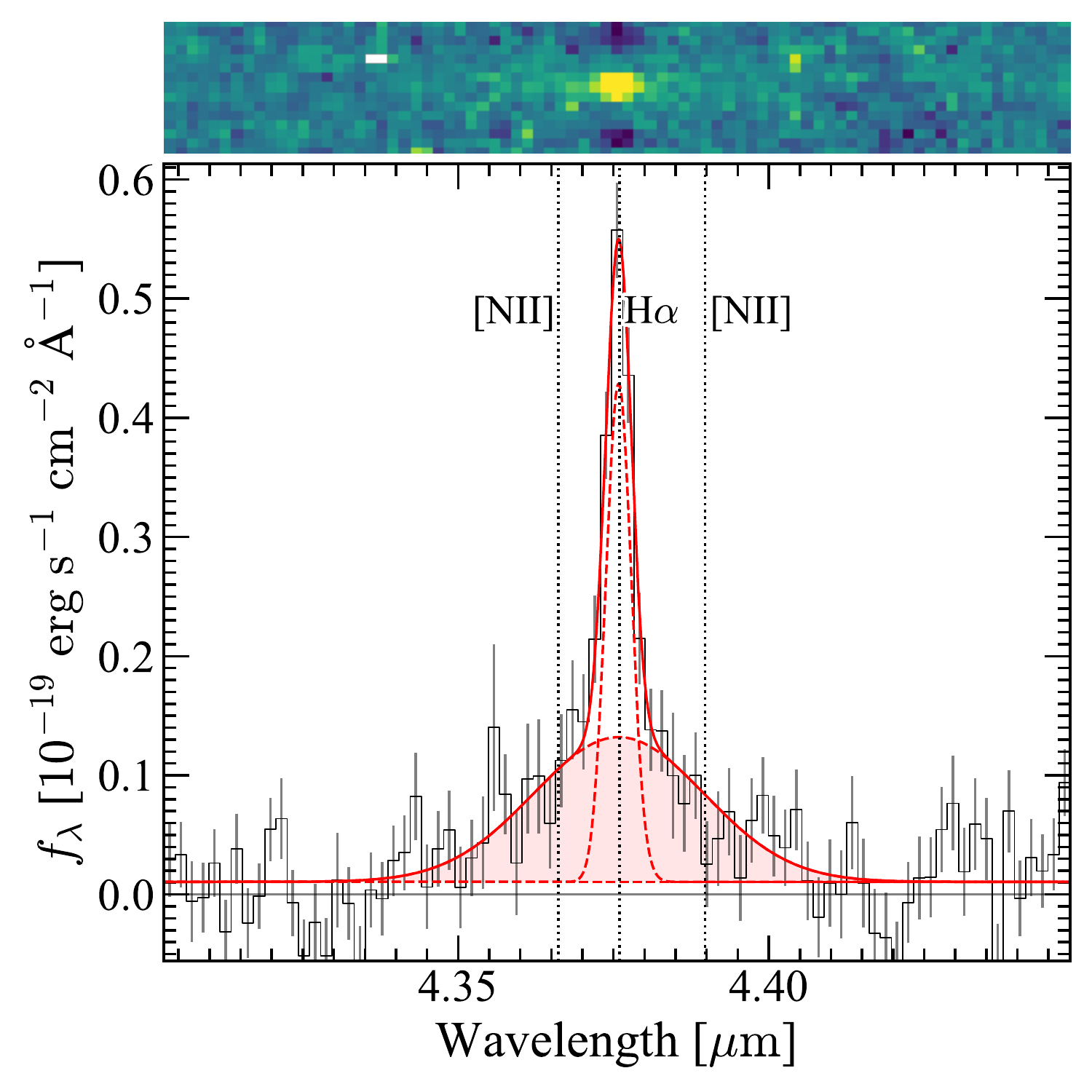}
\end{center}
\end{minipage}
\begin{minipage}{0.227\hsize}
\begin{center}
\includegraphics[width=0.99\hsize, bb=7 9 354 425]{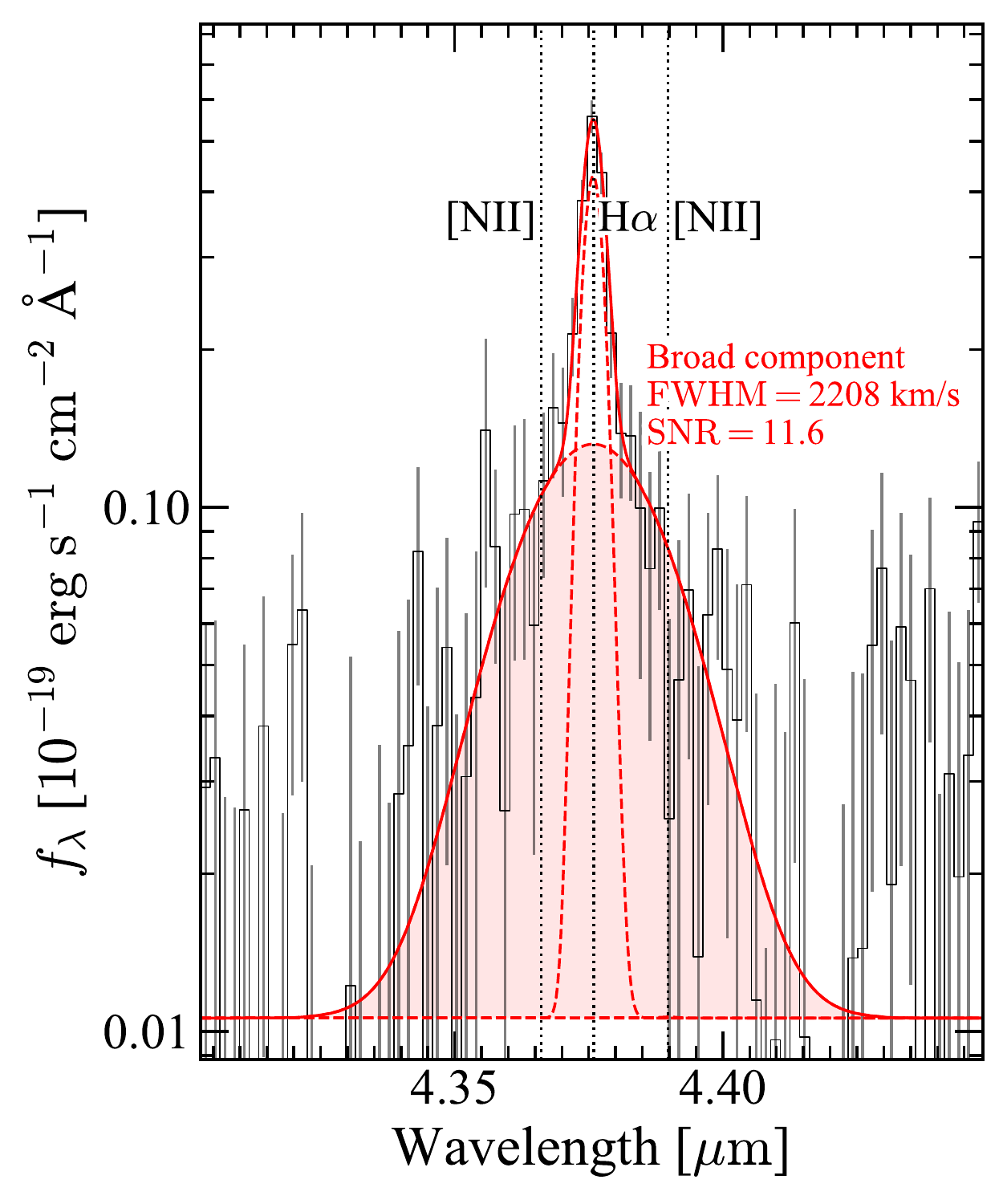}
\end{center}
\end{minipage}
\begin{minipage}{0.47\hsize}
\begin{center}
\includegraphics[width=0.99\hsize, bb=2 9 707 430]{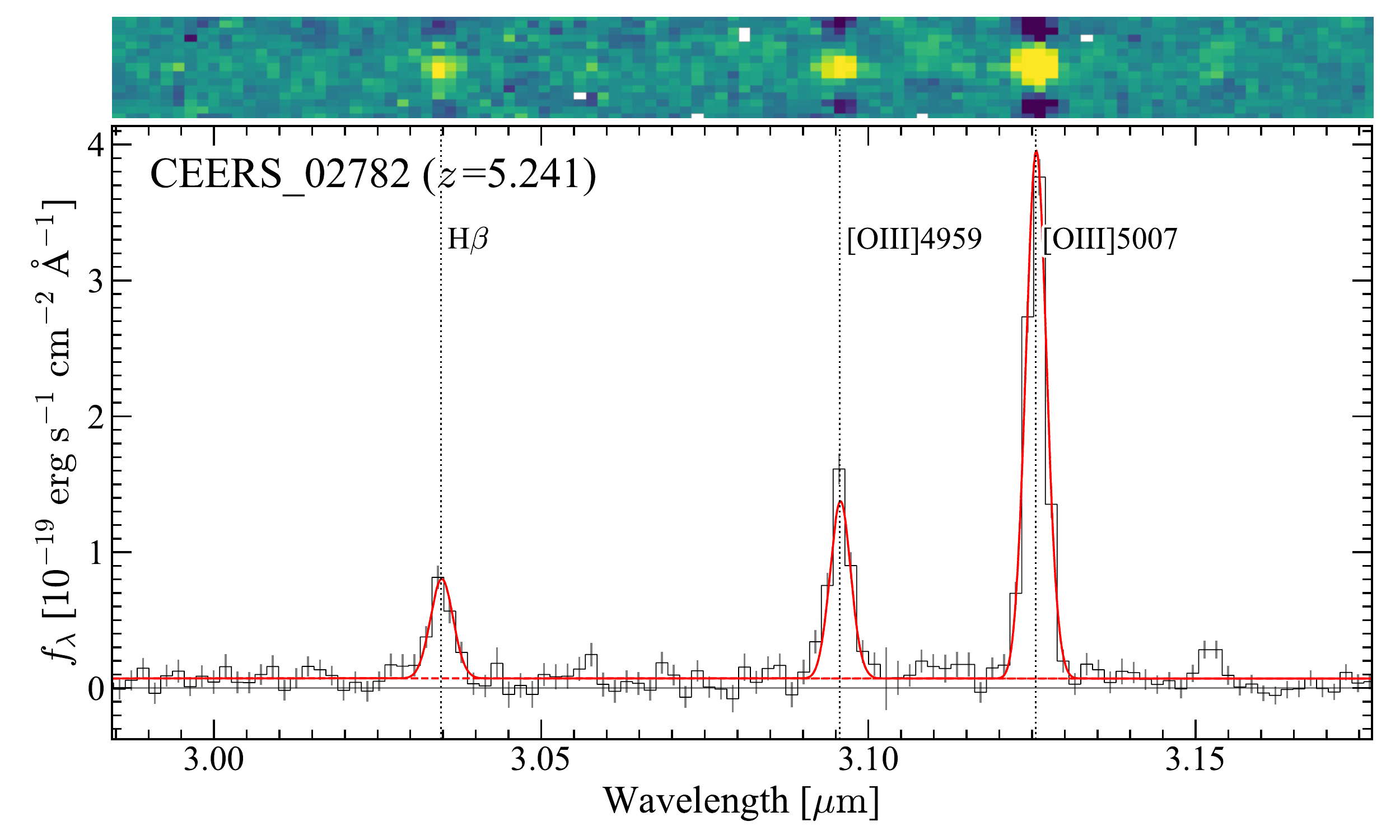}
\end{center}
\end{minipage}
\begin{minipage}{0.277\hsize}
\begin{center}
\includegraphics[width=0.99\hsize, bb=9 9 425 430]{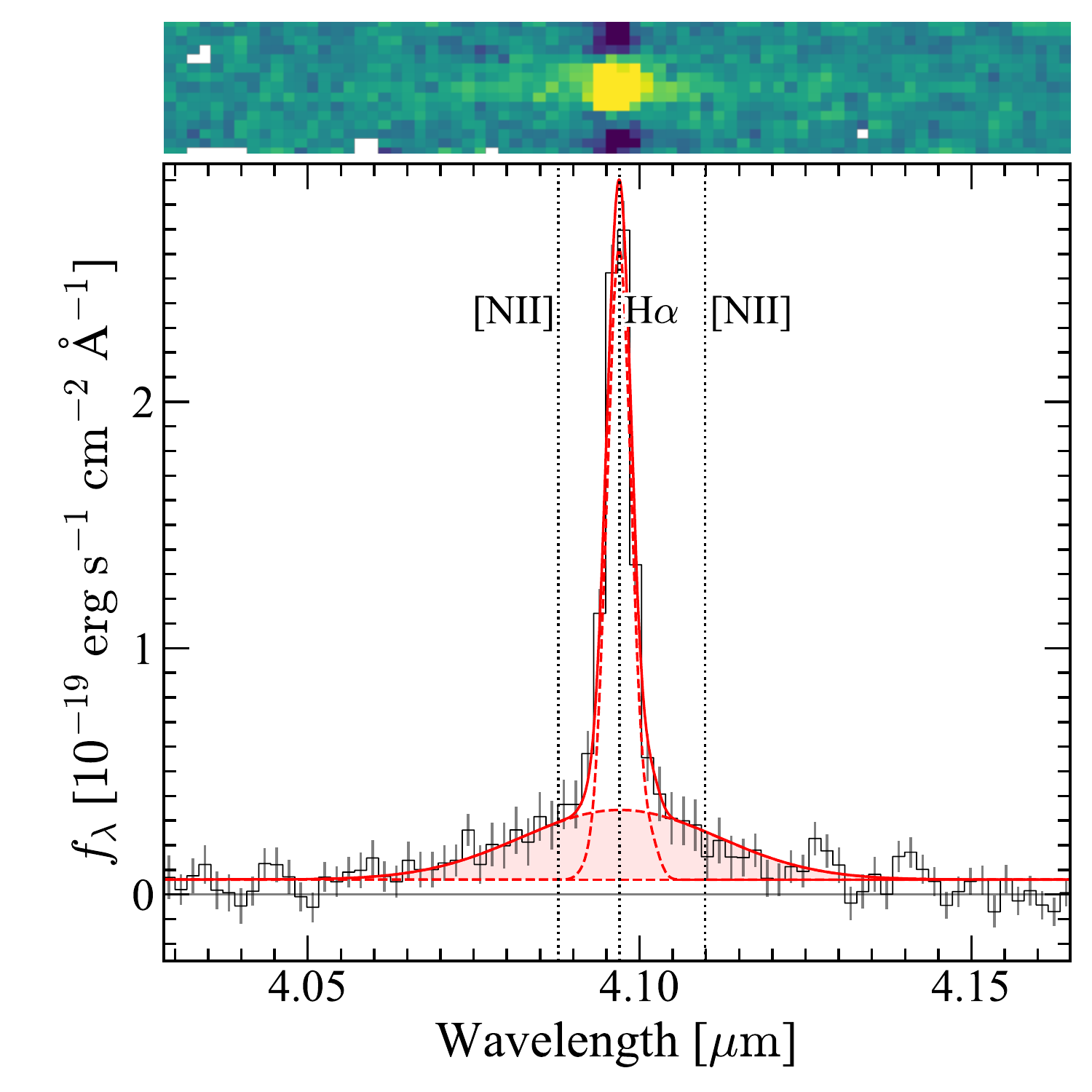}
\end{center}
\end{minipage}
\begin{minipage}{0.227\hsize}
\begin{center}
\includegraphics[width=0.99\hsize, bb=7 9 354 425]{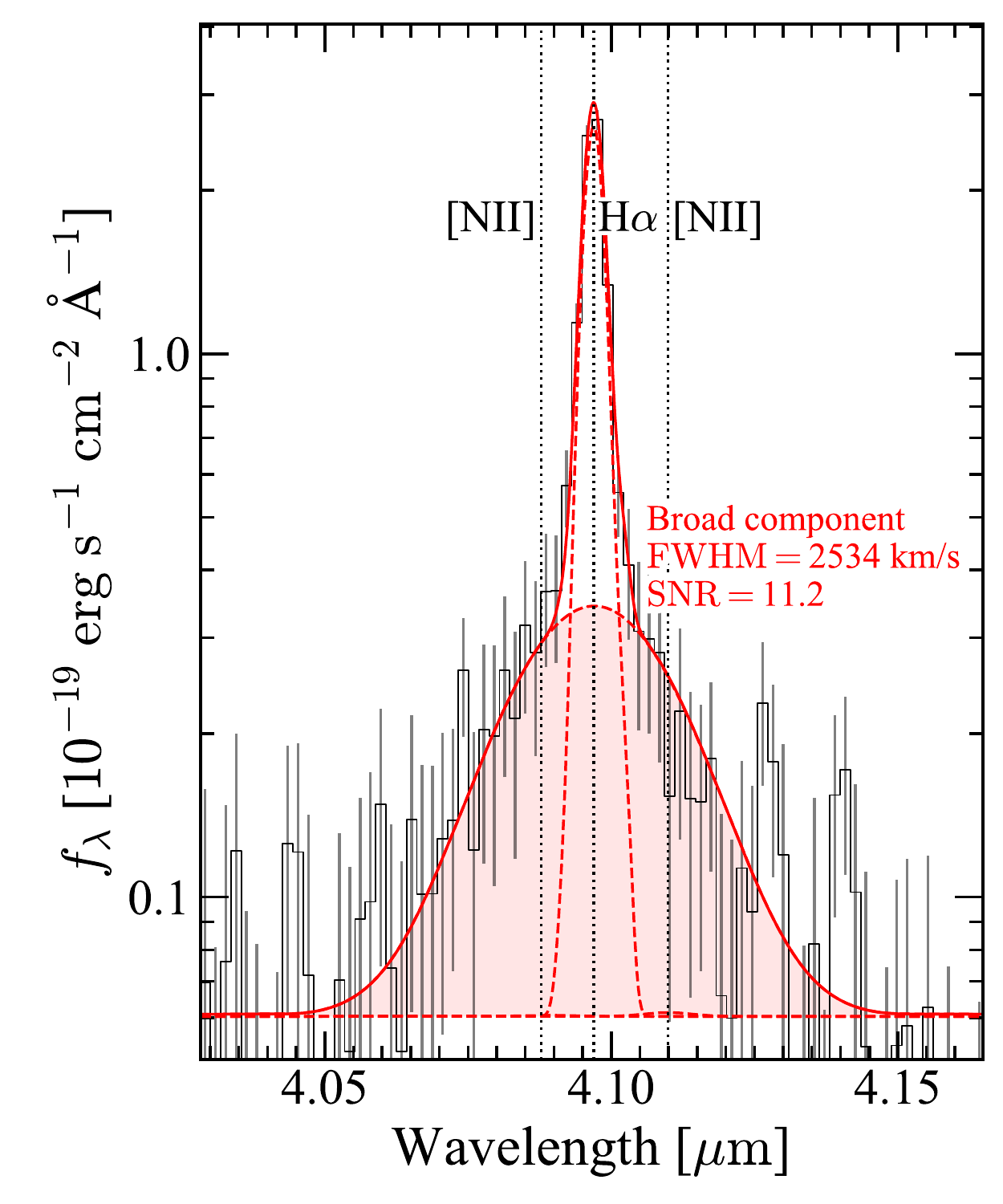}
\end{center}
\end{minipage}
\begin{minipage}{0.47\hsize}
\begin{center}
\includegraphics[width=0.99\hsize, bb=2 9 707 430]{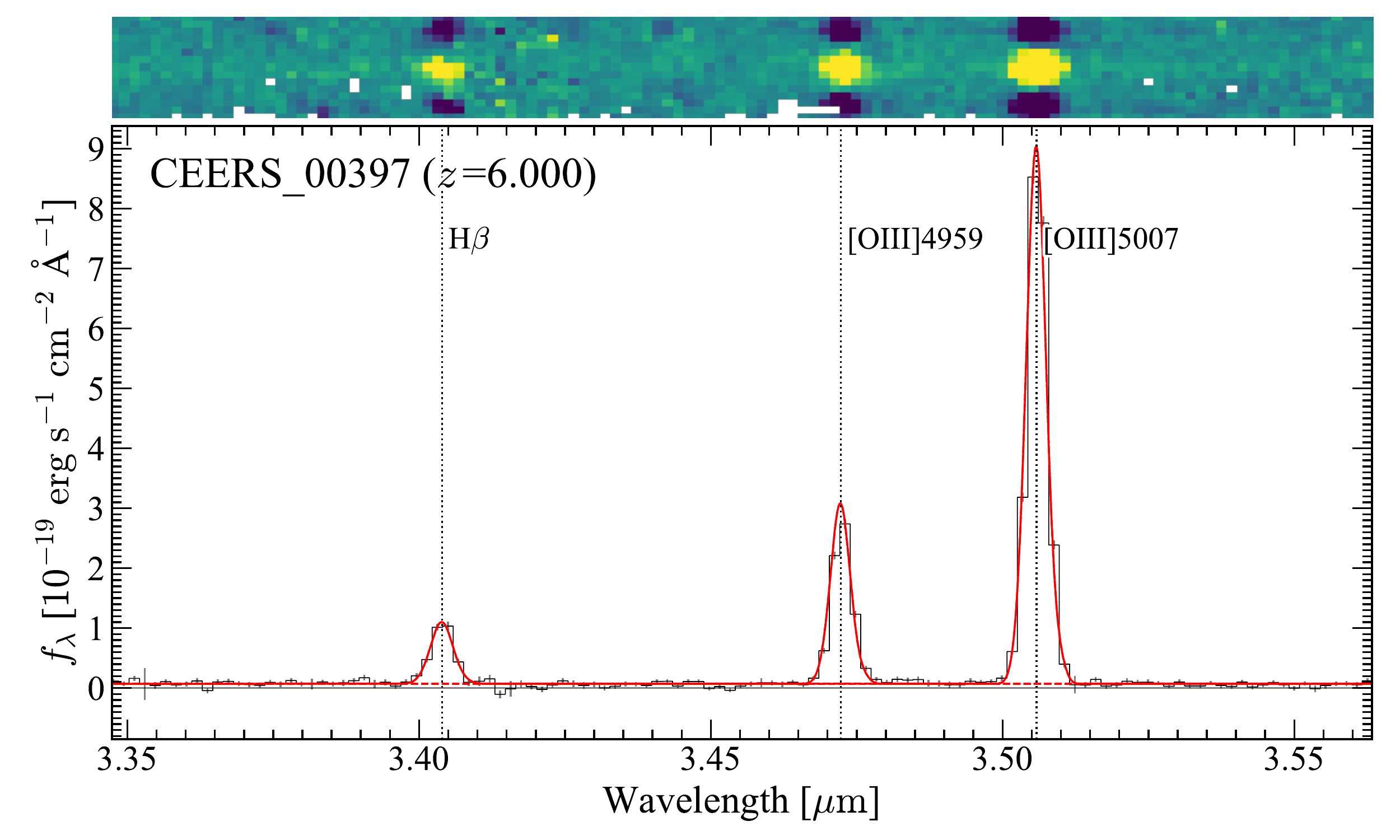}
\end{center}
\end{minipage}
\begin{minipage}{0.277\hsize}
\begin{center}
\includegraphics[width=0.99\hsize, bb=9 9 425 430]{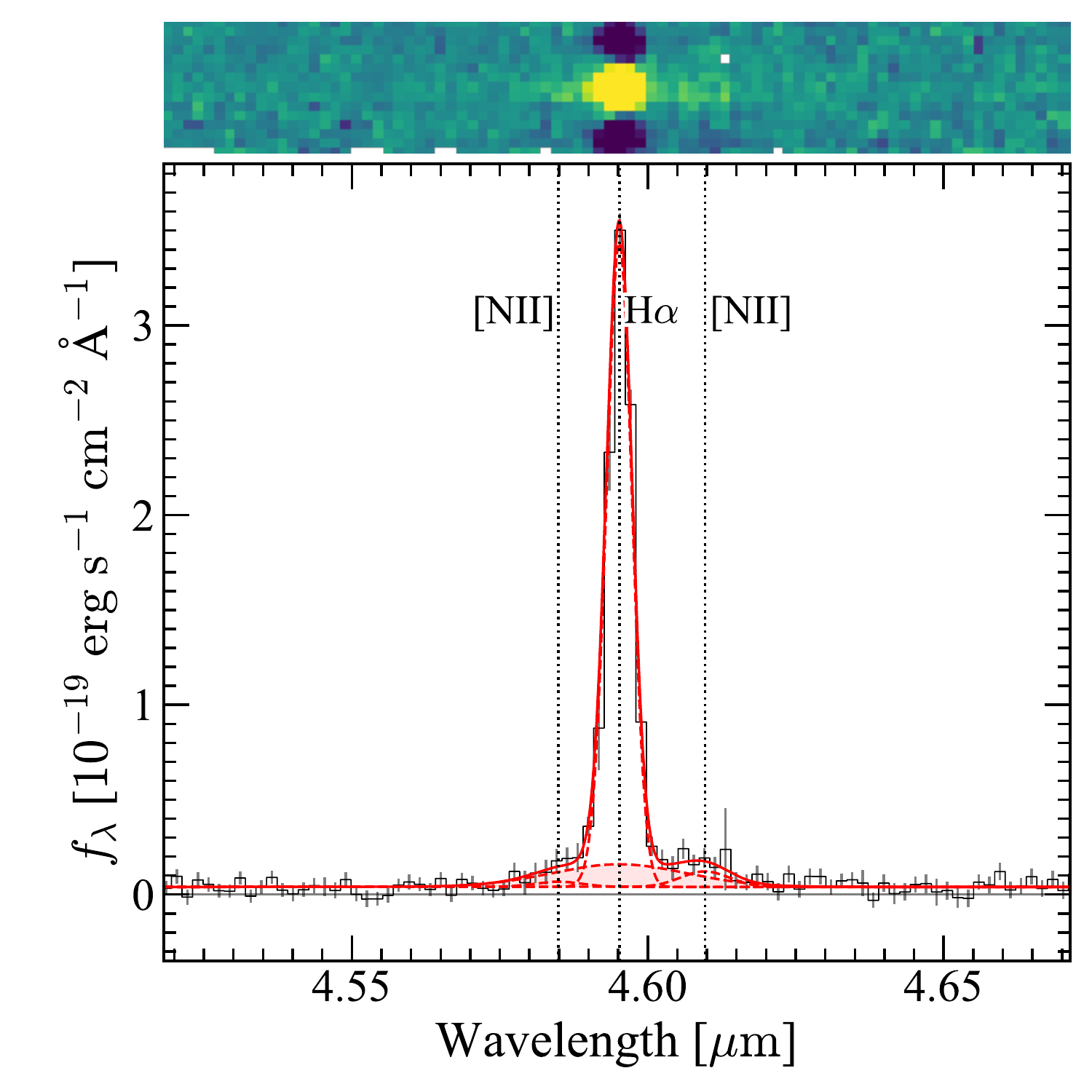}
\end{center}
\end{minipage}
\begin{minipage}{0.227\hsize}
\begin{center}
\includegraphics[width=0.99\hsize, bb=7 9 354 425]{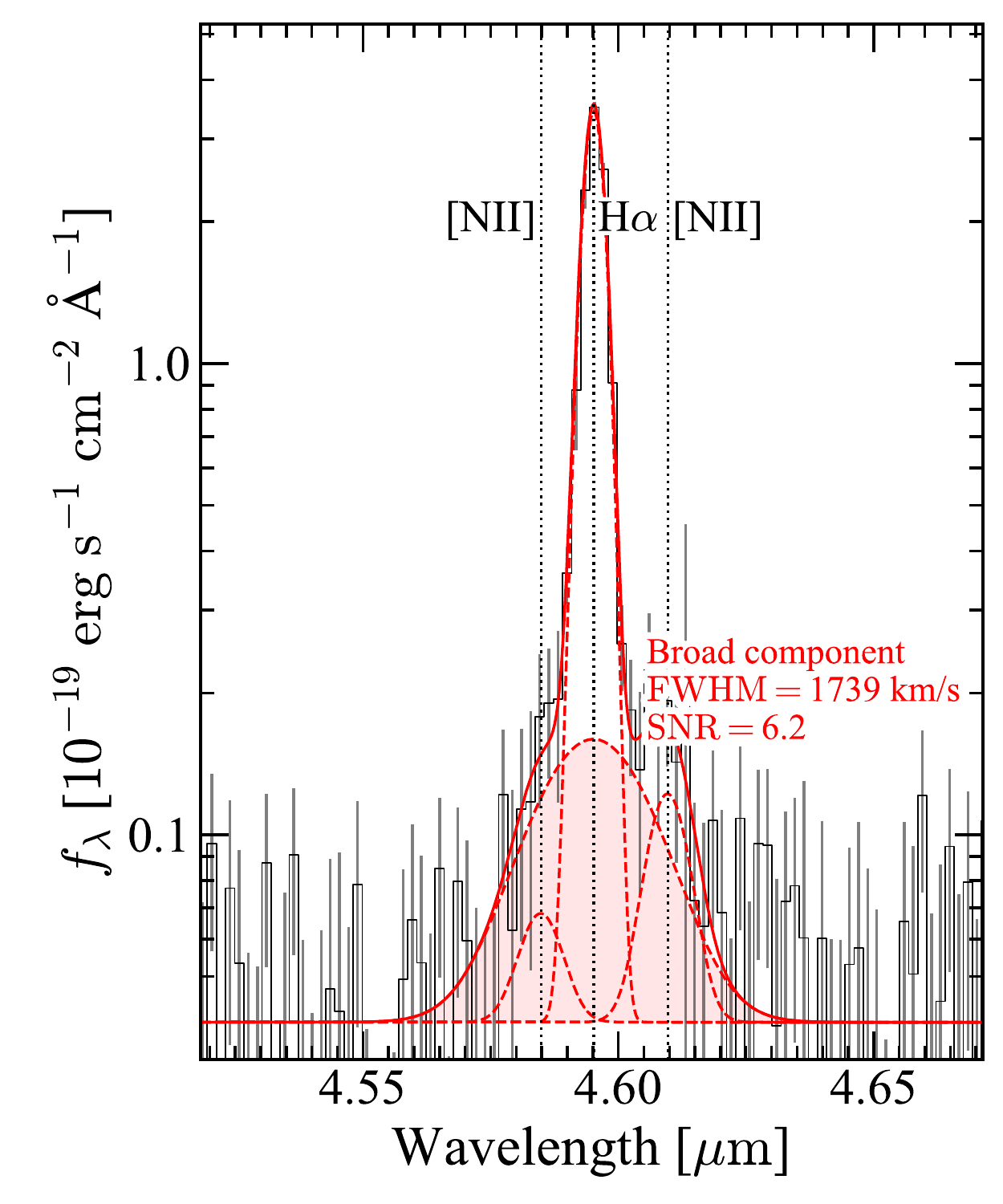}
\end{center}
\end{minipage}
\caption{
Same as Figure \ref{fig_spec_line_1} but for CEERS\_01665, CEERS\_00672, CEERS\_02782, and CEERS\_00397.
}
\label{fig_spec_line_2}
\end{figure*}

\begin{figure*}
\centering
\begin{minipage}{0.47\hsize}
\begin{center}
\includegraphics[width=0.99\hsize, bb=2 9 707 430]{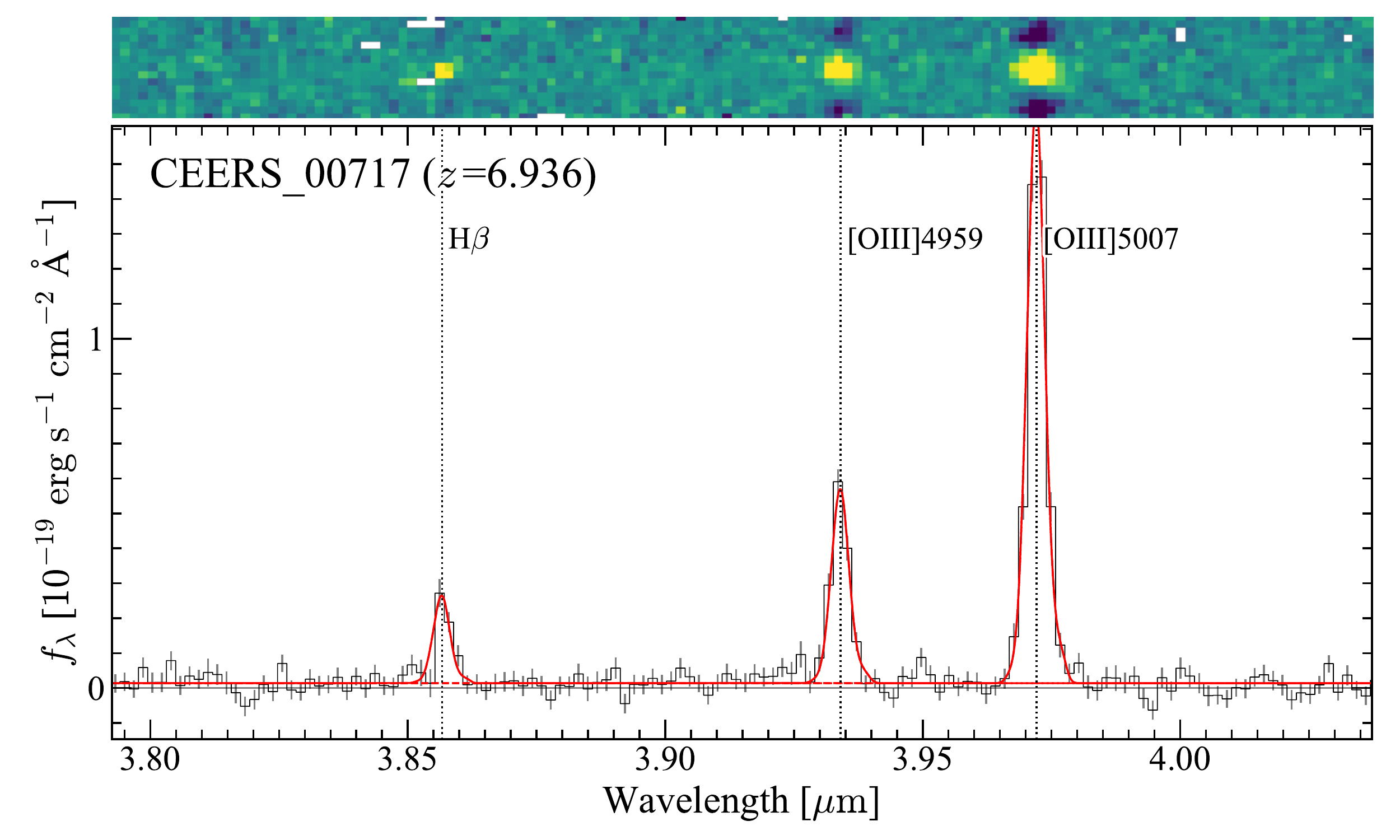}
\end{center}
\end{minipage}
\begin{minipage}{0.277\hsize}
\begin{center}
\includegraphics[width=0.99\hsize, bb=9 9 425 430]{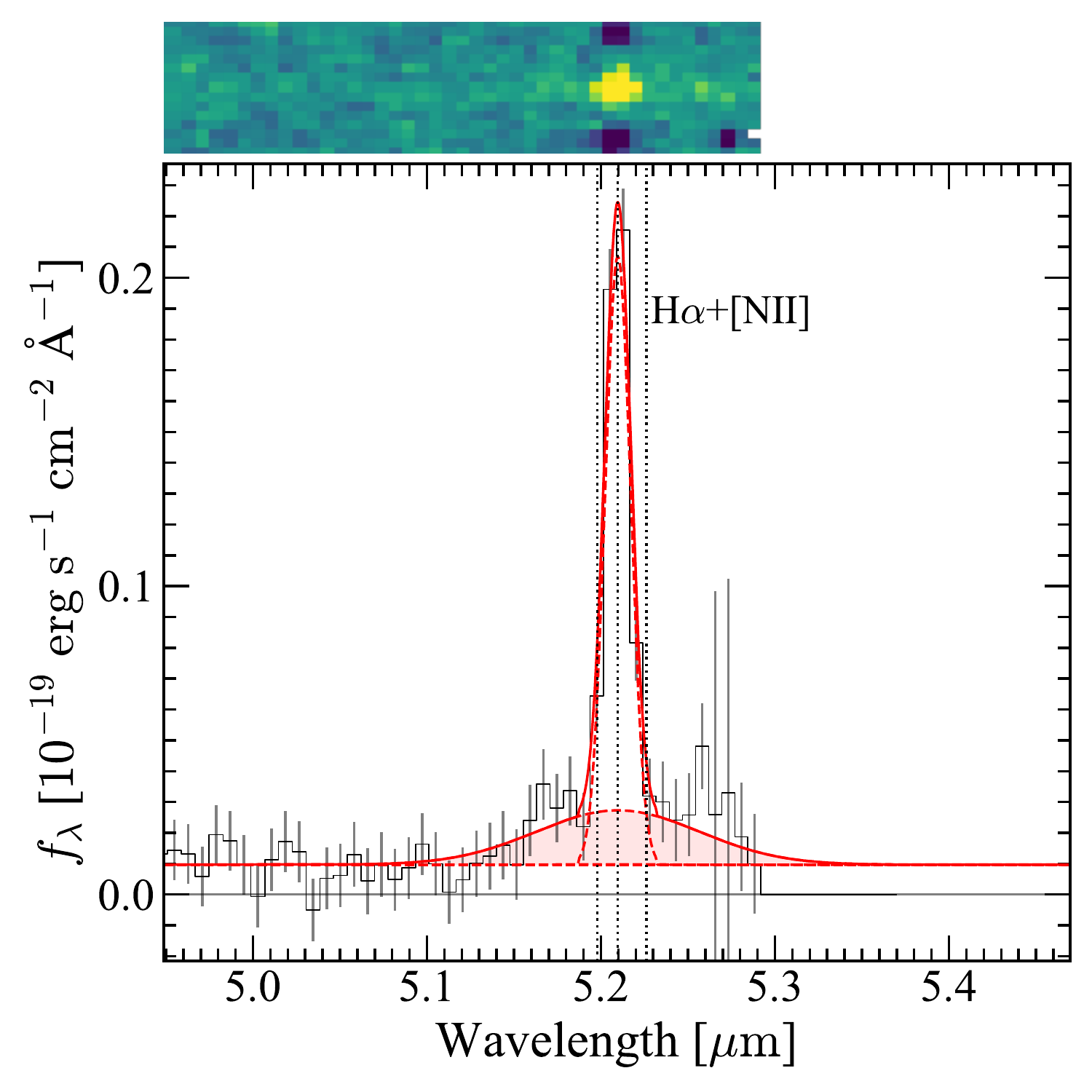}
\end{center}
\end{minipage}
\begin{minipage}{0.227\hsize}
\begin{center}
\includegraphics[width=0.99\hsize, bb=7 9 354 425]{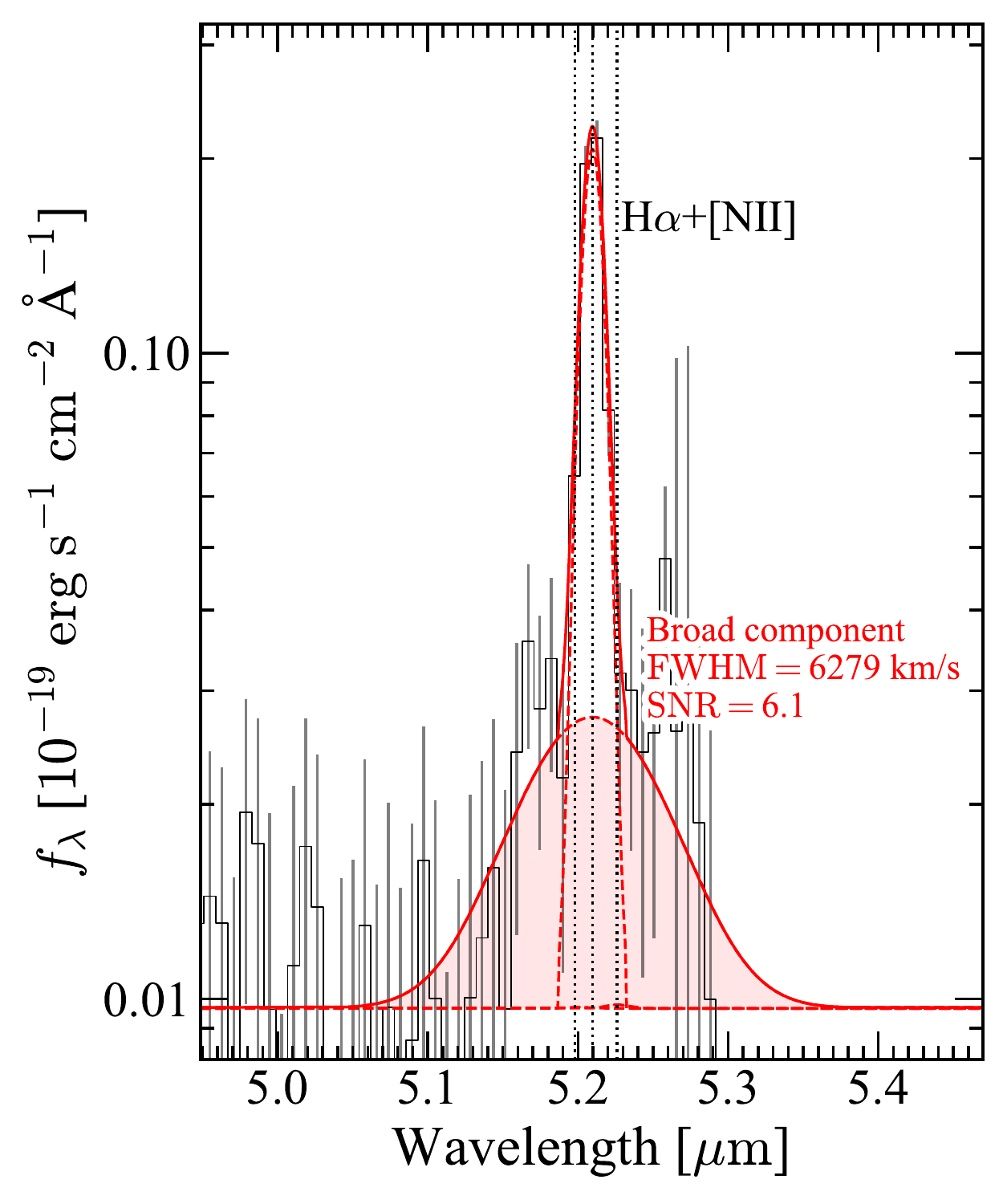}
\end{center}
\end{minipage}
\begin{minipage}{0.47\hsize}
\begin{center}
\includegraphics[width=0.99\hsize, bb=2 9 707 430]{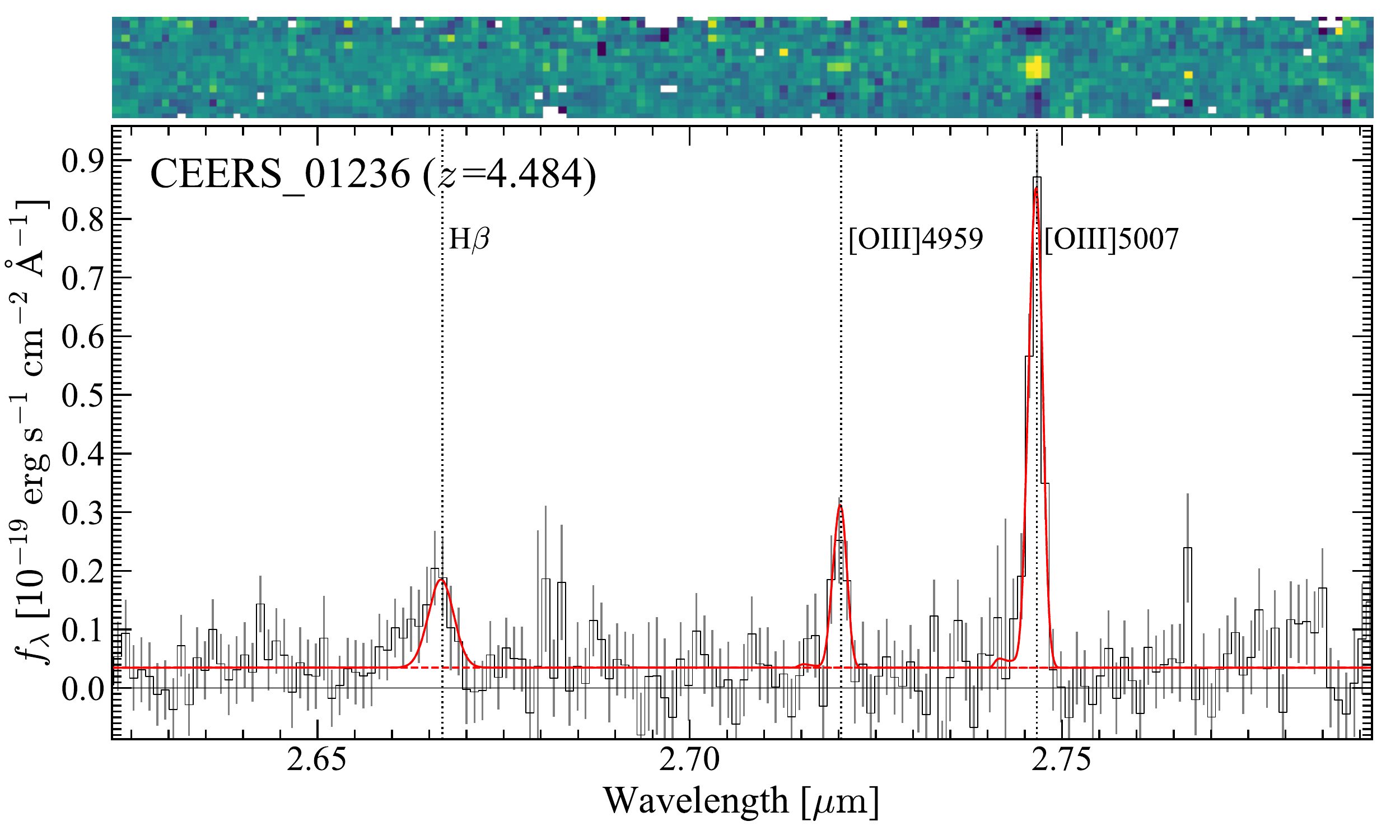}
\end{center}
\end{minipage}
\begin{minipage}{0.277\hsize}
\begin{center}
\includegraphics[width=0.99\hsize, bb=9 9 425 430]{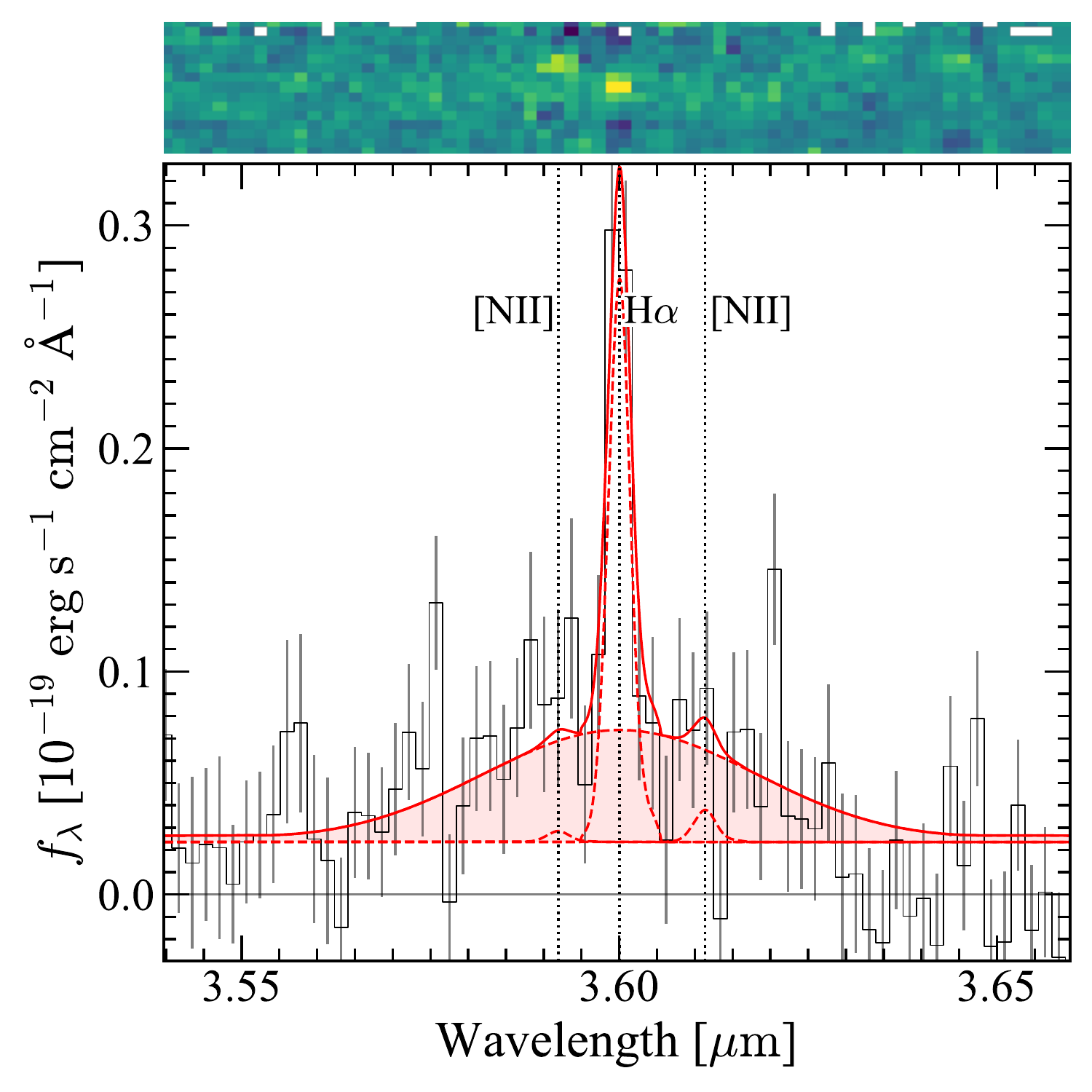}
\end{center}
\end{minipage}
\begin{minipage}{0.227\hsize}
\begin{center}
\includegraphics[width=0.99\hsize, bb=7 9 354 425]{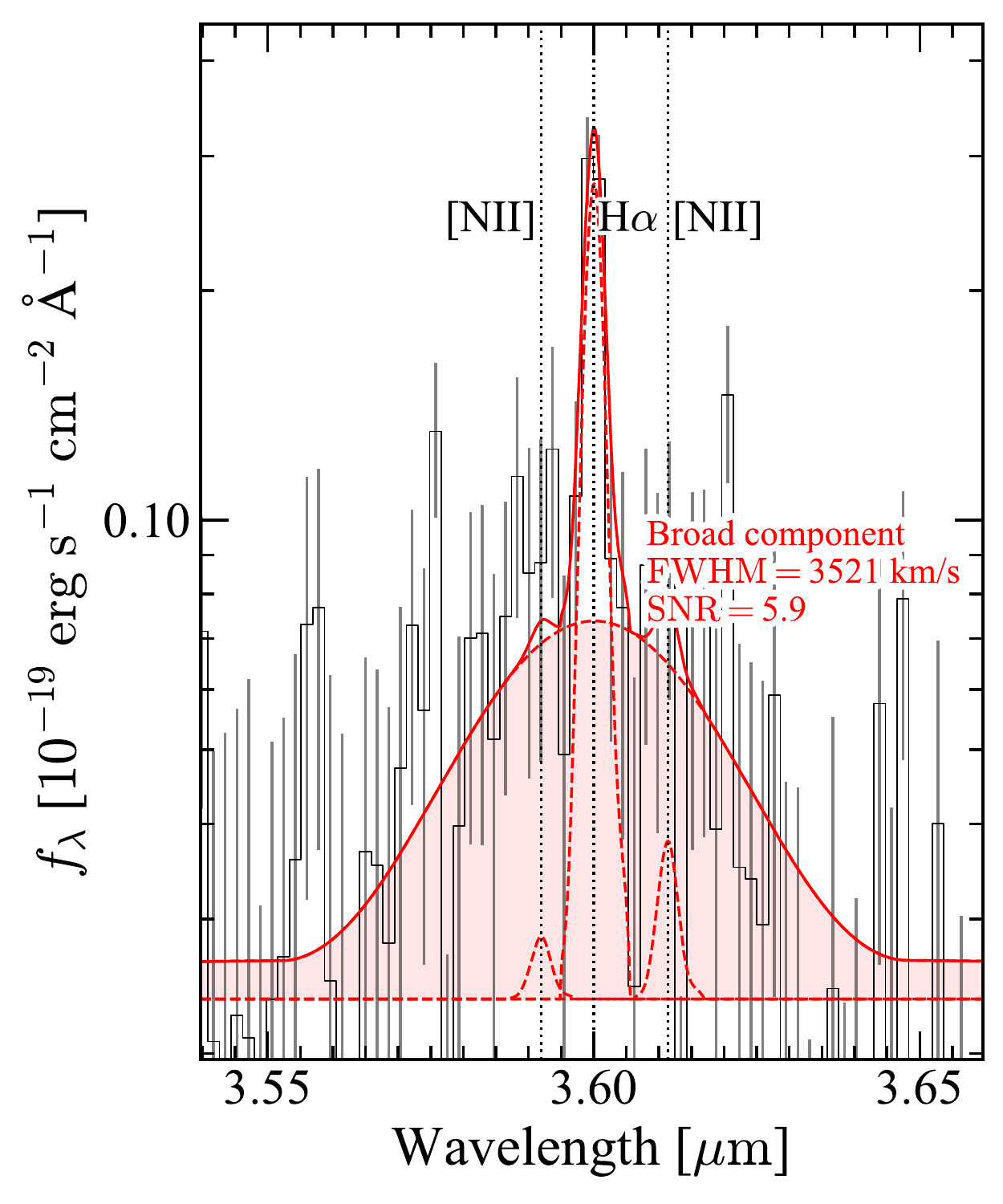}
\end{center}
\end{minipage}
\vspace{-0.2cm}
\caption{
Same as Figure \ref{fig_spec_line_1} but for CEERS\_00717 and CEERS\_01236.
}
\label{fig_spec_line_3}
\end{figure*}

\begin{deluxetable*}{cccccccccc}
\tablecaption{Physical Properties of Our Faint Type-1 AGNs}
\label{tab_N23}
\tablehead{\colhead{Name} & \colhead{R.A.} & \colhead{Decl.} & \colhead{$z_\m{spec}$} & \colhead{$M_\m{UV}$} & \colhead{$12+\m{log(O/H)}$} & \colhead{$\m{E(B-V)}$} & \colhead{$\m{SNR_{H\alpha,broad}}$} & \redc{$\Delta$AIC$^\#$}\\
\colhead{}& \colhead{}& \colhead{}& \colhead{}& \colhead{(mag)}& \colhead{}& \colhead{(mag)}& \colhead{}& \colhead{}}
\startdata
\multicolumn{9}{c}{Our Final Broad-Line AGN Sample}\\
CEERS\_01244 & 14:20:57.76 & $+$53:02:09.8 & $4.478$  & $-19.50^{+0.18}_{-0.18}$ & $7.67^{+0.20}_{-0.23}$ & 0.00 & 37.3 & \redc{789.8}\\
GLASS\_160133 & 00:14:19.27 & $-$30:25:27.8 & $4.015$  & $-18.94^{+0.08}_{-0.09}$ & $7.95^{+0.06}_{-0.05}$ & 0.16 & 32.5 & \redc{2022.4}\\
GLASS\_150029 & 00:14:18.52 & $-$30:25:21.3 & $4.583$  & $-19.20^{+0.10}_{-1.08}$ & $7.70^{+0.09}_{-0.08}$ & 0.17 & 18.1 & \redc{431.1}\\
CEERS\_00746$^\dagger$ & 14:19:14.19 & $+$52:52:06.5 & $5.624$  & $-17.82^{+0.33}_{-0.61}$ & $8.29^{+0.24}_{-0.22}$ & 1.38 & 17.3 & \redc{210.5}\\
CEERS\_01665 & 14:20:42.77 & $+$53:03:33.7 & $4.483$  & $-21.66^{+0.07}_{-0.07}$ & $8.05^{+0.19}_{-0.19}$ & 0.42 & 11.6 & \redc{132.0}\\
CEERS\_00672 & 14:19:33.52 & $+$52:49:58.7 & $5.666$  & $-16.71^{+0.31}_{-1.16}$ & $8.57^{+0.30}_{-0.23}$ & 0.80 & 11.6 & \redc{51.9}\\
CEERS\_02782$^\dagger$ & 14:19:17.63 & $+$52:49:49.0 & $5.241$  & $-19.50^{+0.12}_{-0.00}$ & $7.51^{+0.16}_{-0.16}$ & 0.28 & 11.2 & \redc{78.6}\\
CEERS\_00397 & 14:19:20.69 & $+$52:52:57.7 & $6.000$  & $-21.23^{+0.17}_{-0.24}$ & $7.87^{+0.14}_{-0.14}$ & 0.19 & 6.2 & \redc{51.2}\\
CEERS\_00717 & 14:20:19.54 & $+$52:58:19.9 & $6.936$  & $-21.49^{+0.12}_{-0.12}$ & $7.87^{+0.17}_{-0.18}$ & 0.00 & 6.1 & \redc{23.9}\\
CEERS\_01236 & 14:20:34.87 & $+$52:58:02.2 & $4.484$  & $-19.61^{+0.27}_{-0.23}$ & $7.55^{+0.29}_{-0.28}$ & 0.00 & 5.9 & \redc{23.3}\\
\hline
\multicolumn{9}{c}{Other Possible Candidates with $\m{SNR_{broad}<5}$}\\
CEERS\_01465 & 14:19:33.12 & $+$52:53:17.7 & $5.274$  & $-19.72^{+1.81}_{-0.45}$ & $8.25^{+0.23}_{-0.22}$ & 0.00 & 4.4 & \redc{\nodata}\\
CEERS\_01019$^*$ & 14:20:08.49 & $+$52:53:26.4 & $8.681$  & $-22.44^{+0.45}_{-0.17}$ & $8.04^{+0.17}_{-0.17}$ & 0.00 & 3.1(H$\beta$)$^*$ & \redc{\nodata}\\
\enddata
\tablecomments{The name, spectroscopic redshift ($z_\m{spec}$), UV magnitude ($M_\m{UV}$), metallicity ($12+\m{log(O/H)}$), and extinction ($\m{E(B-V)}$) are taken from \citet{2023arXiv230112825N}.
\\
$^\dagger$ CEERS\_00746 and CEERS\_02782 are reported as CEERS 3210 and CEERS 1670, respectively, in \citet{2023ApJ...954L...4K}.\\
$^*$ CEERS\_01019 is reported as CEERS\_1019 in \citet{2023arXiv230308918L}. The signal-to-noise ratio of the broad H$\beta$ line instead of H$\alpha$ is presented.\\
\redc{$^\#$ Difference between AIC values for the fittings without the broad component and with the broad component.}
}
\end{deluxetable*}

\begin{figure}
\centering
\begin{center}
\includegraphics[width=0.99\hsize, bb=2 8 389 356]{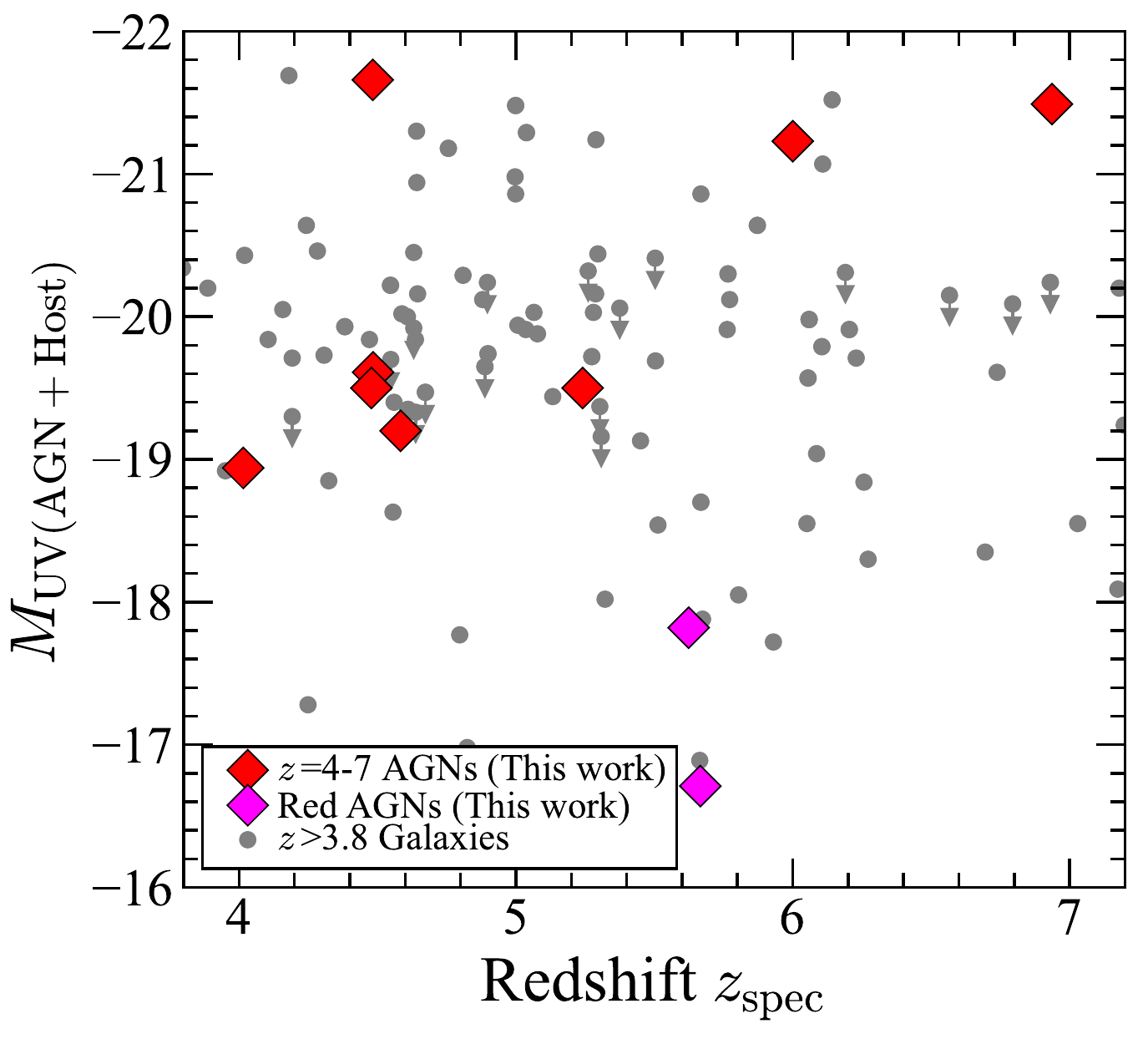}
\end{center}
\caption{
UV magnitude and spectroscopic redshifts of our selected broad-line AGNs (the diamonds).
The magenta diamonds are red AGNs ($\m{E(B-V)}>0.5$) and the red diamonds are other AGNs identified in this study.
The gray circles show star-forming galaxies in \citet{2023arXiv230112825N}.
}
\label{fig_MUV_z}
\end{figure}

\subsection{Our AGN Sample}

Using the selection criteria above, we select a total of 10 galaxies that show broad H$\alpha$ emission lines.
Figures \ref{fig_spec_line_1}-\ref{fig_spec_line_3} show NIRSpec spectra of the selected galaxies, and Table \ref{tab_N23} summarizes the physical properties derived in \citet{2023arXiv230112825N}.
CEERS\_00746 and CEERS\_02782 were already reported as broad-line AGNs in \citet{2023ApJ...954L...4K} as CEERS 3210 and CEERS 1670, respectively, and our measured broad line widths are consistent with those of \citet{2023ApJ...954L...4K} within $1-2\sigma$.
These 10 galaxies show significant broad-line emission only in a permitted line (H$\alpha$), and their forbidden lines, especially {\sc[Oiii]$\lambda$5007}, which is detected with a signal-to-noise ratio higher than H$\alpha$, are well fitted with narrow ($\m{FWHM}<500\ \m{km\ s^{-1}}$) components.
GLASS\_160133 and GLASS\_150029 show outflow component whose line width is $\m{FWHM}=540$ and $140\ \m{km\ s^{-1}}$ in {\sc[Oiii]$\lambda$5007}, respectively.
\redc{We evaluate the fittings using the Akaike Information Criterion (AIC; \citealt{1974ITAC...19..716A}), which is defined by $\m{AIC}=-2\m{log}L+2k$, where $L$ is the maximum likefood and $k$ is the number of free parameters.
As shown in Table \ref{tab_N23}, all of the selected objects are well fitted with the broad component rather than without the broad component with $\Delta\m{AIC}>20$.}
These properties are consistent with the fact that they are broad-line type-1 AGNs.

Figure \ref{fig_MUV_z} shows the UV magnitudes and redshifts of our selected AGNs.
Our selected AGNs are typically faint with a UV magnitude of $-22\lesssim M_\m{UV}\lesssim-17$ mag, much fainter than the low-luminosity quasars found in ground-based observations at similar redshifts \citep[e.g.,][]{2018PASJ...70S..34A,2020ApJ...904...89N,2018ApJ...869..150M}.
None of our AGNs is selected as X-ray AGNs in \citet{2019ApJ...884...19G}, probably because their X-ray emission is too faint to be detected with current X-ray observations.
Our AGNs are widely distributed in the UV magnitude and redshift space and are not biased compared to the star-forming galaxy sample in \citet{2023arXiv230112825N}, \redc{except for a possible gap in $-21<M_\m{UV}<-20\ \m{mag}$ probably due to the small sample size}.
Since we have found 10 AGNs in the sample of 185 galaxies at $z>3.8$, about $5\%$ of the galaxies at $z=4-7$ with $-22\lesssim M_\m{UV}\lesssim-17$ mag are broad-line type-1 AGNs, which is higher than $z\sim0$ galaxies, as discussed in Section \ref{ss_uvlf}.
\redc{Note that this value, ``5\%", should be taken with caution, because the observed AGN fraction depends on the sensitivity (see discussion in Section \ref{ss_uvlf}) and the selection function for the parent spectroscopic sample that could be very complex.}

There are two other galaxies showing a broad permitted line but not selected due to their inefficient signal-to-noise ratio.
CEERS\_01465 at $z_\m{spec}=5.269$ shows a possible broad H$\alpha$ emission line with an FWHM of $3603^{+1136}_{-291}\ \m{km\ s^{-1}}$, but the signal-to-noise ratio of the broad line is $\sim4$, lower than the threshold value of our selection criteria.
The other galaxy is CEERS\_01019 at $z_\m{spec}=8.679$, which was first confirmed spectroscopically with Ly$\alpha$ by \citet{2015ApJ...810L..12Z}.
This galaxy was reported to show a $4.3\sigma$ {\sc Nv}$\lambda$1243 whose ionization potential is 77 eV \citep{2018MNRAS.479.1180M}, and recently, \citet{2023arXiv230308918L} report a broad H$\beta$ emission line with $\m{FWHM}=1196\pm349\ \m{km\ s^{-1}}$ with $\sim2\sigma$ ($3.53/1.51=2.3$), indicating that this galaxy is an AGN.
Our analysis also shows a similar broad H$\beta$ line with $\m{FWHM}=1664^{+475}_{-302}\ \m{km\ s^{-1}}$ with $\sim3\sigma$, consistent with the results in \citet{2023arXiv230308918L}.
Although these two galaxies are good candidates for broad-line AGNs, we do not include them in our final sample due to the moderate signal-to-noise ratio of the broad line.

Broad emission lines are sometimes interpreted as mergers or outflows, but the observed broad emission lines in our AGNs seen only in H$\alpha$ are not likely made by mergers or outflows.
Merging galaxies show broad emission lines both in the permitted and forbidden lines and the emission lines are decomposed with two or more narrow components, typically with $\m{FWHM}<500\ \m{km\ s^{-1}}$ \citep[e.g.,][]{2019PASJ...71...71H,2021A&A...653A.111R}, different from the broad ($\m{FWHM}>1000\ \m{km\ s^{-1}}$) single component only seen in H$\alpha$ in our AGNs.
Galaxies with strong outflows also show broad emission lines, but such broad lines are usually seen in both permitted and forbidden lines \citep[e.g.,][]{2019ApJ...873..102F,2022ApJ...929..134X}.
Moreover, the line width of the broad line in star-forming galaxies is typically $\m{FWHM}\sim300-500\ \m{km\ s^{-1}}$, not larger than $1000\ \m{km\ s^{-1}}$.
Type-2 AGNs sometimes show broad emission lines due to outflows with $\m{FWHM}>1000\ \m{km\ s^{-1}}$, but this broad component is again seen in both {\sc[Oiii]} and H$\alpha$ lines \citep[e.g.,][]{2014ApJ...787...38F,2014ApJ...796....7G}.
One remaining possibility is a strong low-metallicity outflow whose metallicity and ionization parameter are tuned to show a broad component only in H$\alpha$, but such outflow is not observed even in low-metallicity galaxies \citep{2022ApJ...929..134X}.
Given these comparisons, it is reasonable to interpret that our selected galaxies harbor broad-line type-1 AGNs.

\section{Physical Properties of Our AGNs}\label{ss_result}

\begin{figure*}
\centering
\begin{center}
\includegraphics[height=0.1\hsize, bb=1 0 1153 145]{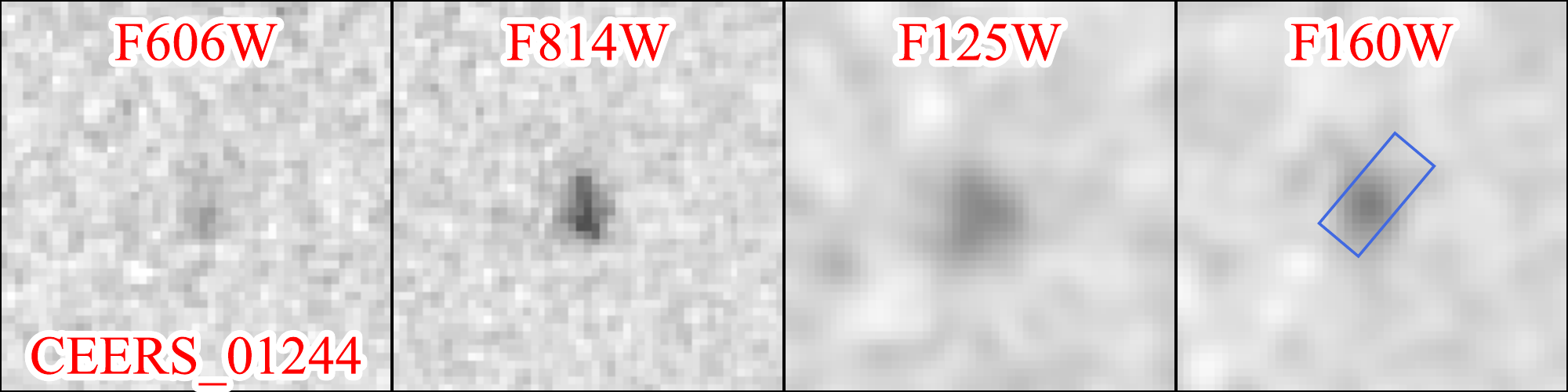}
\end{center}
\vspace{-0.5cm}
\begin{center}
\includegraphics[height=0.1\hsize, bb=1 0 1153 145]{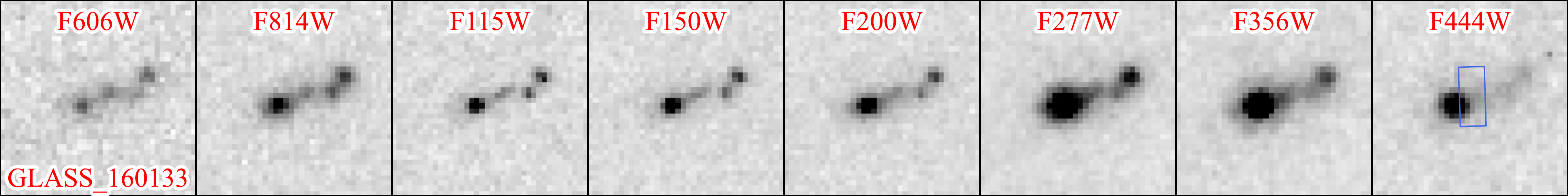}
\end{center}
\vspace{-0.5cm}
\begin{center}
\includegraphics[height=0.1\hsize, bb=1 0 1153 145]{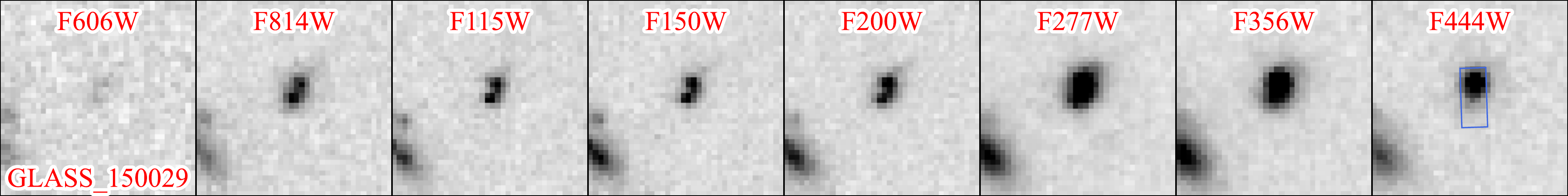}
\end{center}
\vspace{-0.5cm}
\begin{center}
\includegraphics[height=0.1\hsize, bb=1 0 1153 145]{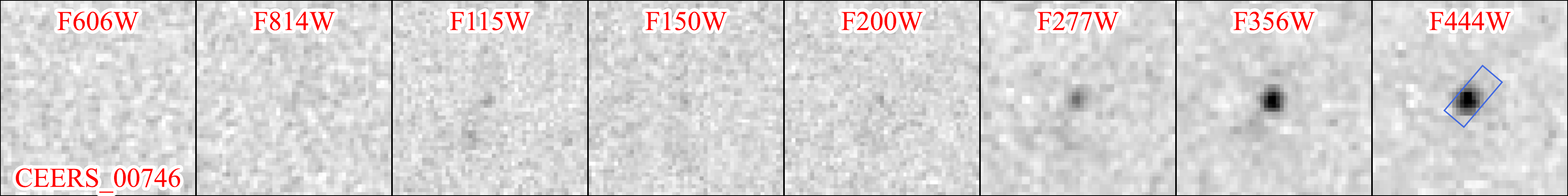}
\end{center}
\vspace{-0.5cm}
\begin{center}
\includegraphics[height=0.1\hsize, bb=1 0 1153 145]{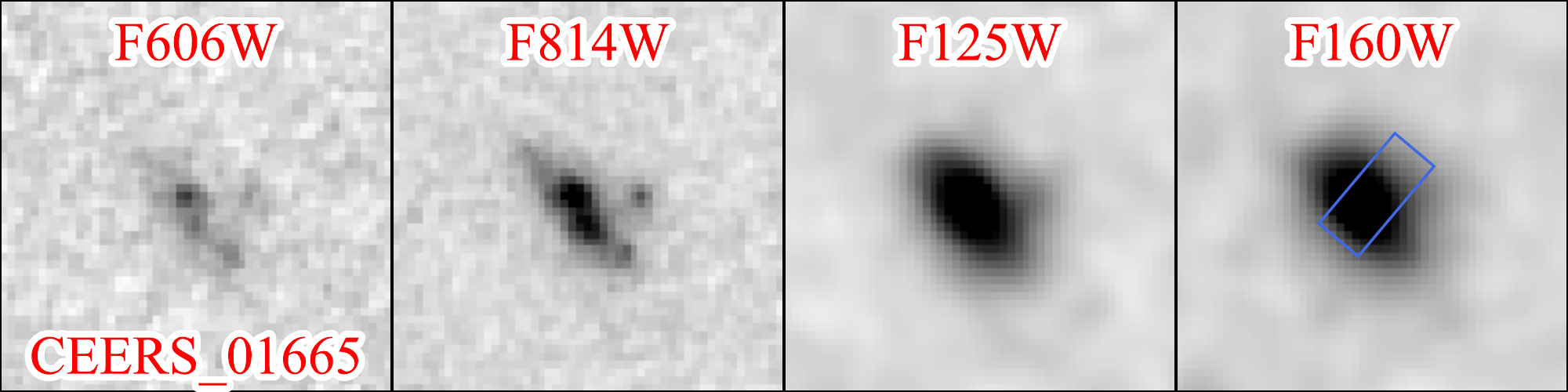}
\end{center}
\vspace{-0.5cm}
\begin{center}
\includegraphics[height=0.1\hsize, bb=1 0 1153 145]{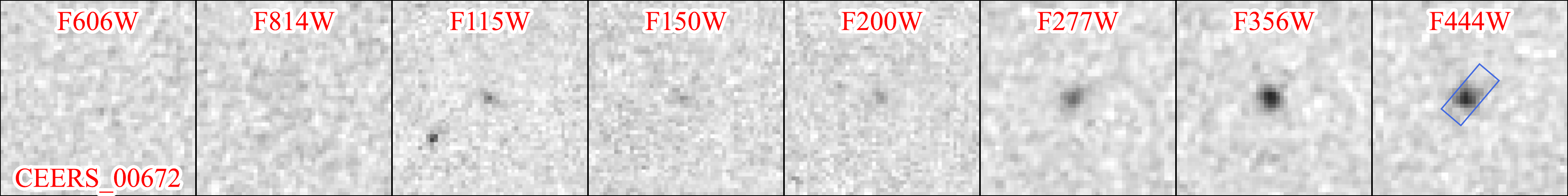}
\end{center}
\vspace{-0.5cm}
\begin{center}
\includegraphics[height=0.1\hsize, bb=1 0 1153 145]{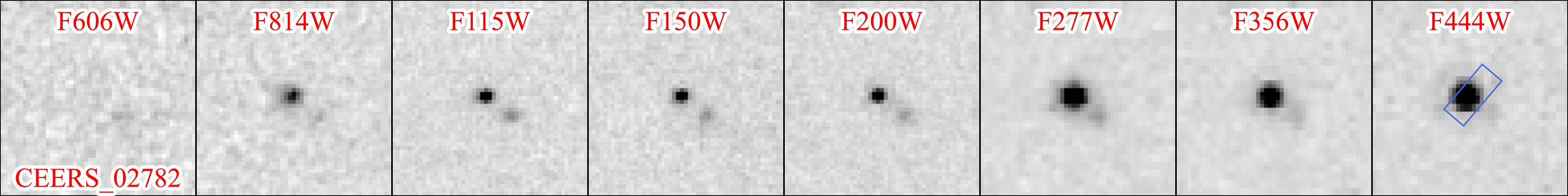}
\end{center}
\vspace{-0.5cm}
\begin{center}
\includegraphics[height=0.1\hsize, bb=1 0 1153 145]{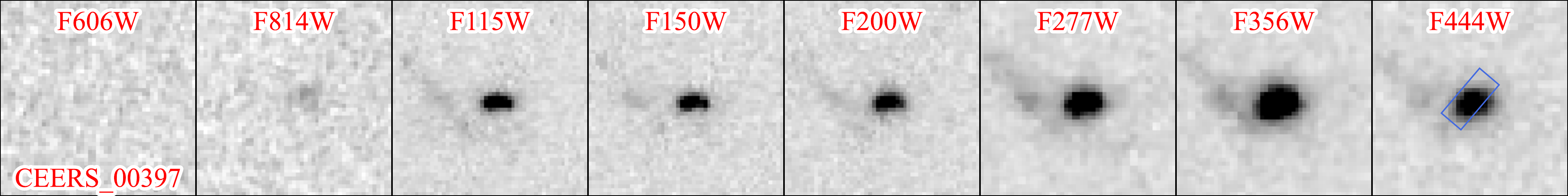}
\end{center}
\vspace{-0.5cm}
\begin{center}
\includegraphics[height=0.1\hsize, bb=1 0 1153 145]{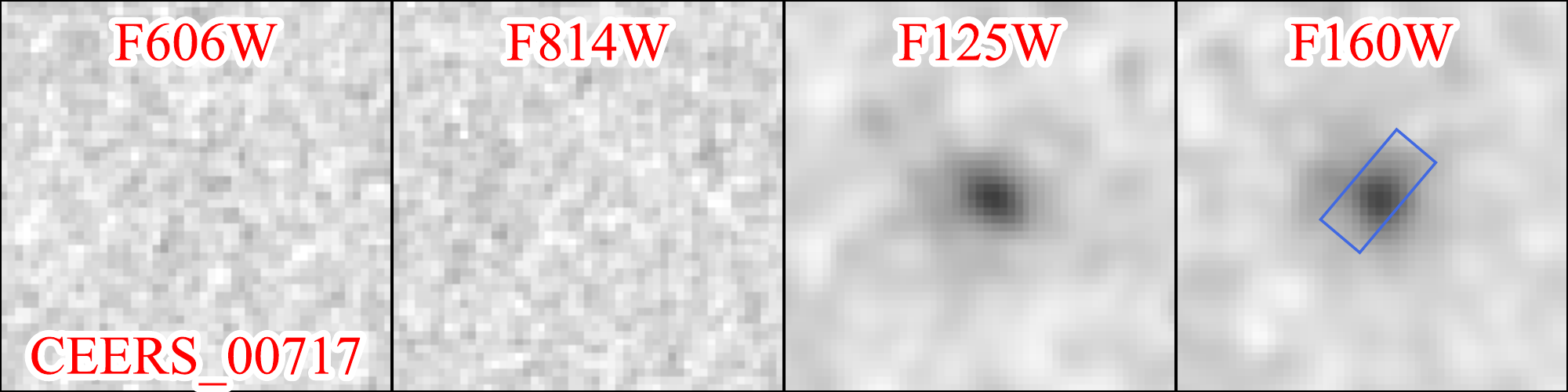}
\end{center}
\vspace{-0.5cm}
\begin{center}
\includegraphics[height=0.1\hsize, bb=1 0 1153 145]{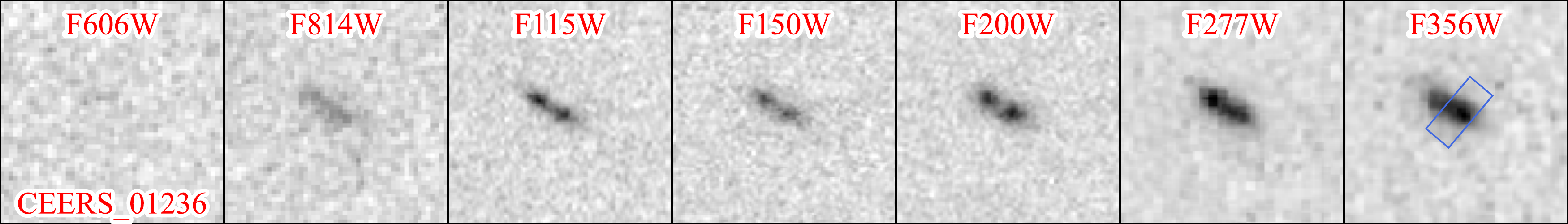}
\end{center}
\caption{
$3\arcsec\times3\arcsec$ images of our faint type-1 AGNs. The HST/ACS F606W, F814W, and the JWST/NIRCam F115W, F150W, F200W, F277W, F356W, and F444W images are shown.
For objects whose NIRCam images are not available, we instead show the HST/WFC3 F125W and F160W images.
The NIRSpec MSA aperture is shown with the blue rectangle in the longest wavelength band.
More than half of the sources show extended morphologies, indicating that the total lights are significantly contributed from their host galaxies.
}
\label{fig_snapshot}
\end{figure*}

\begin{figure*}
\centering
\begin{center}
\includegraphics[width=0.9\hsize, bb=0 0 2025 810]{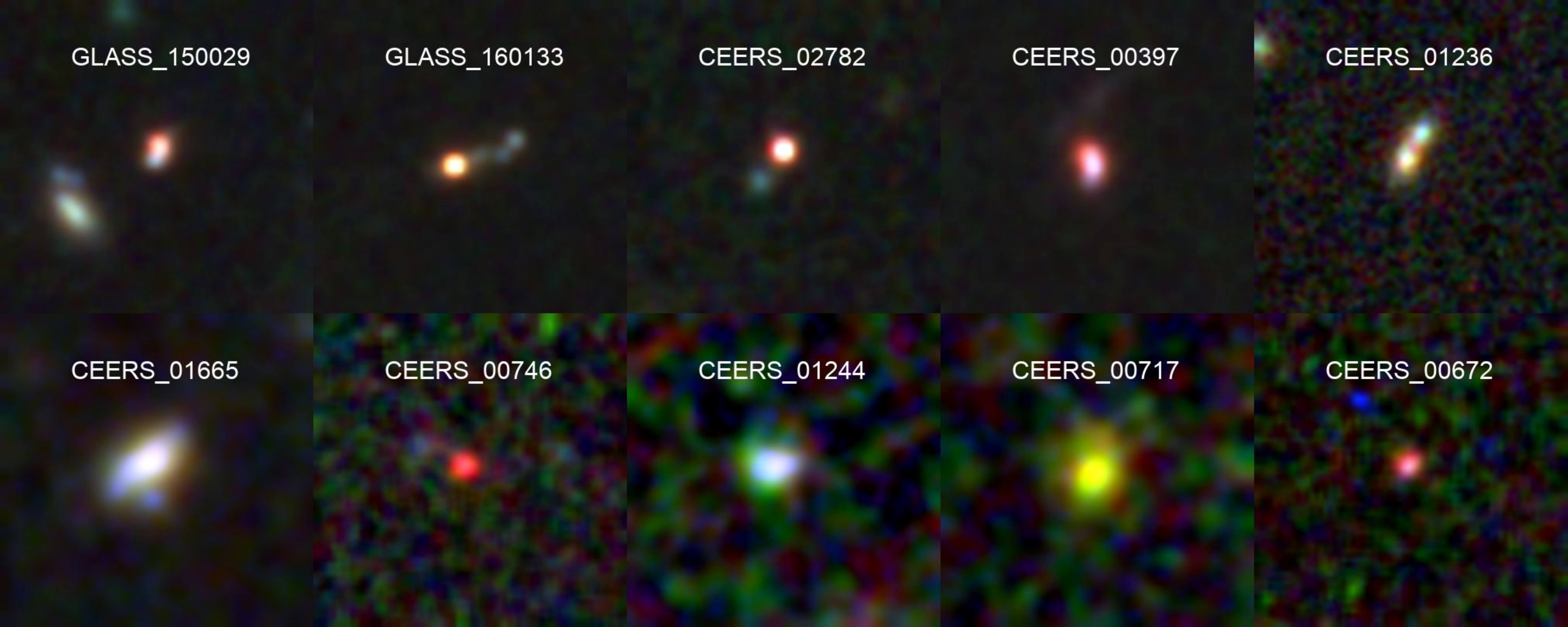}
\end{center}
\caption{
\redc{False color stamps of our faint AGNs.
We use JWST/NIRCam F115W, F200W, and F356W images, while HST/ACS F814W and HST/WFC3 F125W and F160W are used for the objects whose NIRCam images are not available.
Each thumbnail is 2\arcsec$\times$2\arcsec in size.}
}
\label{fig_rgb}
\end{figure*}

\subsection{Morphology}\label{ss_mor}

Figure \ref{fig_snapshot} shows snapshots of our selected AGNs, \redc{and Figure \ref{fig_rgb} present their false color images}.
Remarkably, the selected AGNs have a variety of morphologies, including not only compact point sources but also extended sources.
CEERS\_00746, CEERS\_00672, and CEERS\_02782 show compact morphologies consistent with the point-spread function (PSF).
Among these three sources, CEERS\_00746 (CEERS 3210 in \citealt{2023ApJ...954L...4K}) and CEERS\_00672 show red spectral-energy distributions (SEDs) and large dust attenuation estimated from the Balmer decrement as presented in Section \ref{ss_dust}, indicating that these two sources are dusty red AGNs.
The remaining seven AGNs show extended morphologies, different from the ones reported in \citet{2023ApJ...954L...4K}.
Such a high fraction (70\%) of extended morphologies indicates that the total lights of the faint AGNs with $-22\lesssim M_\m{UV}\lesssim-17$ mag are partly dominated by their host galaxies like Seyfert galaxies, which is discussed in Section \ref{ss_dis_high} \citep[see also][]{2021MNRAS.502..662B}.
\redc{Interestingly, many AGNs show clumpy morphologies that may indicate merger activity, which is consistent with a scenario in which a merger triggers the AGN activity \citep{2006ApJS..163....1H}.}

\begin{figure*}
\centering
\begin{center}
\begin{minipage}{0.49\hsize}
\begin{center}
\includegraphics[width=0.99\hsize, bb=11 8 426 356]{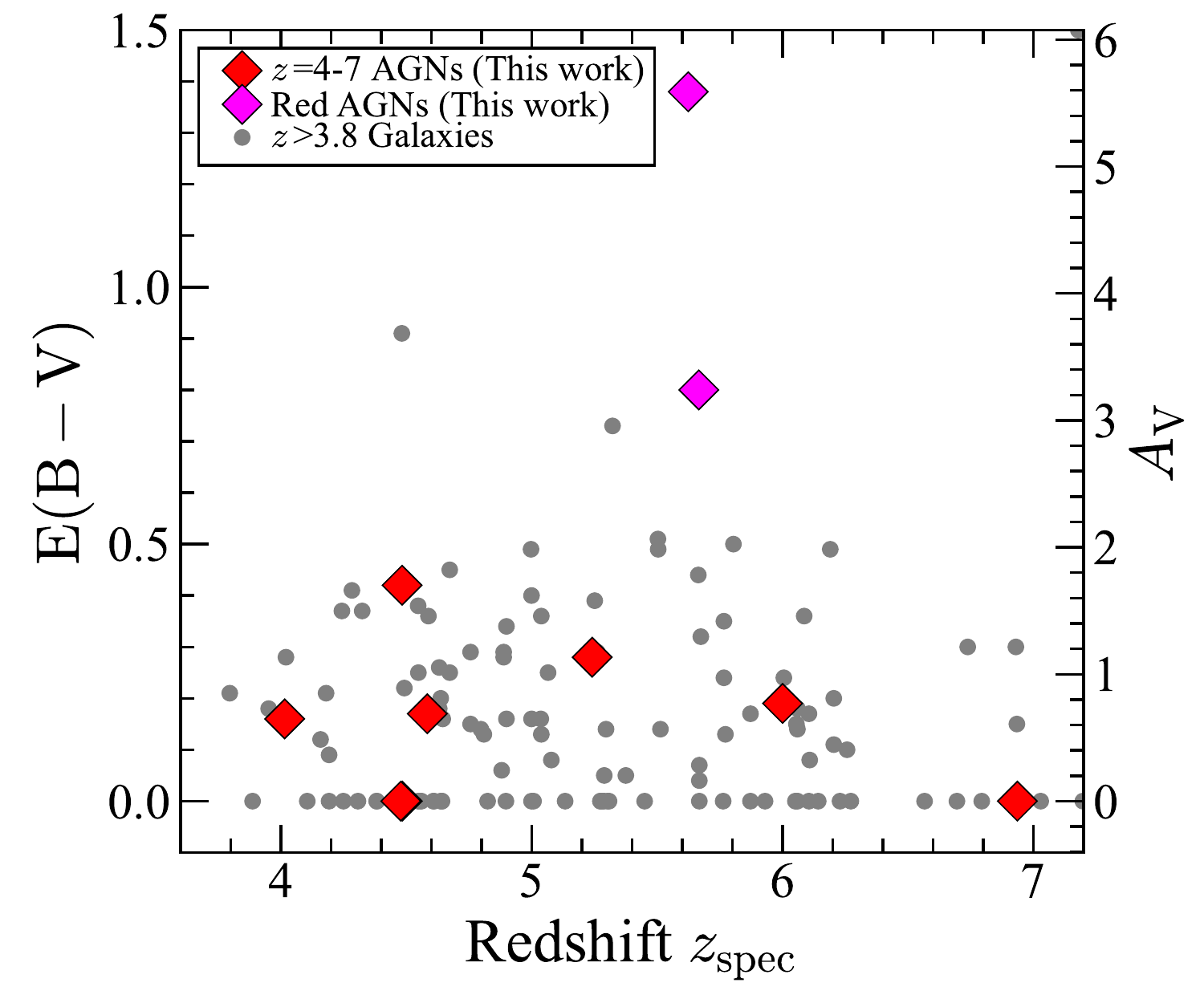}
\end{center}
\end{minipage}
\begin{minipage}{0.49\hsize}
\begin{center}
\includegraphics[width=0.99\hsize, bb=11 8 426 356]{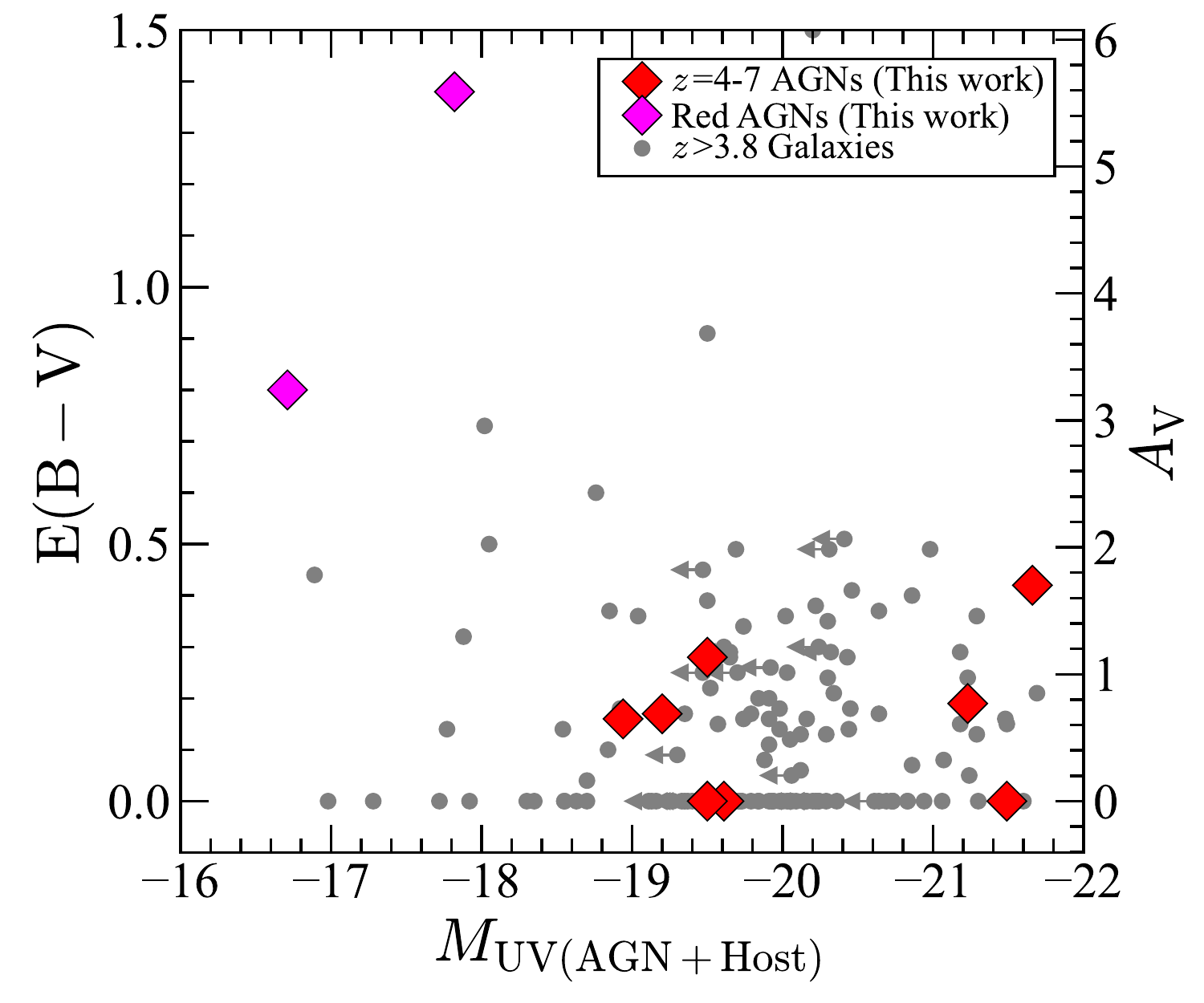}
\end{center}
\end{minipage}
\end{center}
\caption{
Dust attenuation estimated from the Balmer decrement as a function of spectroscopic redshift (left) and UV magnitude (right).
The magenta diamonds are red AGNs ($\m{E(B-V)}>0.5$) and the red diamonds are other AGNs identified in this study.
The gray circles show star-forming galaxies in \citet{2023arXiv230112825N}.
}
\label{fig_EBV}
\end{figure*}

\begin{figure*}
\centering
\begin{center}
\begin{minipage}{0.49\hsize}
\begin{center}
\includegraphics[width=0.99\hsize, bb=16 3 427 281]{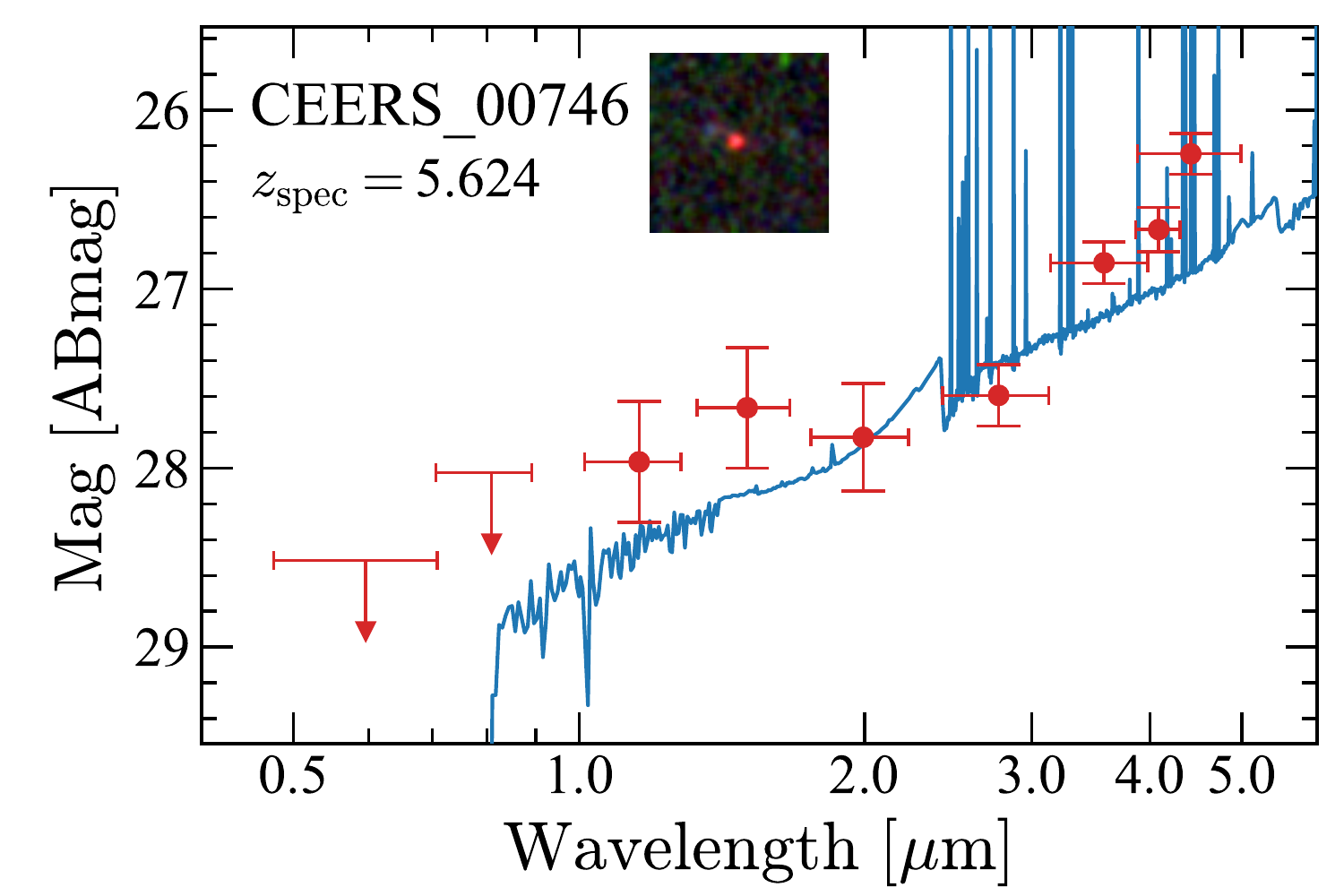}
\end{center}
\end{minipage}
\begin{minipage}{0.49\hsize}
\begin{center}
\includegraphics[width=0.99\hsize, bb=16 3 427 281]{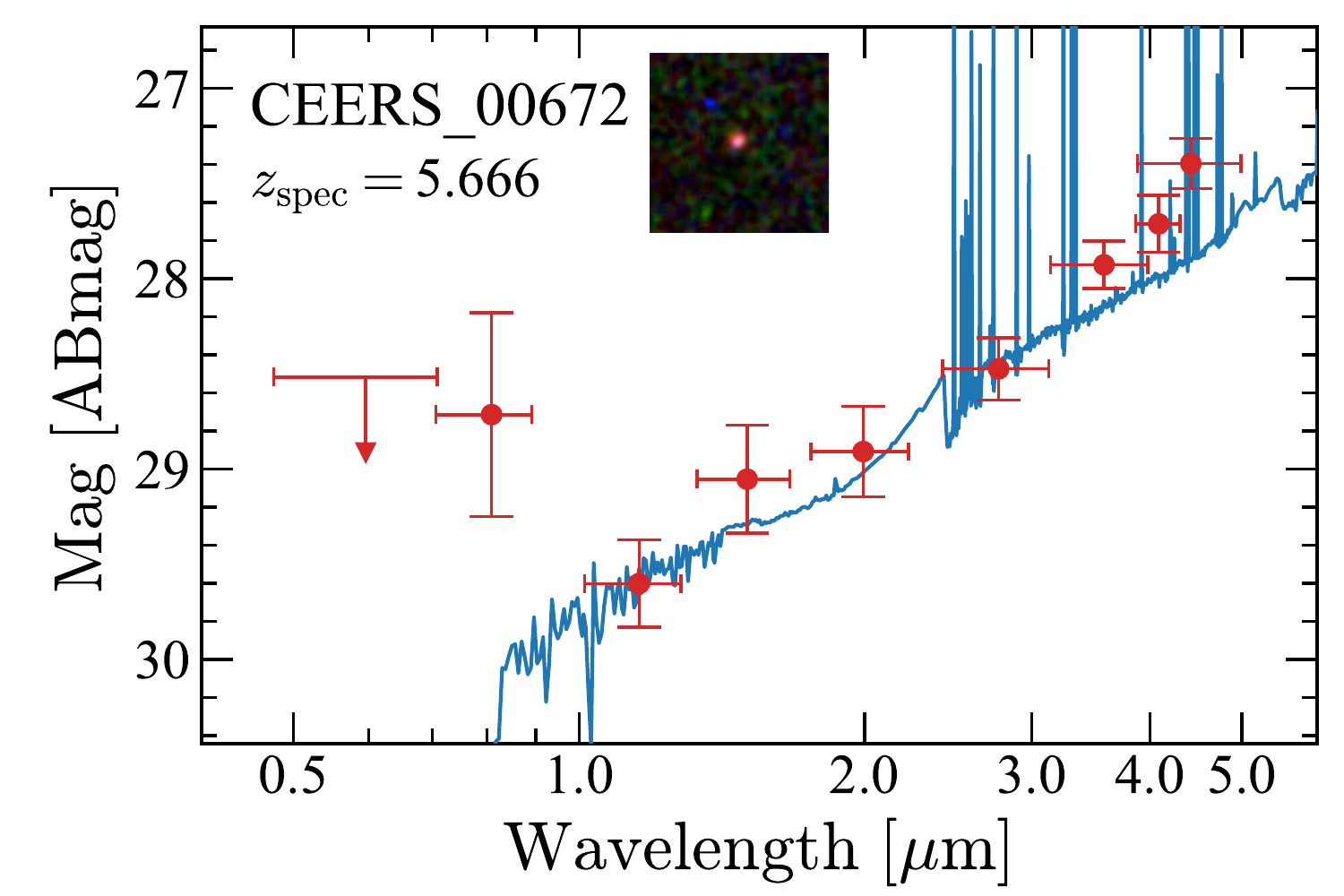}
\end{center}
\end{minipage}
\end{center}
\caption{
SEDs of two Red AGNs, CEERS\_00746 (left) and CEERS\_00672 (right).
CEERS\_00746 is also reported in \citet{2023ApJ...954L...4K}.
The red circles and arrows show the measured magnitudes and 2$\sigma$ upper limits, respectively.
The blue curve is the best-fit SEDs if we use galaxy templates.
\redc{The inset panel shows the false color image in Figure \ref{fig_rgb}.}
}
\label{fig_SED}
\end{figure*}

\subsection{Dust Attenuation}\label{ss_dust}

We evaluate the dust attenuation of our selected AGNs using the color excess, $\m{E(B-V)}$, estimated from the Balmer decrement in \citet{2023arXiv230112825N} assuming the \citet{2000ApJ...533..682C} dust extinction law.
Figure \ref{fig_EBV} shows the dust attenuation of our AGNs as functions of the redshift and the UV magnitude.
Most of our AGNs have negligible or moderate dust attenuation with $\m{E(B-V)}=0.0-0.5$, except for two sources, CEERS\_00746 and CEERS\_00672 with $\m{E(B-V)}=1.4$ and $0.8$, respectively.
CEERS\_00746 was previously reported as a red AGN in \citet{2023ApJ...954L...4K}, and we have newly found an additional red AGN, CEERS\_00672, at $z=5.666$.
The two AGNs show red SEDs as shown in Figure \ref{fig_SED}, consistent with the high dust attenuation inferred from the Balmer decrement.
These findings indicate that the NIRSpec-selected broad-line AGNs include both blue and red AGNs.
The red AGNs are found only in a faint UV-magnitude range with $-18< M_\m{UV}<-16\ \m{mag}$ (Figure \ref{fig_MUV_z}).

\subsection{Emission Lines}

\begin{figure}
\centering
\begin{center}
\includegraphics[width=0.99\hsize, bb=11 11 354 354]{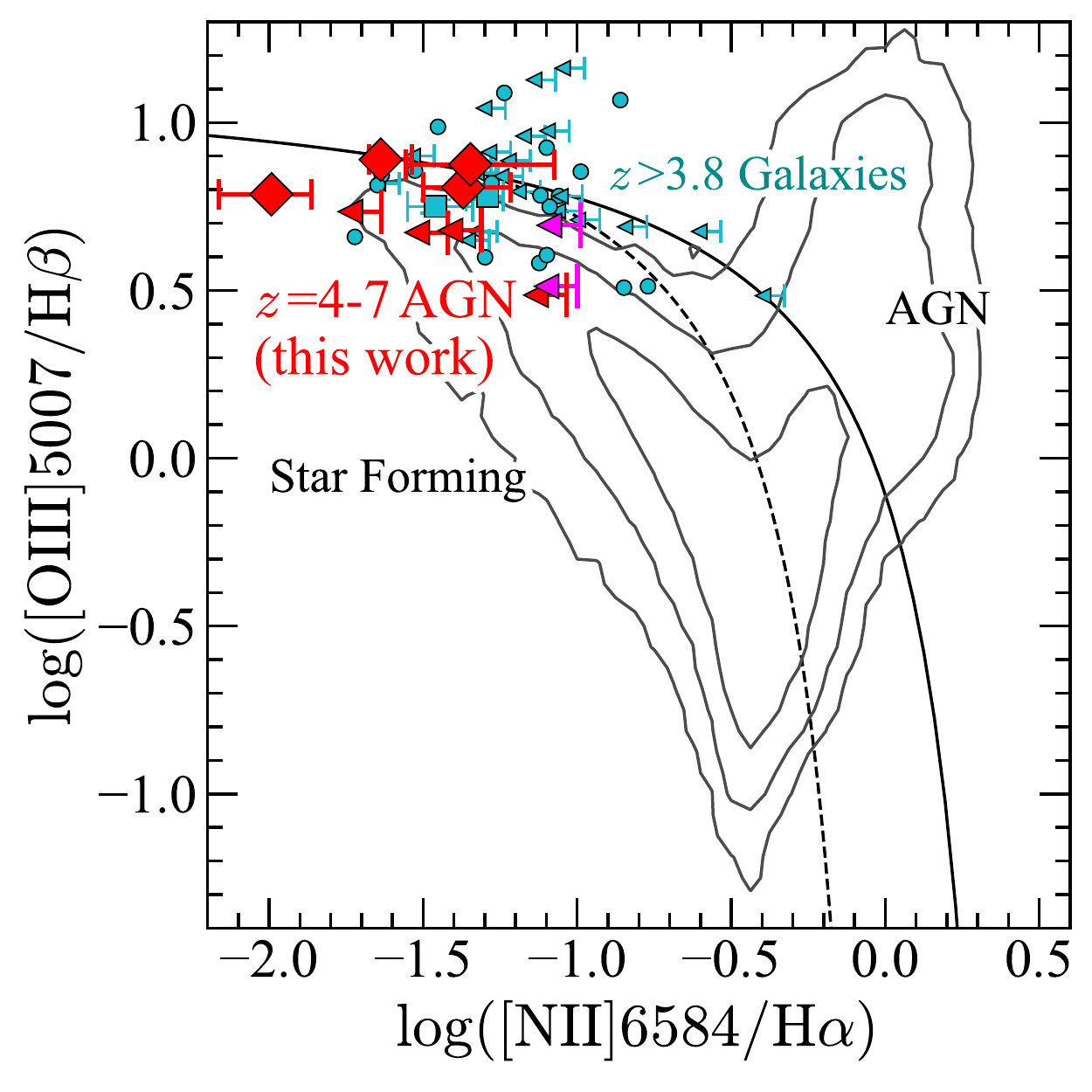}
\end{center}
\caption{
The BPT diagram.
The red symbols show narrow emission line ratios of our galaxies having AGNs at $z=4-7$, and the cyan symbols are other galaxies at $z\gtrsim4$ in \citet{2023arXiv230112825N}.
The black contours show SDSS galaxies at $z\sim0$ \citep{2009ApJS..182..543A}.
The black solid and dashed curves are boundaries between AGNs and star-forming galaxies obtained in \citet{2001ApJ...556..121K} and \citet{2003MNRAS.346.1055K}, respectively.
Our galaxies having AGNs cannot be distinguished from normal star-forming galaxies at $z>4$ with the BPT diagram.
}
\label{fig_BPT}
\end{figure}

In Figure \ref{fig_BPT}, we plot the narrow-line ratios of our AGNs in the BPT diagram \citep{1981PASP...93....5B} that is commonly used to classify galaxies as dominated by emission from AGN or star formation.
Our AGNs are located in the region with high {\sc[Oiii]}/H$\beta$ and low {\sc[Nii]}/H$\alpha$, similar to other galaxies at $z>3.8$ without broad line emission \citep{2023arXiv230112825N,2023arXiv230106696S}, and above the sequence of $z\sim0$ galaxies, as seen in $z=2-3$ star forming galaxies \citep[e.g.,][]{2014ApJ...795..165S,2015ApJ...801...88S,2017ApJ...835...88K}.
These results indicate that our galaxies harboring AGNs cannot be distinguished from normal star-forming galaxies at $z>4$ with the BPT diagram, \redc{as discussed in \citet{2023ApJ...954L...4K}}, because of the low metallicities in these sources.
Indeed, low-metallicity ($Z<Z_\odot$) AGNs with moderately weak {\sc[Nii]} emission are found at $z\sim0$ in \citet{2017ApJ...842...44K}.
As shown in Table \ref{tab_N23}, metallicities of our galaxies having AGNs are typically sub-solar, resulting in the weak {\sc[Nii]} emission in contrast to typical $z\sim0$ AGNs showing strong {\sc[Nii]} emission.

\begin{figure*}
\centering
\begin{center}
\begin{minipage}{0.32\hsize}
\begin{center}
\includegraphics[width=0.99\hsize, bb=9 9 425 424,clip]{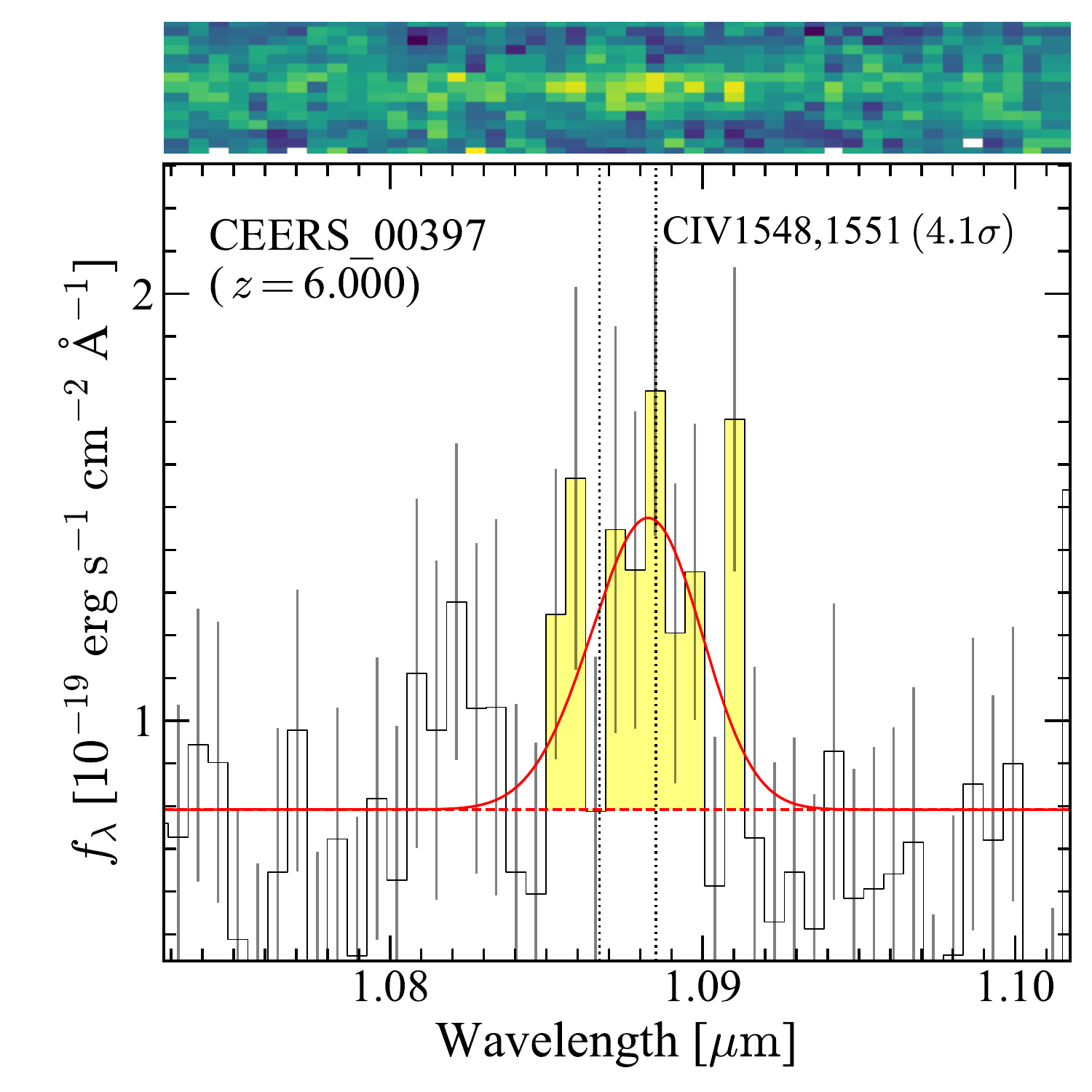}
\end{center}
\end{minipage}
\begin{minipage}{0.32\hsize}
\begin{center}
\includegraphics[width=0.99\hsize, bb=9 9 425 424,clip]{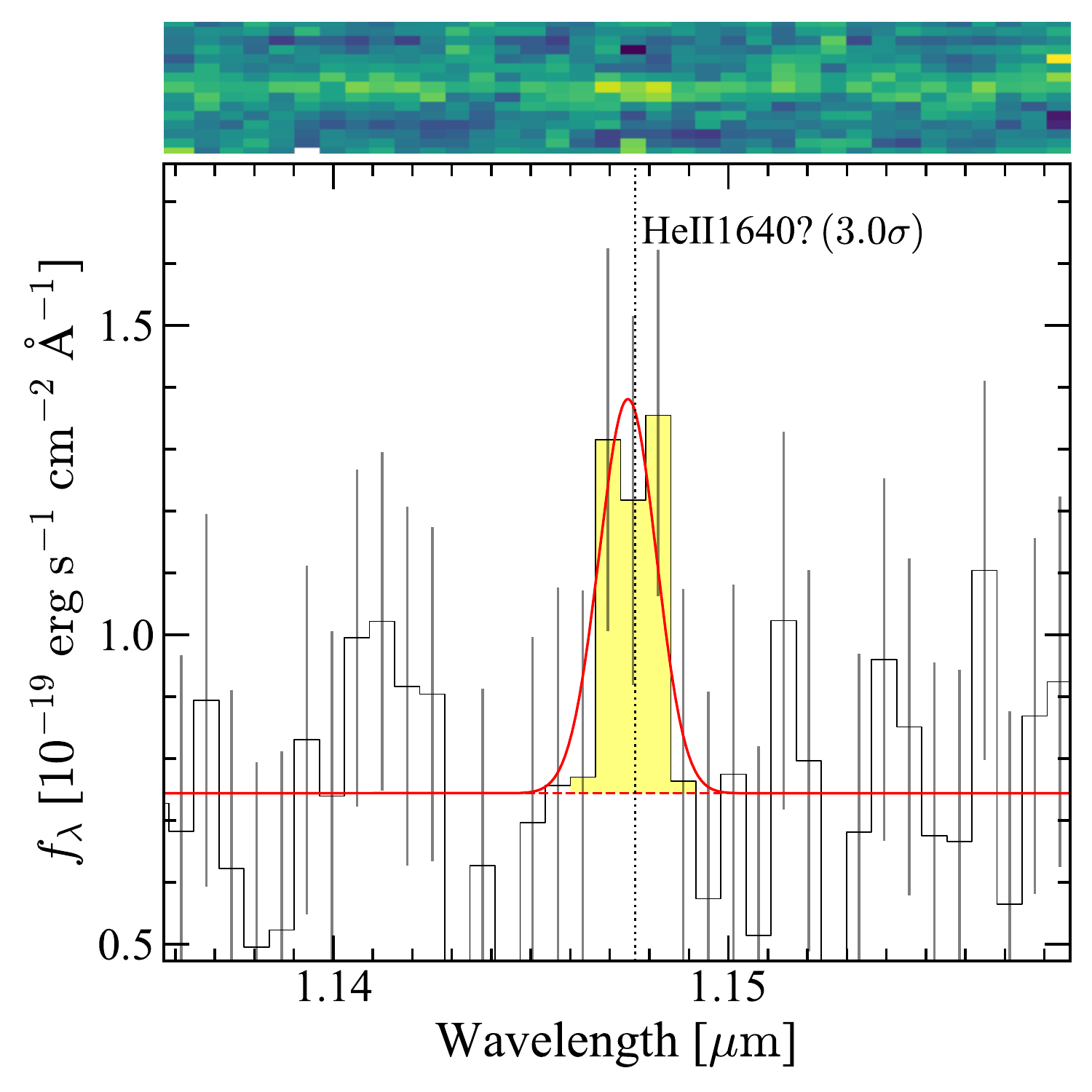}
\end{center}
\end{minipage}
\begin{minipage}{0.32\hsize}
\begin{center}
\includegraphics[width=0.99\hsize, bb=9 9 425 424,clip]{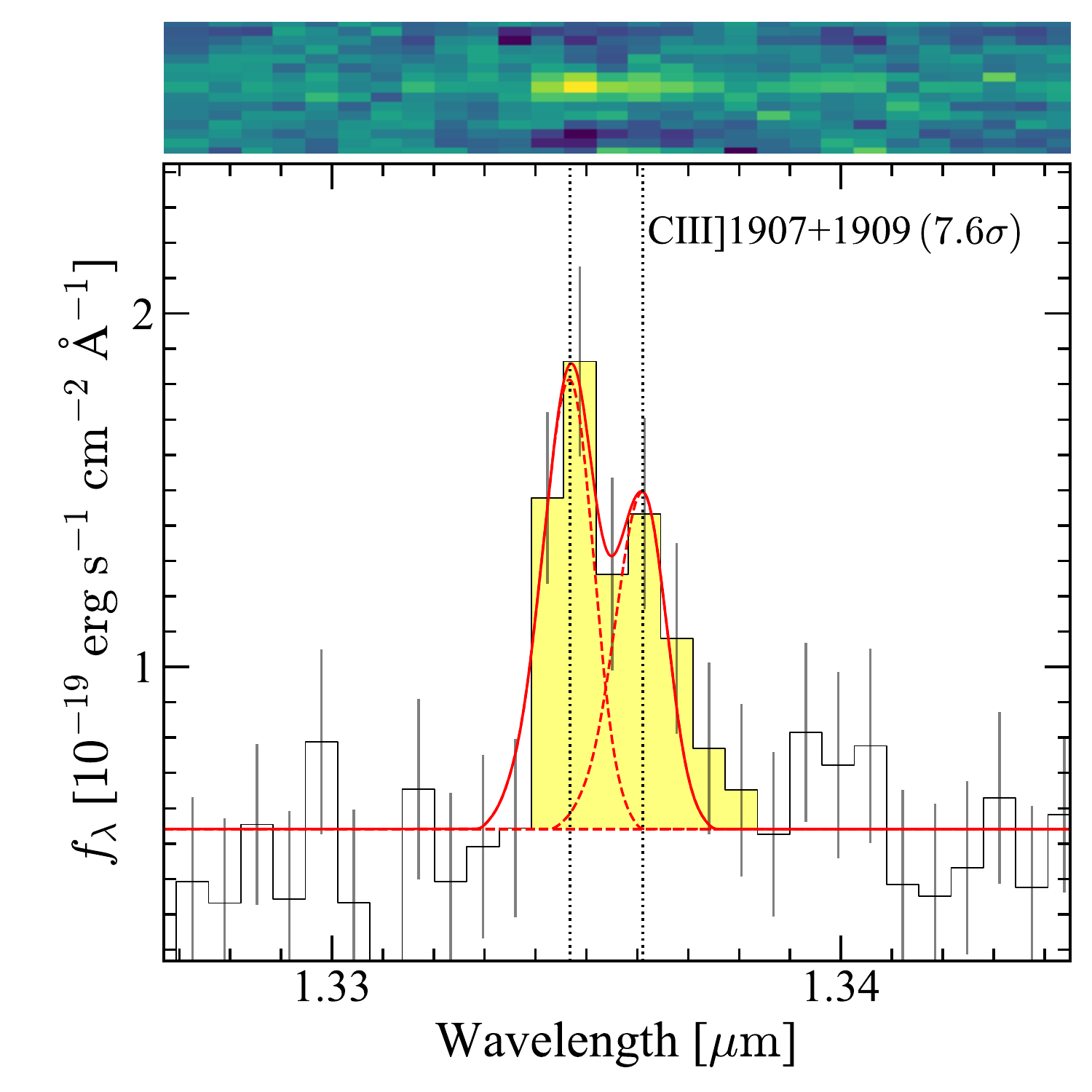}
\end{center}
\end{minipage}
\end{center}
\begin{center}
\begin{minipage}{0.32\hsize}
\begin{center}
\includegraphics[width=0.99\hsize, bb=0 10 359 360,clip]{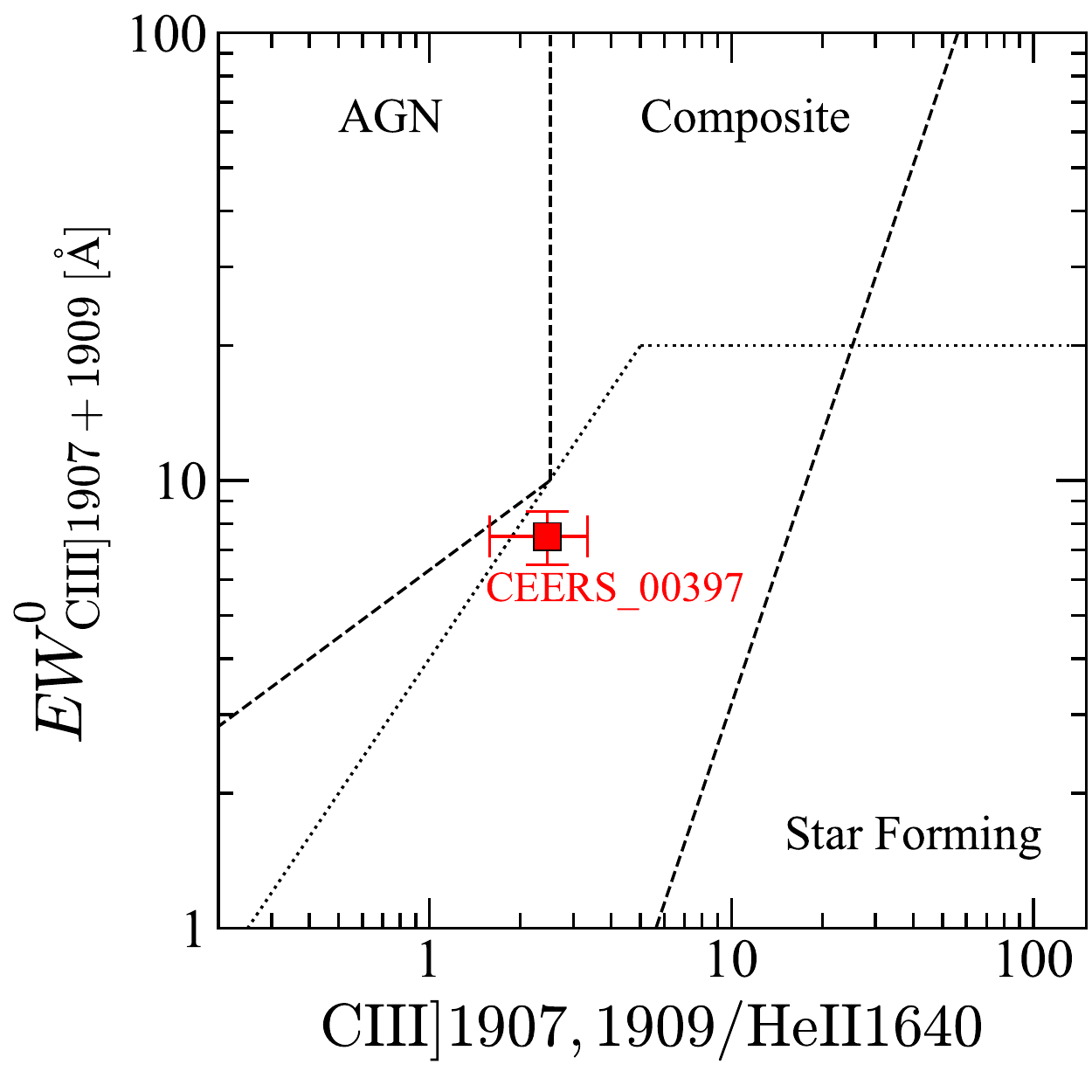}
\end{center}
\end{minipage}
\begin{minipage}{0.32\hsize}
\begin{center}
\includegraphics[width=0.99\hsize, bb=0 10 359 360,clip]{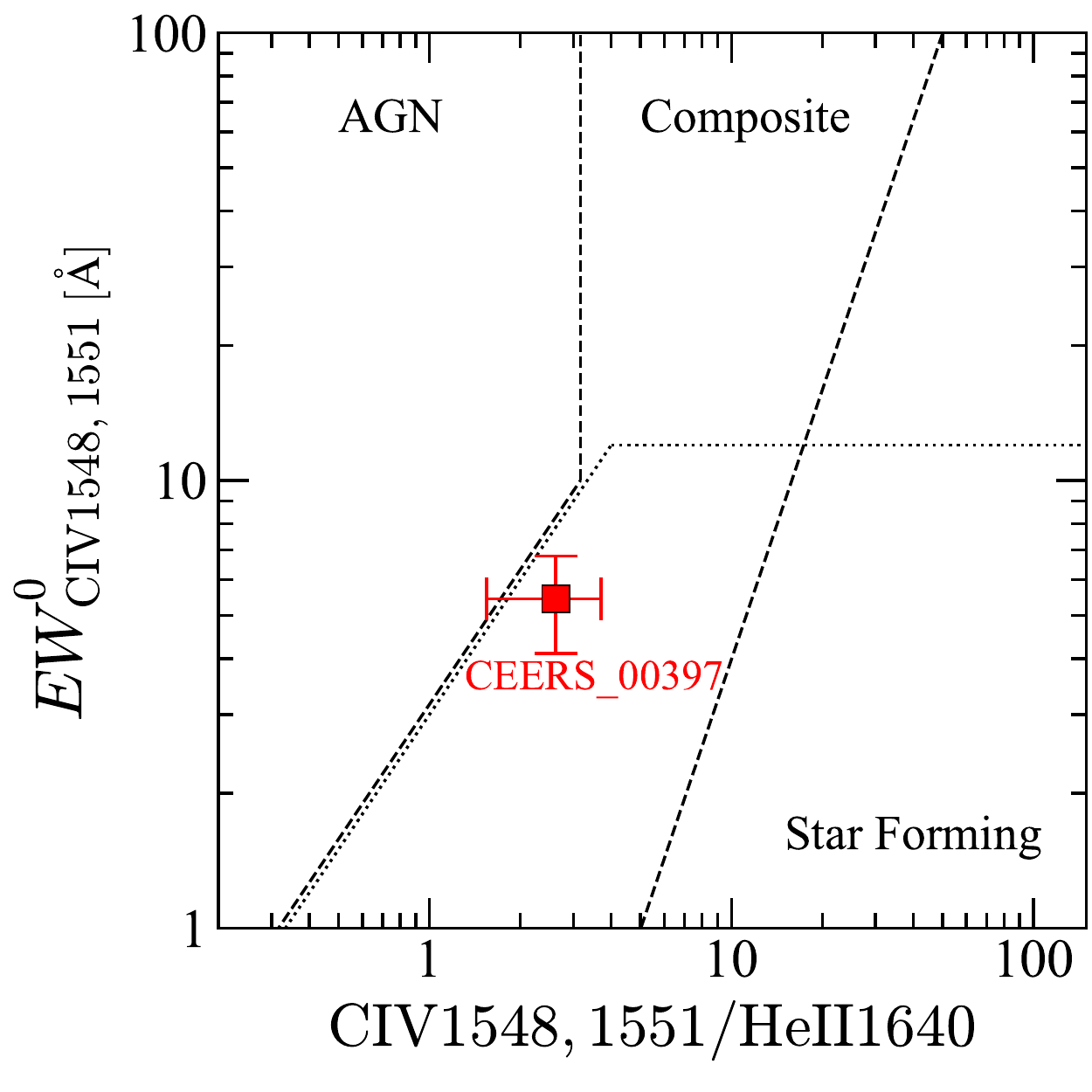}
\end{center}
\end{minipage}
\begin{minipage}{0.32\hsize}
\begin{center}
\includegraphics[width=0.99\hsize, bb=0 5 359 360,clip]{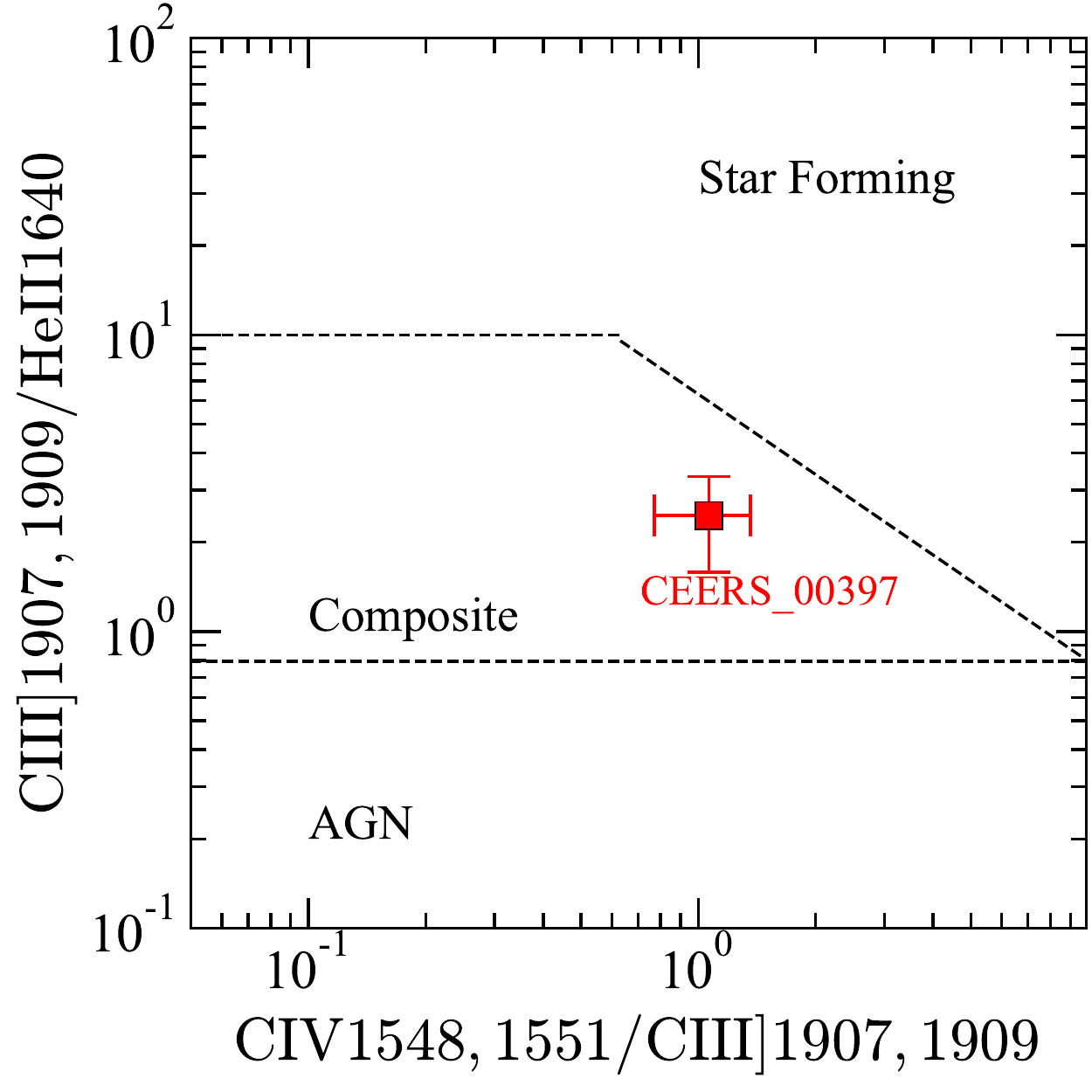}
\end{center}
\end{minipage}
\end{center}
\caption{\redc{(Top:) Rest-frame UV spectrum of CEERS\_00397 at $z=6.000$ around wavelengths of {\sc Civ}$\lambda\lambda$1548,1551 (left), He{\sc ii}$\lambda$1640 (middle), and {\sc Ciii}]$\lambda\lambda$1907,1909 (right).
The spectrum shows the broad ($\m{FWHM}\sim1000\ \m{km\ s^{-1}}$) and redshifted ($+800\ \m{km\ s^{-1}}$) {\sc Civ}, tentative He{\sc ii}, and clear {\sc Ciii}] emission lines.
(Bottom:) The rest-UV line diagnostics to separate AGNs and star-forming galaxies.
The dashed lines are criteria to distinguish AGNs, composites, and star-forming galaxies proposed in \citet{2019MNRAS.487..333H,2022arXiv221202522H}, while the dotted line is to separate AGNs from star-forming galaxies proposed in \citet{2018A&A...612A..94N}.
Our faint AGN, CEERS\_00397, is located in the composite region of \citet{2019MNRAS.487..333H,2022arXiv221202522H} and around the boundary in \citet{2018A&A...612A..94N}, indicating that these rest-UV lines in CEERS\_00397 are partly dominated by its AGN.
}
}
\label{fig_spec_UV}
\end{figure*}

We have investigated the presence of rest-frame UV emission lines and rest-optical high-ionization lines in these galaxies having AGNs.
We do not find significant high-ionization lines such as Ne{\sc V}$\lambda$3426 and He{\sc ii}$\lambda$4686.
\redc{Rest-frame UV {\sc Ciii]}$\lambda\lambda$1907,1909 emission lines are seen in four objects, CEERS\_01244, GLASS\_150029, CEERS\_01665, and CEERS\_00397, with rest-frame equivalent widths (EW)
of $EW^0_\m{CIII]}=7-13\ \m{\AA}$, comparable to both star-forming galaxies and AGNs \citep{2018A&A...612A..94N}.
Among them, CEERS\_00397 also shows broad ($\m{FWHM}\sim1000\ \m{km\ s^{-1}}$) and redshifted ($+800\ \m{km\ s^{-1}}$) {\sc Civ}$\lambda\lambda$1548,1551 and tentative He{\sc ii}$\lambda$1640 lines (the top panels in Figure \ref{fig_spec_UV}).
As shown in the bottom panels in Figure \ref{fig_spec_UV}, rest-UV diagnostics using {\sc Ciii]}, {\sc Civ}, and He{\sc ii} indicate that the AGN activity is partly contributing the these emission line fluxes but not dominating the total fluxes, which is consistent with its Seyfert-like nature.
}

\begin{figure*}
\centering
\begin{center}
\includegraphics[width=0.8\hsize, bb=18 18 425 425]{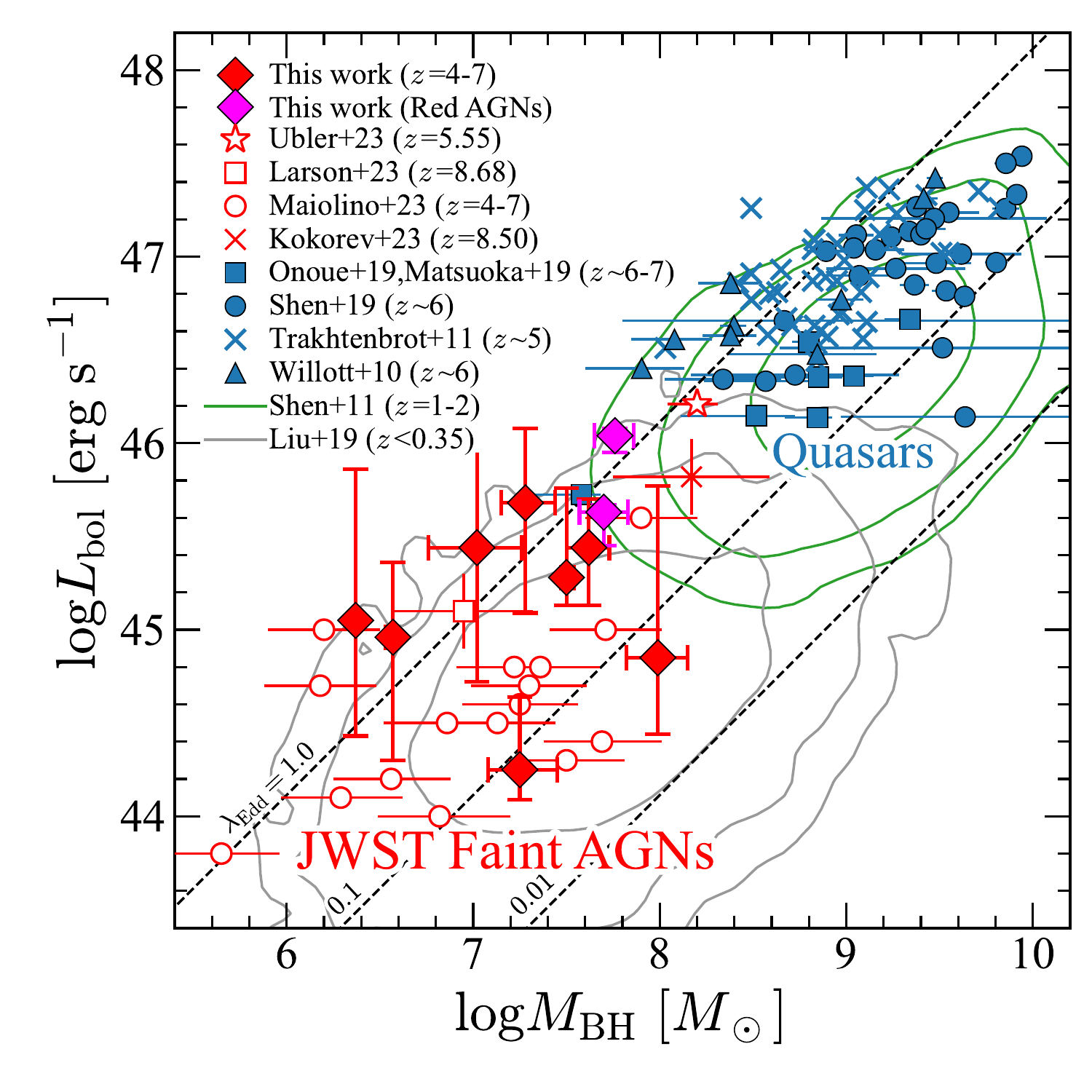}
\end{center}
\caption{
Relation between the bolometric luminosity ($L_\m{bol}$) and black hole mass ($M_\m{BH}$).
The blue symbols show previous measurement for quasars at $z>4$ (square: \citealt{2019ApJ...880...77O} and \citealt{2019ApJ...872L...2M}, circle: \citealt{2019ApJ...873...35S}, triangle: \citealt{2010AJ....140..546W}, cross: \citealt{2011ApJ...730....7T}).
Our AGNs at $z=4-7$ identified with JWST/NIRSpec (the red and magenta diamonds) occupy a unique parameter space with lower $L_\m{bol}$ and $M_\m{BH}$, distinct from the quasars previously identified with the ground-based telescopes.
The red open symbols are faint AGNs identified with JWST/NIRSpec observations (star: \citealt{2023arXiv230206647U}, square: \citealt{2023arXiv230308918L}, \redc{circle: \citealt{2023arXiv230801230M}, cross: \citealt{2023arXiv230811610K}).}
The green and gray contours show the low redshift AGNs at $z=1-2$ \citep{2019ApJ...873...35S} and $z<0.35$ \citep{2019ApJS..243...21L}.
The dashed lines show the bolometric luminosities with the Eddington ratios of $\lambda_\m{Edd}=L_\m{bol}/L_\m{Edd}=1.0$, $0.1$, and $0.01$. 
}
\label{fig_MBH_Lbol}
\end{figure*}

\begin{deluxetable*}{cccccccccc}
\tablecaption{Derived AGN Properties}
\label{tab_MBH}
\tablehead{\colhead{Name} & \colhead{$L_\mathrm{H\alpha,broad}$} & \colhead{$L_\mathrm{broad}/L_\mathrm{narrow}$} & \colhead{$\mathrm{FWHM_\mathrm{H\alpha,broad}}$} & \colhead{$M_\mathrm{BH}$} & \colhead{$L_\mathrm{bol}$} & \colhead{$\lambda_\mathrm{Edd}$} \\
\colhead{}& \colhead{($\mathrm{erg\ s^{-1}}$)}& \colhead{} & \colhead{($\mathrm{km\ s^{-1}}$)}& \colhead{($M_\odot$)}& \colhead{($\mathrm{erg\ s^{-1}}$)} & }
\startdata
CEERS\_01244 & $7.7^{+0.2}_{-0.2}\times10^{42}$& $2.51^{+0.17}_{-0.22}$& $2228^{+75}_{-52}$& $3.2^{+0.3}_{-0.2}\times10^{7}$& $1.9^{+3.8}_{-0.6}\times10^{45}$& $0.46^{+1.04}_{-0.16}$\\
GLASS\_160133 & $1.2^{+0.1}_{-0.1}\times10^{42}$& $0.26^{+0.02}_{-0.01}$& $1028^{+19}_{-13}$& $2.3^{+0.1}_{-0.1}\times10^{6}$& $1.1^{+6.1}_{-0.9}\times10^{45}$& $3.68^{+21.21}_{-2.84}$\\
GLASS\_150029 & $8.2^{+0.7}_{-0.5}\times10^{41}$& $0.22^{+0.02}_{-0.01}$& $1429^{+10}_{-67}$& $3.7^{+0.2}_{-0.3}\times10^{6}$& $9.1^{+13.8}_{-7.1}\times10^{44}$& $1.89^{+3.31}_{-1.49}$\\
CEERS\_00746 & $6.8^{+0.4}_{-0.4}\times10^{43}$& $5.06^{+1.10}_{-1.26}$& $1660^{+157}_{-162}$& $5.8^{+1.5}_{-1.3}\times10^{7}$& $1.1^{+0.1}_{-0.2}\times10^{46}$& $1.47^{+0.60}_{-0.52}$\\
CEERS\_01665 & $6.8^{+0.9}_{-0.6}\times10^{42}$& $0.28^{+0.05}_{-0.05}$& $1794^{+282}_{-171}$& $1.9^{+0.8}_{-0.5}\times10^{7}$& $4.8^{+7.2}_{-3.6}\times10^{45}$& $1.93^{+4.61}_{-1.59}$\\
CEERS\_00672 & $1.8^{+0.2}_{-0.2}\times10^{43}$& $1.95^{+0.45}_{-0.40}$& $2208^{+277}_{-241}$& $5.0^{+1.7}_{-1.3}\times10^{7}$& $4.3^{+0.4}_{-1.4}\times10^{45}$& $0.65^{+0.31}_{-0.33}$\\
CEERS\_02782 & $7.6^{+0.8}_{-0.7}\times10^{42}$& $0.85^{+0.14}_{-0.10}$& $2534^{+260}_{-266}$& $4.2^{+1.2}_{-1.0}\times10^{7}$& $2.8^{+2.3}_{-1.4}\times10^{45}$& $0.51^{+0.71}_{-0.32}$\\
CEERS\_00397 & $2.6^{+0.8}_{-0.4}\times10^{42}$& $0.19^{+0.07}_{-0.05}$& $1739^{+359}_{-317}$& $1.0^{+0.8}_{-0.5}\times10^{7}$& $2.8^{+14.2}_{-2.2}\times10^{45}$& $2.02^{+20.68}_{-1.80}$\\
CEERS\_00717 & $1.2^{+0.3}_{-0.2}\times10^{42}$& $0.57^{+0.17}_{-0.11}$& $6279^{+805}_{-881}$& $9.8^{+4.4}_{-3.2}\times10^{7}$& $7.1^{+51.8}_{-4.3}\times10^{44}$& $0.06^{+0.63}_{-0.04}$\\
CEERS\_01236 & $4.8^{+1.1}_{-0.8}\times10^{41}$& $2.39^{+1.12}_{-0.59}$& $3521^{+649}_{-485}$& $1.8^{+1.0}_{-0.6}\times10^{7}$& $1.8^{+2.6}_{-0.5}\times10^{44}$& $0.08^{+0.20}_{-0.04}$\\
\vspace{-0.5cm}
\enddata
\end{deluxetable*}

\subsection{$M_\m{BH}-L_\m{bol}$ Relation}

We estimate the black hole mass, $M_\m{BH}$, and bolometric luminosity, $L_\m{bol}$, of our AGNs.
The black hole mass is estimated with the following equation calibrated at $z\sim0$ in \citet{2005ApJ...630..122G}:
\begin{align}\label{eq_mbh}
&M_\m{BH}=2.0\times10^6\ M_\odot\notag\\
&\times\left(\frac{L_\m{H\alpha,broad}}{10^{42}\ \m{erg\ s^{-1}}}\right)^{0.55}\left(\frac{\m{FWHM}_\m{H\alpha,broad}}{10^{3}\ \m{km\ s^{-1}}}\right)^{2.06},
\end{align}
where the $L_\m{H\alpha,broad}$ is the extinction-correlated broad-line H$\alpha$ luminosity, and $\m{FWHM}_\m{H\alpha,broad}$ is the FWHM of the broad H$\alpha$ emission line.
We use the broad-line luminosity as an input of Equation (\ref{eq_mbh}) following \citet{2023ApJ...954L...4K}, because it is not clear whether the narrow-line emission line seen in our AGNs originates from the AGN or {\sc Hii} regions in its host galaxy.
The estimated black hole masses are presented in Table \ref{tab_MBH}.
If we include the narrow line component in the H$\alpha$ luminosity, the black hole mass increase by 0.1 dex on average (0.4 dex at the maximum).

Estimating the bolometric luminosity of our AGNs is not straightforward, because the continuum luminosities observed with the HST and JWST images are possibly significantly contributed by the lights from their host galaxies.
Therefore, we instead estimate the bolometric luminosity from the H$\alpha$ luminosity.
Our best estimates of the bolometric luminosity, $L_\m{bol}$, comes from the following equation between the luminosity at the rest-frame $5100\ \m{\AA}$, $L_{5100}$, and the H$\alpha$ luminosity including both the broad and narrow components \citep{2005ApJ...630..122G}:
\begin{align}\label{eq_lbol}
L_\m{5100}=10^{44} \left(\frac{L_\m{H\alpha}}{5.25\times10^{42}\ \m{erg\ s^{-1}}}\right)^{\frac{1}{1.157}}\ \m{erg\ s^{-1}},
\end{align}
and the relation between $L_{5100}$ and $L_\m{bol}$ with the bolometric correction in \citet{2006ApJS..166..470R}, $L_\m{bol}=10.33\times L_\m{5100}$.
\redc{Since it is not clear whether the narrow-line emission in our AGNs originates from the AGN or star formation in host galaxies, we set the lower limit of the bolometric luminosity using Equation (\ref{eq_lbol}) with the broad-line H$\alpha$ luminosity as an input assuming that the narrow component comes from the host galaxy} and the bolometric correction of $L_\m{bol}=9.8\times L_\m{5100}$ in \citet{2004MNRAS.352.1390M}.
As the upper limit, we use the following equation in \citet{2009MNRAS.399.1907N} calibrated with type-2 AGNs:
\begin{align}
\m{log}&\left(\frac{L_\m{bol}}{\m{erg\ s^{-1}}}\right)=\m{log}\left(\frac{L_\m{H\beta}}{\m{erg\ s^{-1}}}\right)\notag\\
&+3.48+\m{max}[0.0,0.31(\m{log([OIII]/H\beta)}-0.6)],
\end{align}
where the H$\beta$ luminosity is estimated from the extinction-corrected narrow-line H$\alpha$ line assuming Case B recombination.
The estimated bolometric luminosities are presented in Table \ref{tab_MBH}.

Figure \ref{fig_MBH_Lbol} shows the estimated $M_\m{BH}$ and $L_\m{bol}$ of our AGNs.
The black hole masses and bolometric luminosities of our AGNs are $M_\m{BH}\sim10^6-10^8\ M_\odot$ and $L_\m{bol}\sim10^{44}-10^{46}\ \m{erg\ s^{-1}}$, respectively, indicating that our AGNs have lower black hole masses and lower bolometric luminosities than quasars at $z\sim4-7$ identified in the ground-based observations \citep{2011ApJ...730....7T,2019ApJ...872L...2M,2019ApJ...873...35S,2019ApJ...880...77O}.
The two red AGNs show relatively higher $M_\m{BH}$ and $L_\m{bol}$, indicating that these AGNs might be in the transition phase between the faint AGNs with low $M_\m{BH}$ and low-luminosity quasars.

The black hole masses of our AGNs are comparable to those of $z\sim0$ AGNs in \citet{2019ApJS..243...21L}, but our AGNs show bolometric luminosities higher than those of $z\sim0$ AGNs in \citet{2019ApJS..243...21L} on average, resulting in higher Eddington ratios ($\lambda_\m{Edd}=L_\m{bol}/L_\m{Edd}$).
This distribution of higher $\lambda_\m{Edd}$ of our AGNs compared to those in $z\sim0$ may be due to selection bias, because AGNs with faint broad-line H$\alpha$ emission may not be selected in our selection criteria with the signal-to-noise ratio threshold.
Deeper NIRSpec spectroscopy is needed to investigate whether the AGNs at $z>4$ with $M_\m{BH}\sim10^6-10^8\ M_\odot$ have systematically higher  $\lambda_\m{Edd}$ than $z\sim0$ AGNs or not.

\subsection{$M_\m{BH}-M_*$ Relation}

\subsubsection{AGN-Host Decomposition}\label{ss_agnhost}

\begin{figure}
\centering
\begin{center}
\includegraphics[width=0.99\hsize, bb=0 0 433 145]{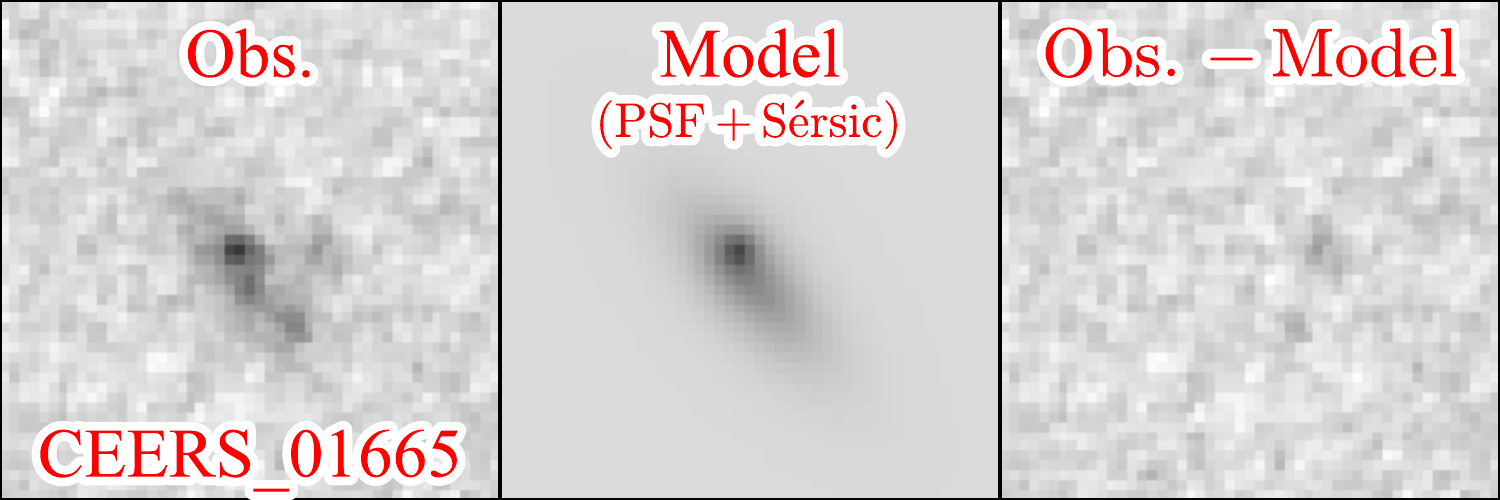}
\\
\vspace{0.2cm}
\includegraphics[width=0.99\hsize, bb=0 0 433 145]{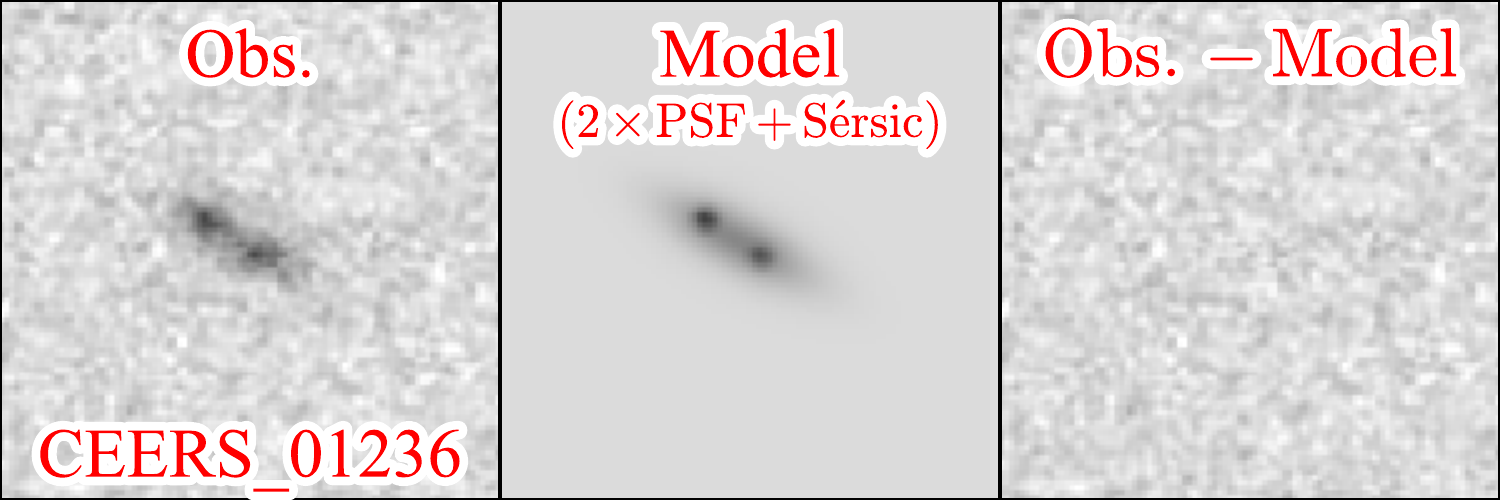}
\end{center}
\caption{
Examples of the AGN-host decomposition analysis.
The top and bottom panels show the images of CEERS\_01665 and CEERS\_01236 in the F606W and F150W bands, respectively.
The left, middle, and right panels show the observed images, models, and the residuals, respectively.
CEERS\_01665 (CEERS\_01236) is well fitted by a model with one PSF and the S\'ersic profile ($2\times$PSF and the S\'ersic profile). 
}
\label{fig_HostAGN}
\end{figure}

To estimate the stellar mass of AGN's host galaxies, we conduct the AGN-host decomposition analysis using the high-resolution JWST and HST images, in the same manner as \citet{2023arXiv230302929Z}.
We conduct the two-dimensional decomposition by fitting the images of our AGNs in all bands with 1) a PSF profile only, 2) a PSF and a S\'ersic profile, and 3) two PSFs and one S\'ersic profile. 
For each band, we generate the PSF by selecting and stacking bright stars in the same fields.
For the S\'ersic profile, we restrict the S\'ersic index $n$ to the range of $n=1-4$.
We find that CEERS\_00746 and CEERS\_00672 are well-fitted with a PSF only model, and their host galaxies are not seen.
CEERS\_02782 is also well fitted with a PSF, but a tentative residual is seen, possibly indicative of its host galaxy.
The other sources are fitted with a PSF and a S\'ersic profile, except for CEERS\_01236, which is well fitted with a model of two PSFs and a S\'ersic profile, indicating that this source could be a dual AGN.
Examples of our results are shown in Figure \ref{fig_HostAGN}, and fitted models are summarized in Table \ref{tab_host}.
For GLASS\_160133, we cannot obtain a reasonable fitting solution with either PSF, PSF$+$S\'ersic, or $\mathrm{2{\times}PSF}+$S\'ersic models.

\begin{deluxetable}{ccc}
\tablecaption{Stellar Mass of Host Galaxies}
\label{tab_host}
\tablehead{\colhead{Name} & \colhead{$\m{log}M_*$} & \colhead{Fitting} \\
\colhead{}& \colhead{$(M_\odot)$}& \colhead{}}
\startdata
CEERS\_01244 & $8.63^{+0.63}_{-1.03}$ & PSF$+$S\'ersic\\
GLASS\_160133$^\dagger$ & $<8.82$ & \nodata\\
GLASS\_150029 & $9.10^{+0.31}_{-0.37}$ & PSF$+$S\'ersic\\
CEERS\_00746 & $<9.11$ & PSF\\
CEERS\_01665 & $9.92^{+0.51}_{-0.68}$ & PSF$+$S\'ersic\\
CEERS\_00672 & $<9.01$ & PSF\\
CEERS\_02782 & $<9.35$ & PSF\\
CEERS\_00397 & $9.36^{+0.36}_{-0.45}$ & PSF$+$S\'ersic\\
CEERS\_00717 & $9.61^{+0.77}_{-1.18}$ & PSF$+$S\'ersic\\
CEERS\_01236 & $8.94^{+0.29}_{-0.54}$ & $\mathrm{2{\times}PSF}+$S\'ersic\\
\enddata
{$^\dagger$ We cannot obtain a reasonable fitting solution for GLASS\_160133.
Thus we use the stellar mass estimated from the SED fitting to total lights as the upper limit for this object.}
\end{deluxetable}

\subsubsection{SED fitting}\label{ss_sed}

To estimate the stellar mass, we conduct SED fittings for host galaxy components using {\tt prospector} \citep{2021ApJS..254...22J}. 
The procedure of the SED fitting is the same as that of \citet{2023ApJS..265....5H}, except for the fixed redshifts based on the NIRSpec emission line measurements.
We use the stellar population synthesis package, Flexible Stellar Population Synthesis (FSPS; \citealt{2009ApJ...699..486C,2010ApJ...712..833C}) for stellar SEDs, and include nebular emission from the photoionization models of Cloudy \citep{2017ApJ...840...44B}.
We assume the \citet{2003PASP..115..763C} stellar initial mass function (IMF) \redc{of $0.1-100\ M_\odot$}, the intergalactic medium (IGM) attenuation model of \citet{1995ApJ...441...18M}, the \citet{2000ApJ...533..682C} dust attenuation law, and a fixed metallicity of 0.2 $Z_\odot$.
We choose a flexible star formation history as adopted in \citet{2023ApJS..265....5H} \redc{with a continuity prior}.
The estimated stellar masses are presented in Table \ref{tab_host}.
\redc{The systematic uncertainty on the stellar mass due to the fixed prior is typically $\sim0.2$ dex \citep{2019ApJ...876....3L}, which is smaller than the statistical uncertainty.}
For sources whose host galaxies are not seen (i.e., CEERS\_00746, CEERS\_00672, and CEERS\_02782) and GLASS\_160133, we estimate the stellar mass by fitting the total lights of that source with the galaxy templates using {\tt prospector}, and use the derived value as the upper limit.

\begin{figure*}
\centering
\begin{center}
\includegraphics[width=0.8\hsize, bb=11 10 425 425]{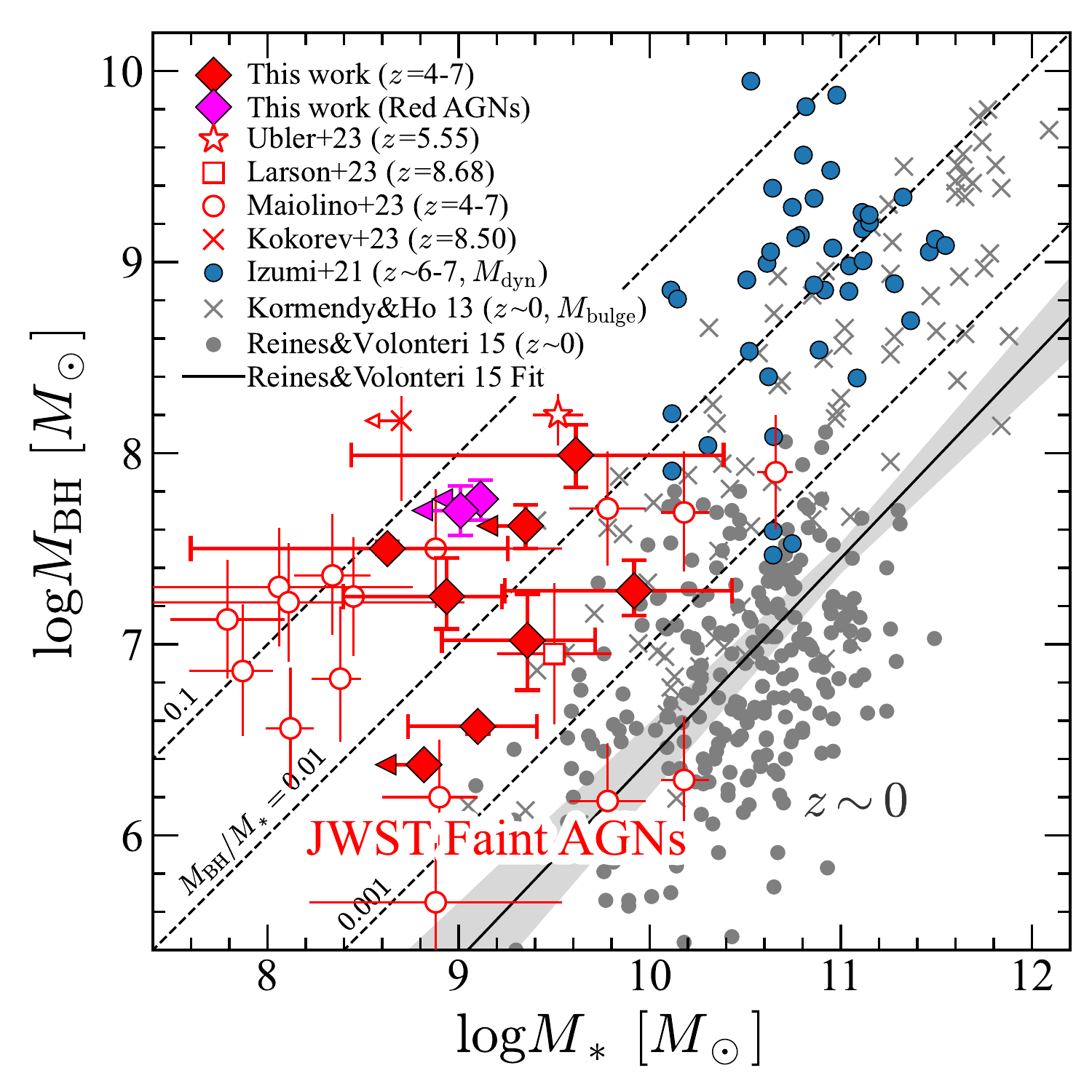}
\end{center}
\caption{
Relation between the black hole mass ($M_\m{BH}$) and host's stellar mass ($M_*$).
The red and magenta diamonds are our AGNs at $z=4-7$.
The red open symbols are faint AGNs identified with JWST/NIRSpec observations (star: \citealt{2023arXiv230206647U}, square: \citealt{2023arXiv230308918L}, \redc{circle: \citealt{2023arXiv230801230M}, cross: \citealt{2023arXiv230811610K}).}
The blue circles show $z>6$ quasar samples compiled in \citet{2021ApJ...914...36I} (see also \citealt{2018PASJ...70...36I,2019PASJ...71..111I}).
The gray crosses and circles are $z\sim0$ AGNs in \citet{2013ARA&A..51..511K} and \citet{2015ApJ...813...82R}, respectively.
The black solid line with the shaded region is the relation at $z\sim0$ in \citet{2015ApJ...813...82R}, and the dashed lines correspond to $M_\m{BH}/M_*=0.1$, $0.01$, and 0.001.
Our $z=4-7$ AGNs show systematically higher $M_\m{BH}$ (lower $M_*$) compared to $z\sim0$ galaxies.
}
\label{fig_MBH_Ms}
\end{figure*}

\subsubsection{Result}

Figure \ref{fig_MBH_Ms} shows the stellar masses and black hole masses of our AGNs.
Compared to the AGNs at $z\sim0$ \citep{2015ApJ...813...82R}, our AGNs at $z\sim4-7$ have similar black hole masses but systematically lower stellar masses.
Similar results are obtained in previous studies with a smaller number of AGNs \citep{2023ApJ...954L...4K,2023arXiv230206647U,2023ApJ...954L...4K}, but we confirm this trend of lower $M_*$ (higher $M_\m{BH}$) with a sample of 10 AGNs at $z=4-7$.
Our results at $z=4-7$ showing lower $M_*$ (higher $M_\m{BH}$) than $z\sim0$ AGNs indicate that the black hole grows faster than its host galaxy at high redshift.
The fast black hole growth is also suggested by previous studies at $z\sim2$ \citep[e.g.,][]{2023arXiv230302929Z}.
Such over-massive black holes compared to their host stellar masses are indeed predicted in some theoretical simulations \citep[e.g.,][]{2021ApJ...907...74T,2022MNRAS.511..616T,2022ApJ...927..237I,2022ApJ...935..140H,2023arXiv230308150Z}.

\begin{deluxetable}{ccc}
\tablecaption{Fraction and Number Densities of Our Broad-Line AGNs}
\label{tab_nd}
\tablehead{\colhead{$M_\m{UV}$} & $f_\m{AGN}$ & \colhead{$\Phi$}  \\
\colhead{(mag)}& & \colhead{$\m{Mpc^{-3}\ mag^{-1}}$}}
\startdata
\multicolumn{3}{c}{$z\sim4\ (z_\m{ave}=4.365)$} \\
$-21.5$ & $0.07^{+0.15}_{-0.06}$ & $8.9^{+21.0}_{-8.7}\times10^{-6}$\\
$-18.5$ & $0.15^{+0.15}_{-0.09}$ & $9.0^{+9.9}_{-8.9}\times10^{-4}$\\
\hline
\multicolumn{3}{c}{$z\sim5\ (z_\m{ave}=4.912)$} \\
$-21.5$ & $<0.10$ & $<1.2\times10^{-5}$\\
$-18.5$ & $0.04^{+0.05}_{-0.02}$ & $1.4^{+1.9}_{-1.4}\times10^{-4}$\\
\hline
\multicolumn{3}{c}{$z\sim6\ (z_\m{ave}=5.763)$} \\
$-21.5$ & $0.07^{+0.17}_{-0.06}$ & $3.5^{+8.2}_{-3.3}\times10^{-6}$\\
$-18.5$ & $0.04^{+0.10}_{-0.04}$ & $1.5^{+3.5}_{-1.5}\times10^{-4}$\\
\hline
\multicolumn{3}{c}{$z\sim7\ (z_\m{ave}=6.936)$} \\
$-21.5$ & $0.07^{+0.17}_{-0.06}$ & $1.6^{+3.7}_{-1.4}\times10^{-6}$\\
$-18.5$ & $<0.19$ & $<5.1\times10^{-4}$\\
\enddata
\tablecomments{Errors are $1\sigma$ and upper limits are $2\sigma$.
}
\end{deluxetable}

\begin{deluxetable*}{ccccccc}
\tablecaption{Fit Parameters for the AGN Luminosity Function, Cosmic emissivity, and Photoionization Rate}
\label{tab_LFpar}
\tablehead{\colhead{Redshift} & \colhead{$M^*_\m{UV}$} & \colhead{$\m{log}\phi^*$} & \colhead{$\alpha$} & \colhead{$\beta$} & \colhead{$\epsilon_{912}$} & \colhead{$\Gamma$}  \\
\colhead{}& \colhead{(mag)}& \colhead{$\m{Mpc^{-3}}$} & \colhead{} & \colhead{} & \colhead{$\m{erg\ s^{-1}\ Mpc^{-3}\ Hz}$} & \colhead{$\m{s^{-1}}$} }
\startdata
$z\sim4\ (z_\m{ave}=4.365)$ &$-21.10$ & $-4.05^{+0.30}_{-2.05}$ & $-1.87$ & $-4.95$ & $4.9^{+4.9}_{-4.9}\times10^{24}$ & $8.5^{+8.4}_{-8.4}\times10^{-13}$ \\
$z\sim5\ (z_\m{ave}=4.912)$ &$-21.39$ & $-5.05^{+0.35}_{-1.43}$ & $-1.94$ & $-4.96$ & $7.6^{+9.3}_{-7.3}\times10^{23}$ & $8.4^{+10.3}_{-8.1}\times10^{-14}$ \\
$z\sim6\ (z_\m{ave}=5.763)$ &$-21.23$ & $-4.87^{+0.34}_{-1.50}$ & $-2.14$ & $-5.03$ & $1.3^{+1.5}_{-1.2}\times10^{24}$ & $9.8^{+11.9}_{-9.5}\times10^{-14}$ \\
$z\sim7\ (z_\m{ave}=6.936)$ &$-20.82$ & $-4.77^{+0.36}_{-0.96}$ & $-2.05$ & $-4.83$ & $8.4^{+11.0}_{-7.5}\times10^{23}$ & $4.7^{+6.1}_{-4.2}\times10^{-14}$ \\
\enddata
\tablecomments{$M^*_\m{UV}$, $\alpha$, and $\beta$ are fixed to values in \citet{2022ApJS..259...20H}.
Errors are $1\sigma$.
}
\end{deluxetable*}

\subsection{UV Luminosity Function}\label{ss_uvlf}

We estimate the number density of our broad-line AGNs detected with NIRSpec.
Since the selection function including the target selection completeness with NIRSpec/MSA is not known, we cannot precisely calculate the number density of our AGNs.
Instead, we present rough estimates based on the spectroscopic results.
We divide our AGN sample and the entire galaxy sample in \citet{2023arXiv230112825N} into magnitude and redshift bins and calculate the fraction of the AGNs to the entire sources in each bin.
Table \ref{tab_nd} summarizes our calculated AGN fraction at each redshift and magnitude bin.
Using this AGN fraction, $f_\m{AGN}(z,M_\m{UV})$, we estimate the number density of the AGNs, $\Phi_\m{AGN}$, from the galaxy UV luminosity function fitted with the double-power law form, $\Phi$, as
\begin{equation}
\Phi_\m{AGN}(z,M_\m{UV})=f_\m{AGN}(z,M_\m{UV})\times\Phi(z,M_\m{UV}),
\end{equation}
where the fitting parameters of the galaxy luminosity function are taken from \citet{2022ApJS..259...20H}, which are consistent with \citet{2020MNRAS.494.1771A,2022arXiv220709342A}.
The error of the number density includes both the Poisson error \citep{1986ApJ...303..336G} and the cosmic variance.
We estimate the cosmic variance following the procedures in \citet{2004ApJ...600L.171S}, using the bias value of faint quasars at $z\sim4$ \citep{2018PASJ...70S..33H} and the effective volume assuming the survey area of $72\ \m{arcmin^2}$ ($=8$ NIRSpec pointings).
\redc{Using the effective volume calculated here, we can also evaluate the lower limit of the number density based on the observed number of AGNs in each bin.
If the $1\sigma$ lower limit calculated with $f_\m{AGN}$ is higher than the volume-based value, we replace the $1\sigma$ lower limit with the volume-based lower limit.}

\begin{figure*}
\centering
\begin{center}
\begin{minipage}{0.49\hsize}
\begin{center}
\includegraphics[width=0.95\hsize, bb=9 6 429 347]{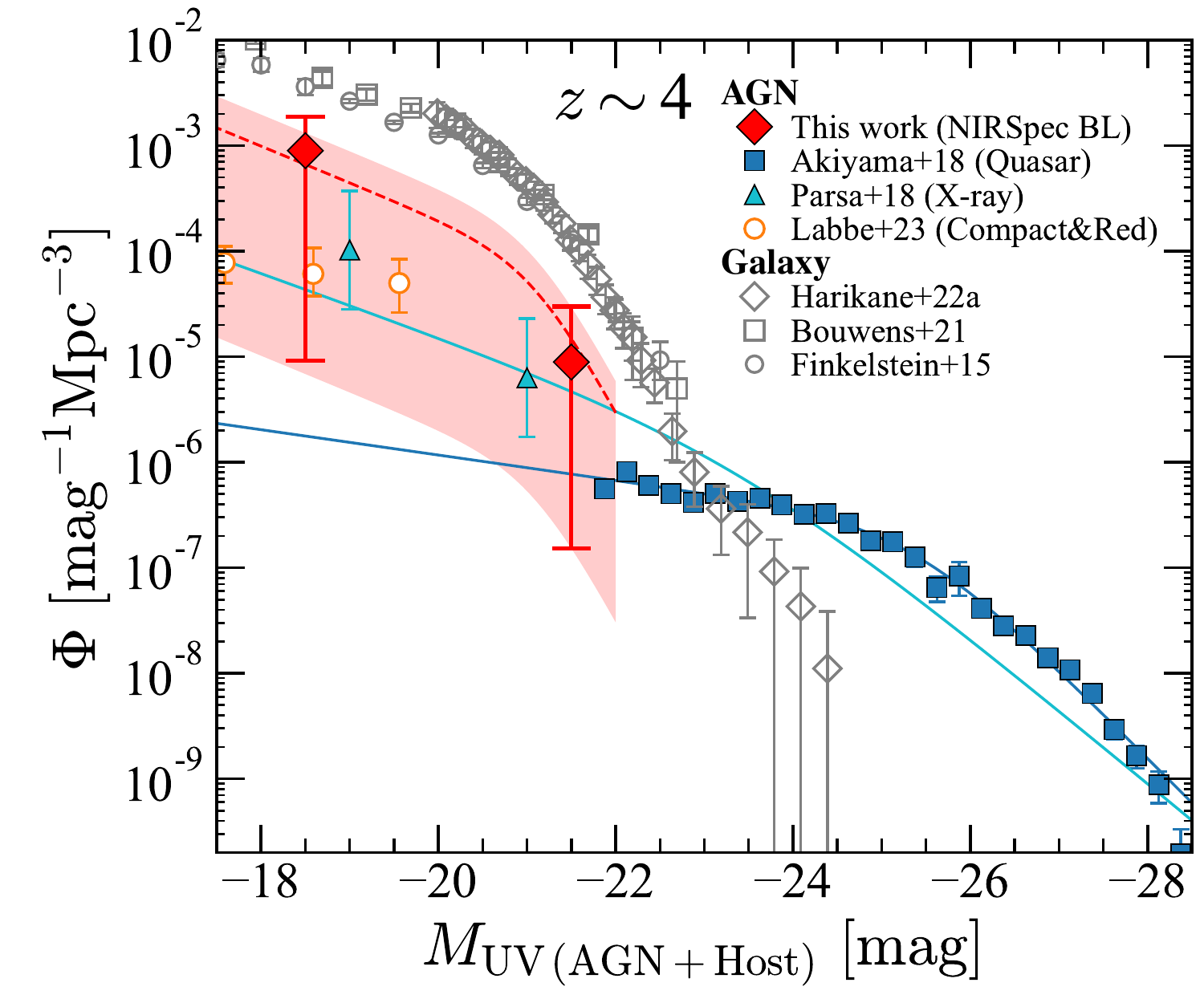}
\end{center}
\end{minipage}
\begin{minipage}{0.49\hsize}
\begin{center}
\includegraphics[width=0.95\hsize, bb=9 6 429 347]{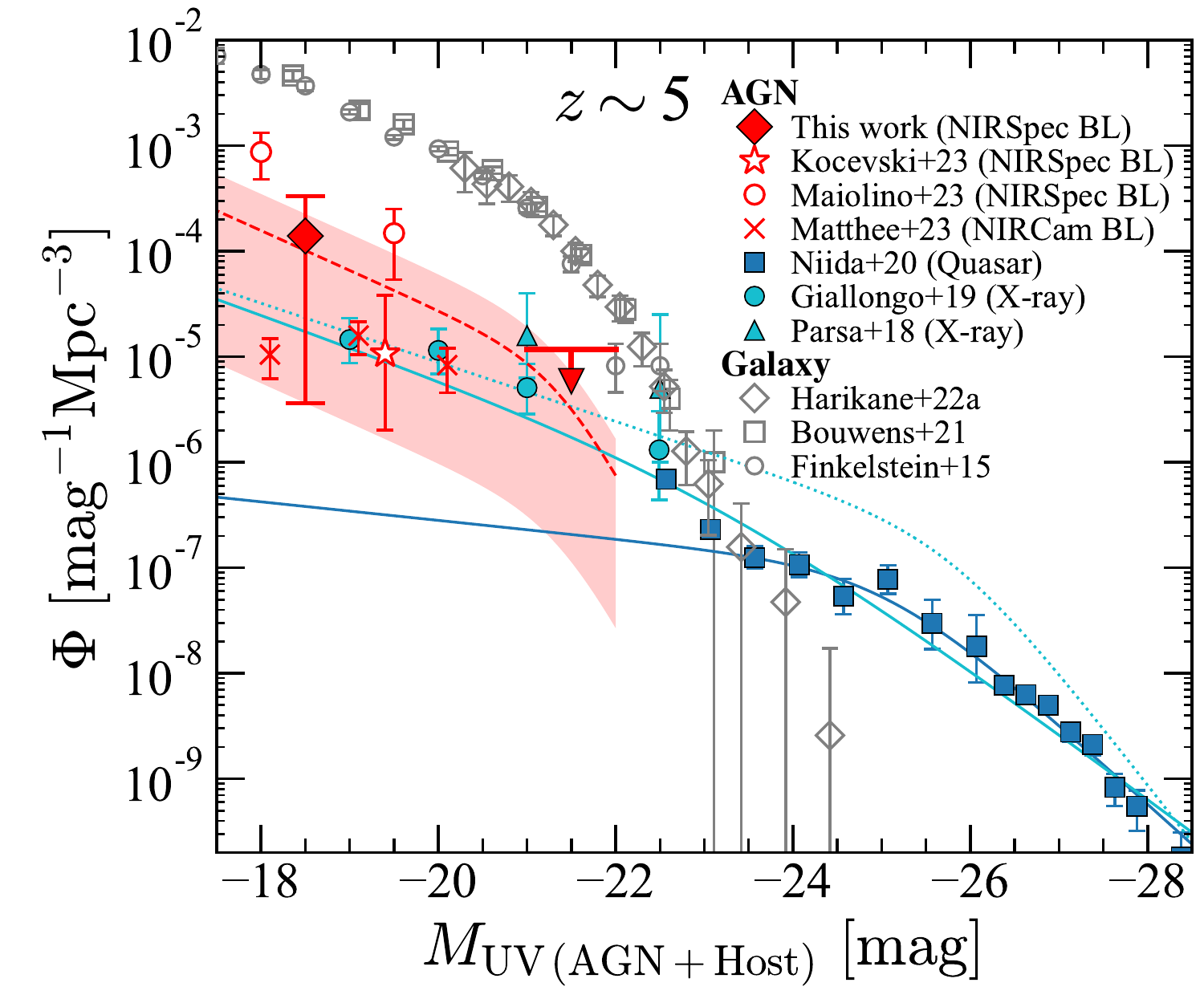}
\end{center}
\end{minipage}
\end{center}
\begin{center}
\begin{minipage}{0.49\hsize}
\begin{center}
\includegraphics[width=0.95\hsize, bb=9 6 429 347]{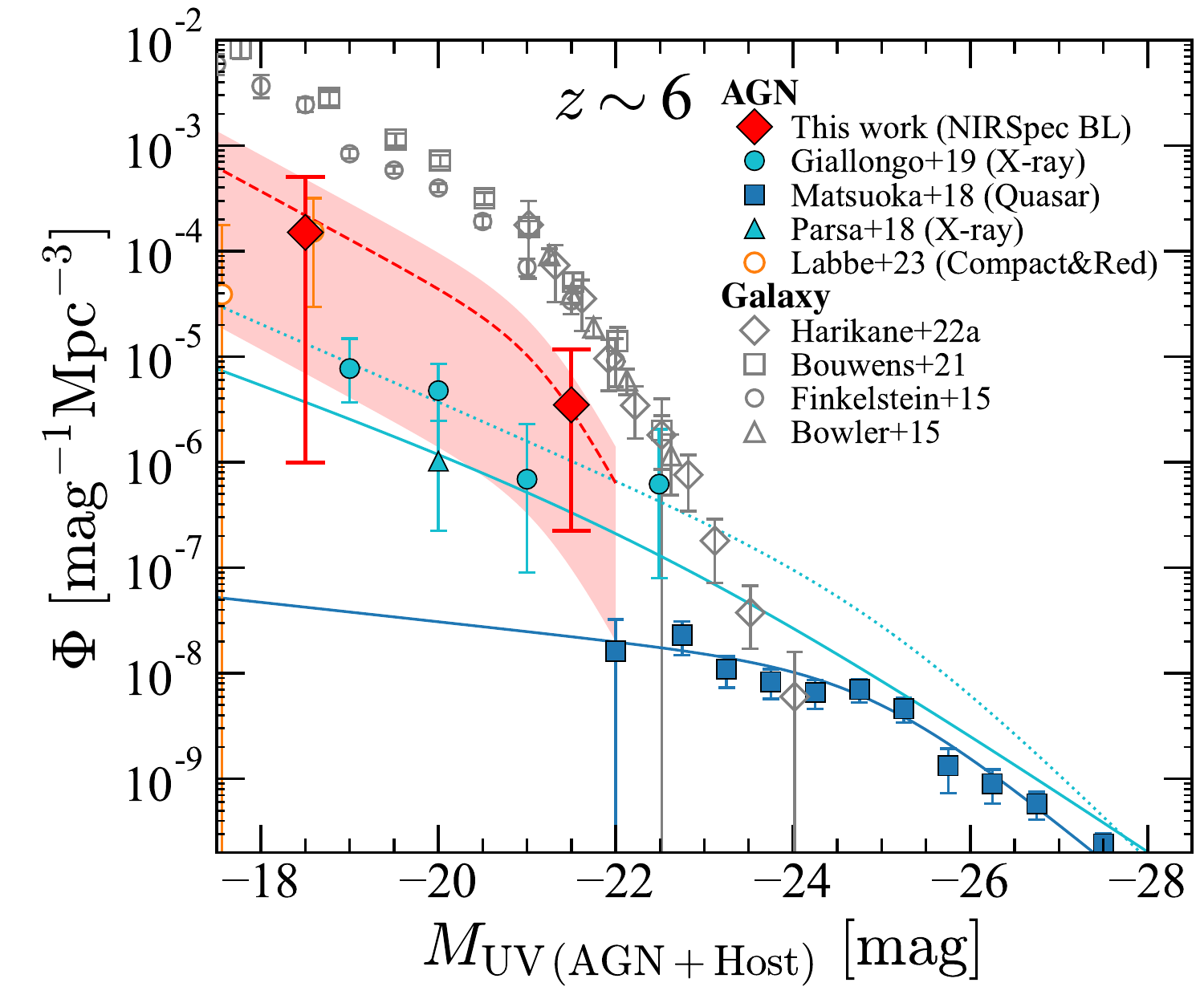}
\end{center}
\end{minipage}
\begin{minipage}{0.49\hsize}
\begin{center}
\includegraphics[width=0.95\hsize, bb=9 6 429 347]{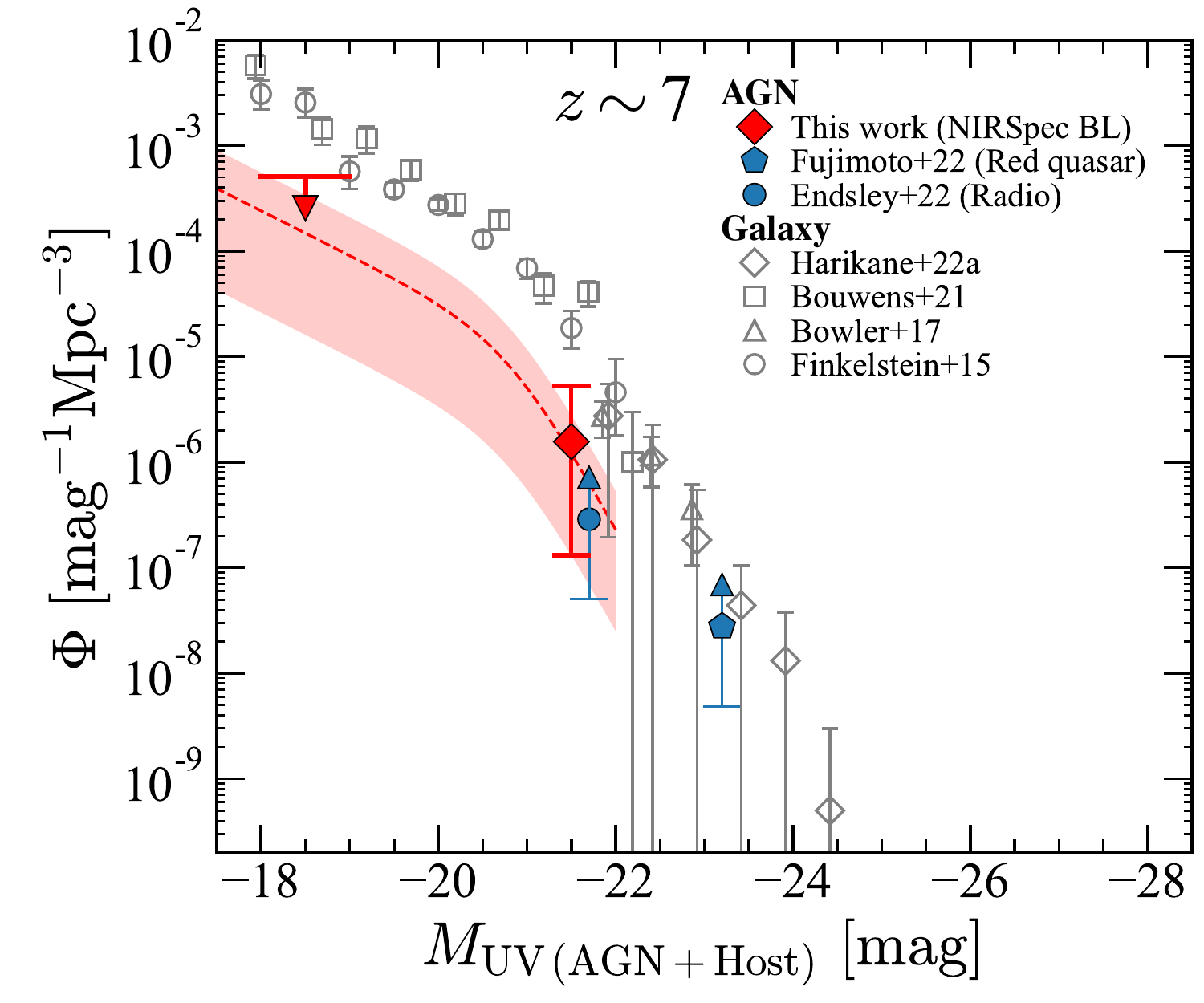}
\end{center}
\end{minipage}
\end{center}
\caption{
UV luminosity functions at $z\sim4$, $5$, $6$, and $7$.
The red diamonds show the number densities of our broad-line (BL) AGNs identified with NIRSpec (Table \ref{tab_nd}), and the red dashed lines with the shaded regions are the best-fit functions (Table \ref{tab_LFpar}).
The cyan triangles and circles are X-ray selected AGNs in \citet{2018MNRAS.474.2904P} and \citet{2019ApJ...884...19G}, respectively, and the blue squares show the number densities of quasars at $z\sim4$, $5$, and $6$ \citep{2018PASJ...70S..34A,2020ApJ...904...89N,2018ApJ...869..150M}.
\redc{In the $z\sim5$ panel, the red open star and circles are number density estimates of $z\sim5$ faint BL AGNs detected with NIRSpec in \citet{2023ApJ...954L...4K} and \citet{2023arXiv230801230M}, respectively, while the red crosses are estimates for BL AGNs found in NIRCam Grism observations in \citet{2023arXiv230605448M}.
The orange open circles at $z\sim4$ and $6$ are number densities of red and compact AGN candidates in \citet{2023arXiv230607320L}.
The results of \citet{2023arXiv230605448M} and \citet{2023arXiv230607320L} are horizontally shifted by $-0.1$ dex for visualization purposes.}
The blue pentagon and circle at $z\sim7$ are lower limits for a red quasar \citep{2022Natur.604..261F} and the radio-selected AGN \citep{2022arXiv220600018E}, respectively, whose number densities are estimated in \citet{2022Natur.604..261F}.
The gray symbols show the number densities of galaxies (diamond: \citealt{2022ApJS..259...20H}, square: \citealt{2021AJ....162...47B}, triangle: \citealt{2015MNRAS.452.1817B,2017MNRAS.466.3612B}, circle: \citealt{2015ApJ...810...71F}).
The cyan solid and dashed lines are the best-fit double power-law functions based on the X-ray selected AGNs in  \citet{2018MNRAS.474.2904P} and \citet{2019ApJ...884...19G}, respectively, and the blue lines are the best-fit functions and their extrapolations based on the quasars \citep{2018PASJ...70S..34A,2020ApJ...904...89N,2018ApJ...869..150M}.
The number densities of our AGNs newly identified in the NIRSpec spectra are higher than the extrapolations of the quasar luminosity function. 
}
\label{fig_UVLF}
\end{figure*}

\begin{figure}
\centering
\begin{center}
\includegraphics[width=0.99\hsize, bb=4 4 389 350]{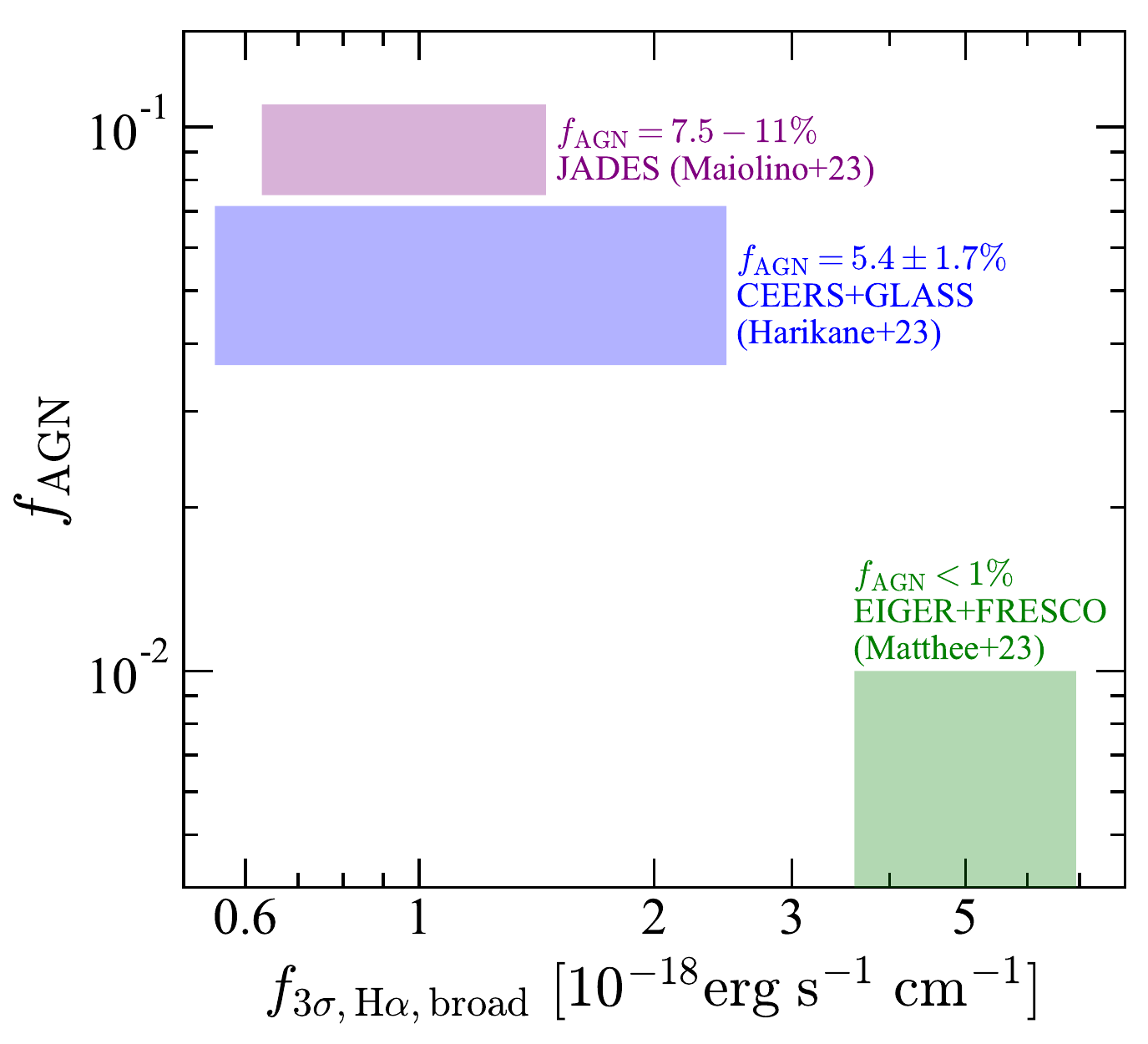}
\end{center}
\caption{
\redc{Broad-line AGN fraction and $3\sigma$ sensitivity for an H$\alpha$ broad emission line in each study.
The sensitivity is for a broad-line width of $\m{FWHM}=2000\ \m{km\ s^{-1}}$.
The purple, blue, and green shaded regions indicate AGN fractions and the 16th$-$84th percentile of the broad-line sensitivities for reported AGNs in JADES \citep{2023arXiv230801230M}, CEERS+GLASS (this study), and EIGER+FRESCO \citep{2023arXiv230605448M}, respectively.
The low AGN fraction from NIRCam Grism observations in \citet{2023arXiv230605448M} is due to the shallower depth compared to NIRSpec observations in this study and \citet{2023arXiv230801230M}.}
}
\label{fig_fAGN_ef}
\end{figure}

Figure \ref{fig_UVLF} shows the calculated number densities of our AGNs at $z\sim4-7$ with previous measurements for AGNs and galaxies.
Tables \ref{tab_nd} and \ref{tab_LFpar} present our number density estimates and parameters of the best-fit double power-law function, 
\begin{align}
&\Phi(M_\m{UV})=\frac{\m{ln}10}{2.5}\phi^*\notag\\
&\times\left[10^{0.4(\alpha+1)(M_\m{UV}-M_\m{UV}^*)}+10^{0.4(\beta+1)(M_\m{UV}-M_\m{UV}^*)}\right]^{-1}.
\end{align}
The number density of our broad-line type-1 AGNs detected with NIRSpec is higher than an extrapolation of the quasar luminosity functions \citep{2018PASJ...70S..34A,2020ApJ...904...89N,2018ApJ...869..150M}, because these quasar studies only select compact objects.
The best estimates of the number densities are higher than those of the X-ray selected AGNs \citep{2018MNRAS.474.2904P,2019ApJ...884...19G}, probably due to the limited sensitivity in X-ray observations.
\redc{Our best estimates are higher than \citet{2023arXiv230605448M} and lower than \citet{2023arXiv230801230M}, because of the difference in sensitivities and selection functions.
As shown in Figure \ref{fig_fAGN_ef}, the typical sensitivity of NIRCam Grism spectra in \citet{2023arXiv230605448M} is lower than those of the NIRSpec data in this study and \citet{2023arXiv230801230M}, resulting in the low AGN fraction in \citet{2023arXiv230605448M}.
The slightly higher AGN fraction in \citet{2023arXiv230801230M} compared to this study may be due to the slightly higher typical sensitivity in the JADES data, or a biased selection to AGN candidates in the NIRSpec MSA target selection as discussed in \citet{2023arXiv230801230M}.
}

\section{Discussion}\label{ss_dis}

\begin{figure*}
\centering
\begin{center}
\begin{minipage}{0.49\hsize}
\begin{center}
\includegraphics[width=0.9\hsize, bb=10 8 425 354]{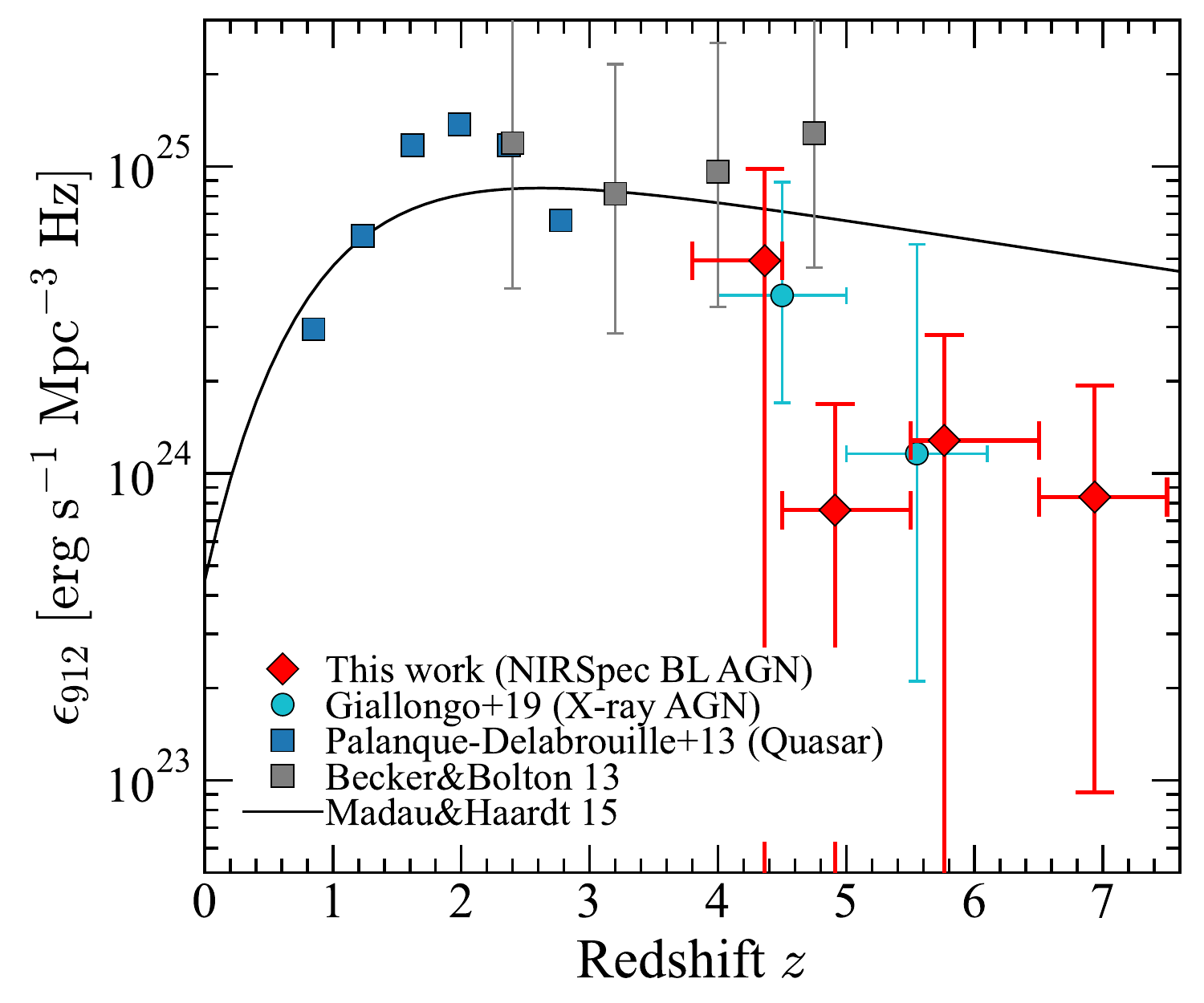}
\end{center}
\end{minipage}
\begin{minipage}{0.49\hsize}
\begin{center}
\includegraphics[width=0.9\hsize, bb=6 8 425 354]{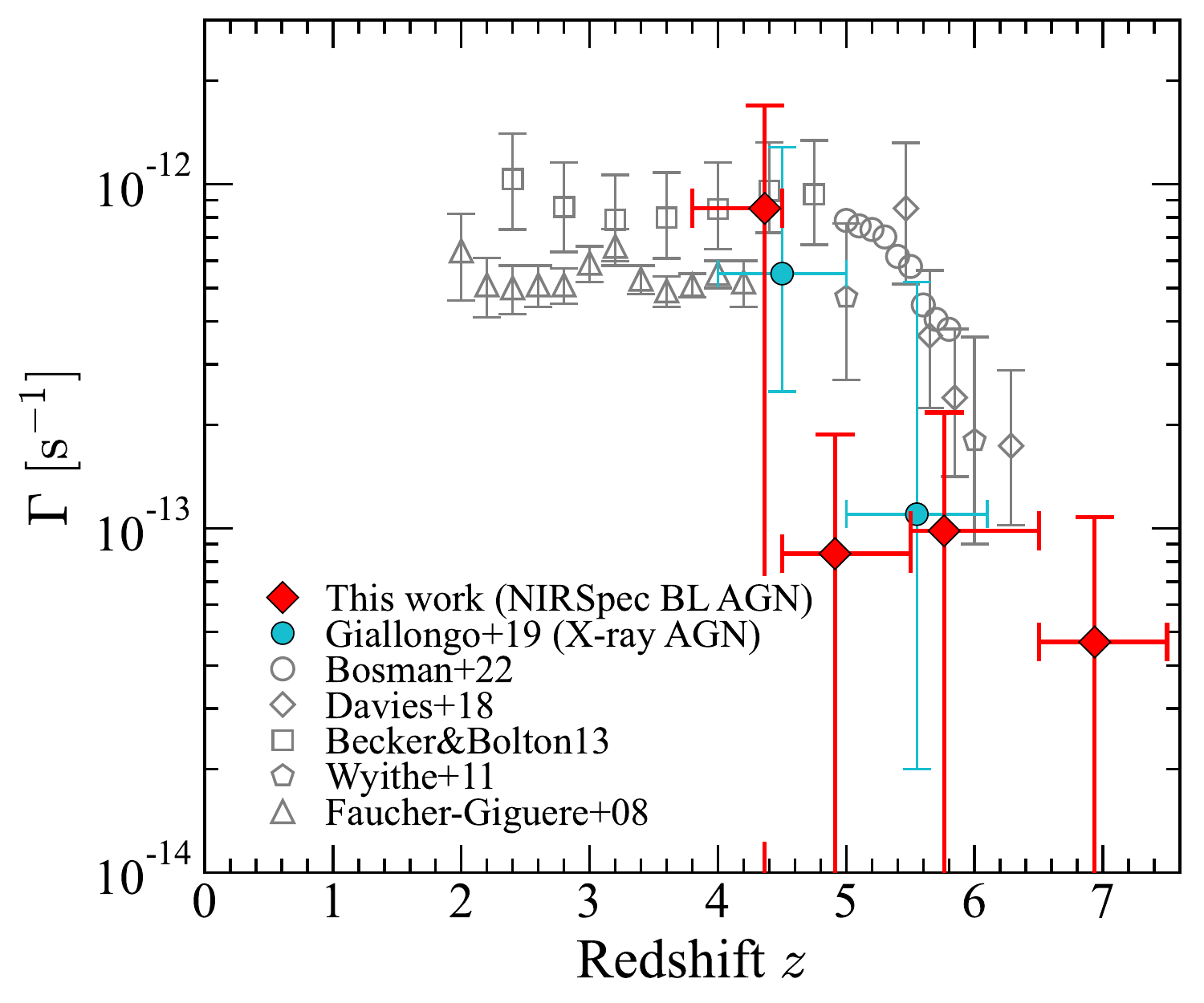}
\end{center}
\end{minipage}
\end{center}
\caption{
(Left:) Cosmic ionizing emissivity density, $\epsilon_{912}$, contributed by AGNs as a function of redshift.
The red diamonds show estimates for our AGNs identified with the NIRSpec spectra assuming the escape fraction of $50\%$.
The cyan circles and blue squares are measurements from X-ray AGNs \citep{2019ApJ...884...19G} and quasars \citep{2013A&A...551A..29P}, respectively.
The gray squares are measurements in \citet{2013MNRAS.436.1023B}, and the black curve is the model in \citet{2015ApJ...813L...8M}.
(Right:) Cosmic photoionization rate, $\Gamma$, a function of redshift.
The red diamonds show estimates for our AGNs, and the cyan circles are measurements in \citet{2019ApJ...884...19G}.
The gray open symbols show measurements from the Ly$\alpha$ forest analysis in quasar spectra (circle: \citealt{2022MNRAS.514...55B}, diamond: \citealt{2018ApJ...855..106D}, square: \citealt{2013MNRAS.436.1023B}, pentagon: \citealt{2011MNRAS.412.1926W}, triangle: \citealt{2008ApJ...688...85F}).
}
\label{fig_e_g}
\end{figure*}

\subsection{Contribution to Cosmic Reionization}

To discuss the contribution of our AGNs to cosmic reionization, we estimate the cosmic ionizing emissivity density, $\epsilon_{912}$, and the cosmic photoionization rate, $\Gamma$, following \citet{2019ApJ...884...19G}.
First, we estimate the rest-frame UV emissivity at $1450\ \m{\AA}$, $\epsilon_{1450}$, by integrating the UV luminosity function at each redshift down to $M_\m{UV}=-18$ mag, the same limit as \citet{2019ApJ...884...19G}.
Since an AGN does not dominate all of the rest-frame UV light as discussed in Section \ref{ss_mor}, we multiply the UV luminosity density with the fraction of light from an AGN to all light including both an AGN and its host galaxy, which is estimated to be $\sim50\%$ \redc{in rest-frame UV (1450 \AA)} based on the decomposition analysis in Section \ref{ss_agnhost}.
Then we calculate $\epsilon_{912}$ from $\epsilon_{1450}$ using the SED presented in \citet{2015MNRAS.449.4204L} and an ionizing photon escape fraction.
The escape fraction of faint AGNs is highly uncertain, with a variety of the values reported ($f_\m{esc}=0.3-0.8$) by previous studies \citep[e.g.,][]{2017MNRAS.465..302M,2018A&A...613A..44G,2019A&A...632A..45R,2022MNRAS.509.1820I}.
In this calculation, we assume $f_\m{esc}=0.5$.
The photoionization rate, $\Gamma$, is estimated from $\epsilon_{912}$ using an equation presented in \citet{2015MNRAS.449.4204L}.
We increase the AGN emissivity by a factor of 1.2 to include the contribution by radiative recombination in the IGM following \citet{2018MNRAS.473..560D}.
Calculated emissivities and photoionization rates are summarized in Table \ref{tab_LFpar}.

Figure \ref{fig_e_g} presents the cosmic ionizing emissivity density and the photoionization rate from our AGNs with previous measurements.
The emissivities of our AGNs at $z\sim4-6$ are comparable to those of X-ray selected AGNs in \citet{2019ApJ...884...19G}, although our AGNs selected with broad emission lines may not be the same population from X-ray selected AGNs.
The cosmic photoionization rates from our AGNs are also comparable to those of X-ray-selected AGNs.
At $z\sim6$, the photoionization rate of our AGNs is lower than the measurement from the Ly$\alpha$ forest analysis \citep[e.g.,][]{2022MNRAS.514...55B,2018ApJ...855..106D,2011MNRAS.412.1926W}.
This indicates that such faint AGNs contribute to cosmic reionization, while the total contribution is not large, up to $\sim50\%$ at $z\sim6$, because of their faint nature.

\subsection{Nature of the NIRSpec-Detected AGNs and Redshift Evolution}\label{ss_dis_high}

Our broad-line type-1 AGNs have low-mass black holes with $M_\m{BH}\sim10^6-10^8\ M_\odot$.
Given that more than half of the AGNs show extended morphologies, the total lights are partly dominated by lights from their host galaxies rather than those of the AGNs, suggesting that these AGNs are type-1 Seyfert galaxies.
Some of them show weak broad H$\alpha$ emission lines with a broad-to-narrow-line luminosity ratio of $L_\m{broad}/L_\m{narrow}\sim0.2-0.6$, suggesting that they are intermediate-type AGNs such as type 1.5, 1.8, and 1.9 Seyfert galaxies \citep{1981ApJ...249..462O,1990agn..conf..161B,1992ApJS...79...49W}.
Indeed, such a weak broad component observed in our AGNs is similarly seen in some Seyfert 1.5-1.9 galaxies \citep[e.g.,][their Figure 1]{2012MNRAS.423..600S}.
The moderate line width of the broad component ($<2000\ \m{km\ s^{-1}}$) in some of our AGNs are comparable to those of narrow-line Seyfert 1 galaxies.

Although our AGNs are similar to intermediate-type AGNs seen in local universe, the fraction of such AGNs may increase toward higher redshifts.
Our analysis indicates that about 5\% of the galaxies at $z\sim4-7$ harbor faint type-1 AGNs, while studies of local AGNs implies that only 1-2\% of galaxies with similar bolometric luminosities ($L_\m{bol}\sim10^{44}-10^{46}\ \m{erg\ s^{-1}}$, or $L_{6166}\sim10^{43}-10^{44}\ \m{erg\ s^{-1}}$) are type-1 AGNs \citep{2012MNRAS.423..600S}.
Spectroscopic studies for $z\sim3$ Lyman break galaxies indicate that $\sim1\%$ of them are type-1 AGNs.
Although a larger sample of broad-line AGNs at $z>4$ is needed, our study finding the high fraction ($\sim5\%$) implies a possible redshift evolution of the type-1 AGN fraction.

Some of our AGNs show weak (but statistically significant) broad H$\alpha$ emission lines compared to their narrow components with low broad-to-narrow line flux ratios, which are detected in the deep NIRspec spectra but will be missed in shallow spectroscopic observations.
Although the broad emission lines are detected, the low broad-to-narrow line flux ratios suggest that these AGNs may be \redc{significantly contributed from {\sc Hii} regions in their host galaxies, or} partly obscured, implying a high obscured fraction in the universe at $z>4$.
Indeed, several X-ray observations indicate a higher fraction of obscured AGNs in the higher redshift universe at $z\sim0-4$ \citep[e.g.,][]{2008A&A...490..905H,2014ApJ...786..104U,2017ApJS..232....8L,2018ApJ...854...33Z,2020A&A...639A..51I,2023ApJ...943..162P}.
From recent ALMA observations, \citet{2022A&A...666A..17G} predict that AGNs at $z>3$ are significantly obscured by their host galaxies' interstellar medium, which may reasonably explain the observed high obscured fraction at high redshifts.
This indicates there may be many obscured AGNs (type-2 AGNs) in the $z>4$ universe possibly more than the type-1 AGNs we have identified, consistent with recent findings of heavily obscured AGNs at $z\sim7$ \citep{2022Natur.604..261F,2022arXiv220600018E}.

\section{Summary}\label{ss_summary}
In this paper, we conduct a systematic search for broad-line emission in the JWST/NIRSpec deep spectra of a total of 185 galaxies at $z_\m{spec}=3.8-8.9$.
We identify 10 faint type-1 AGNs at $z_\m{spec}=4.015-6.936$, two of which are reported in \citet{2023ApJ...954L...4K}, allowing us to conduct the first statistical census of such faint AGNs in the early universe.
Our major findings are summarized below.

\begin{enumerate}
\item
Our faint AGNs show a broad emission line with $\mathrm{FWHM}\simeq1000-6000\ \mathrm{km\ s^{-1}}$ only in the permitted H$\alpha$ line (Figures \ref{fig_spec_line_1}-\ref{fig_spec_line_3}).
Their forbidden {\sc[Oiii]}$\lambda$5007 line is detected with a higher signal-to-noise ratio than H$\alpha$, but show only narrow emission with $\m{FWHM}\lesssim500\ \m{km\ s^{-1}}$.
These spectral features are consistent with the fact that these objects are type-1 AGNs.
The broad emission in some objects is very weak with a broad line-to-narrow line ratio of $L_\m{H\alpha,broad}/L_\m{H\alpha,narrow}=0.2-0.6$, comparable to those seen in local intermediate-type AGNs.

\item 
High-resolution JWST/NIRCam and HST/ACS and WFC3 images reveal that seven AGNs in our sample show extended morphologies, indicating significant contributions to the total lights from their host galaxies (Figure \ref{fig_snapshot}).
The remaining three objects are dominated by compact emission consistent with the PSF, and two of them show red SEDs implying that they are dusty compact AGNs (Figures \ref{fig_EBV} and \ref{fig_SED}).

\item
Our selected AGNs show narrow emission line ratios of high {\sc[Oiii]}/H$\beta$ and low {\sc[Nii]}/H$\alpha$, similar to star-forming galaxies at $z\gtrsim4$.
These AGNs cannot be distinguished from normal star-forming galaxies with the classical BPT diagram (Figure \ref{fig_BPT}), probably due to their low metallicities.

\item
Our faint AGNs have low-mass black holes with $M_\mathrm{BH}\sim10^6-10^8\ M_\odot$, remarkably lower than those of low-luminosity quasars previously identified at $z>4$ with ground-based telescopes (Figure \ref{fig_MBH_Lbol}).
Our AGNs at $z=4-7$ show higher bolometric luminosities than AGNs at $z\sim0$ with similar black hole masses on average, possibly due to the selection bias.
Deeper NIRSpec spectroscopy is needed to confirm this trend.

\item 
We estimate the stellar masses of AGN's host galaxies with careful AGN-host decomposition analyses.
The estimated stellar masses are systematically lower than the black hole mass-stellar mass relation at $z\sim0$ (Figure \ref{fig_MBH_Ms}).
The lower stellar mass (higher black hole mass) at high redshifts implies fast black hole growth, which is consistent with predictions from theoretical simulations.

\item 
We estimate that $\sim5\%$ of galaxies at $z=4-7$ are type-1 AGNs on average, whose fraction is higher than that in local galaxies with similar luminosities.
The estimated number density of our faint AGNs is higher than an extrapolation of the quasar UV luminosity functions and is comparable to those of X-ray selected AGNs (Figure \ref{fig_UVLF}).
Estimates of the ionizing emissivity and photoionization rate by the faint AGNs indicate that such faint AGNs contribute to cosmic reionization, while the total contribution is not large, up to $\sim50\%$ at $z\sim6$, because of their faint nature (Figure \ref{fig_e_g}). 

\end{enumerate}
This study demonstrates that NIRSpec spectroscopy is an efficient and powerful tool to find faint type-1 AGNs with low mass black holes that are embedded in their host galaxies.
Future large NIRSpec observations will uncover a large number of high redshift faint AGNs, allowing us to investigate the galaxy-SMBH co-evolution in the early universe. 

\acknowledgments
\redc{We thank the anonymous referee for careful reading and valuable comments that improved the clarity of the paper.}
We are grateful to Takuma Izumi, Roberto Maiolino, Ryan Sanders, and John Silverman for useful discussions.
This work is based on observations made with the NASA/ESA/CSA James Webb Space Telescope. The data were obtained from the Mikulski Archive for Space Telescopes at the Space Telescope Science Institute, which is operated by the Association of Universities for Research in Astronomy, Inc., under NASA contract NAS 5-03127 for JWST. 
These observations are associated with programs ERS-1324 (GLASS), ERS-1345 (CEERS), GO-2561 (UNCOVER), and ERO-2736.
The authors acknowledge the teams of JWST commissioning, ERO, GLASS, UNCOVER, and CEERS for developing their observing programs with a zero-exclusive-access period.
\redc{The JWST data presented in this paper were obtained from the Mikulski Archive for Space Telescopes (MAST) at the Space Telescope Science Institute. The specific observations analyzed can be accessed via \dataset[10.17909/qp74-bk09]{https://doi.org/10.17909/qp74-bk09}.}
This publication is based upon work supported by the World Premier International Research Center Initiative (WPI Initiative), MEXT, Japan, KAKENHI (20H00180, 21J20785, 21K13953, 21H04467) through Japan Society for the Promotion of Science, and JSPS Core-to-Core Program (grant number: JPJSCCA20210003).
This work was supported by the joint research program of the Institute for Cosmic Ray Research (ICRR), University of Tokyo, and a grant from the Hayakawa Satio Fund awarded by the Astronomical Society of Japan.\\

\software{Galfit \citep{2002AJ....124..266P}, Prospector \citep{2021ApJS..254...22J}, SExtractor \citep{1996A&AS..117..393B}}

\bibliographystyle{apj}
\bibliography{apj-jour,reference}

\end{document}